\documentclass[aps,twocolumn,prc,superscriptaddress,showpacs,nofootinbib,floatfix,amssymb,amsfonts,amsmath]{revtex4-2}
\pdfoutput=1
\UseRawInputEncoding

\usepackage{graphicx}
\usepackage{dcolumn}
\usepackage{bm}
\usepackage{xcolor}
\usepackage{amsmath}    
\usepackage{amsfonts}   
\usepackage{amssymb}
\usepackage{graphicx}   

\begin{document}
\title{Charge radii in covariant density functional theory: a global view}

\author{U.\ C.\  Perera}
\affiliation{Department of Physics and Astronomy, Mississippi
State University, MS 39762}

\author{A.\ V.\ Afanasjev}
\affiliation{Department of Physics and Astronomy, Mississippi
State University, MS 39762}

\author{P. Ring}
\affiliation{Fakult\"at f\"ur Physik, Technische Universit\"at M\"unchen,
D-85748 Garching, Germany}

\date{\today}

\begin{abstract}

  A systematic global investigation of differential charge radii has been performed
within the CDFT framework for the first time. Theoretical results obtained with  conventional
covariant energy density functionals and the separable pairing interaction of Ref.\ \cite{TMR.09}
are compared with experimental differential charge radii  in the regions of the nuclear chart in which
available experimental data crosses the neutron  shell closures at $N = 28, 50, 82$ and 126.  The
analysis of absolute differential radii of different isotopic chains and their relative properties
indicate clearly that such properties are reasonably well described in model calculations in the
cases when the mean-field approximation is justified. However, while the observed clusterization
of differential charge radii of different isotopic chains is well described above the $N=50$ and
$N=126$ shell closures, it is more difficult to reproduce it above the $N=28$ and $N=82$ shell
closures because of possible deficiencies in the underlying single-particle structure. The impact of
the latter has been evaluated for spherical shapes and it was shown that the relative energies of
the single-particle states and the patterns of their occupation with increasing neutron number
have an appreciable impact on the evolution of the $\delta \left < r^2 \right>^{N,N'}$ values. These
factors also limit the predictive power of model calculations in the regions of high densities of the
single-particle states of different origin. It is shown that the kinks in the charge radii at neutron shell
closures are due to the underlying single-particle structure and due to weakening or collapse of pairing at
these closures. The regions of the nuclear chart in which the correlations beyond mean-field are expected 
to have an impact on charge radii are indicated; the analysis shows that the assignment of a calculated 
excited prolate minimum to the experimental ground state allows to understand the trends of the evolution 
of differential charge radii with neutron number in many cases of shape coexistence even at the mean-field 
level. It is usually assumed that pairing is a dominant contributor to odd-even staggering (OES) in charge radii. 
Our analysis paints a more complicated picture. It suggests a new mechanism in which the fragmentation of 
the single-particle content of the ground state in odd-mass nuclei due to particle-vibration coupling provides 
a significant contribution to OES in charge radii.

\end{abstract}

\maketitle

\section{Introduction}
\label{Introduc}

Together with nuclear masses, the charge radii are among the most fundamental properties 
of atomic nuclei.  The essential information on the saturation density of symmetric nuclear 
matter is imprinted into them. They also depend on the properties of nuclear forces and 
nuclear many-body dynamics.

Significant experimental efforts have been dedicated
over the decades for the measurement of charge radii \cite{AM.13,CMP.16}.
While the changes of the charge radii within the isotopic chain are measured with high precision
using laser spectroscopy, the situation with the measurements of
absolute values of root-mean-square ($\it rms$) charge radii $r_{ch}$ is less satisfactory because of lower precision
of their determination in muonic spectra and electronic scattering experiments
and the impossibility of such experiments in radioactive elements
\cite{AM.13,CMP.16}.  For example, for nuclei with proton number  $Z > 83$, (uranium is the
exception), there are no experimental data for the absolute nuclear charge
radii \cite{AM.13}. Theoretical calculations within different density functional
theories (DFTs) provide a quite accurate global description of experimental
charge radii presented in the compilation of Ref.\ \cite{AM.13}: the rms deviations
of calculated $r_{ch}$ from experimental ones are at the level of $\approx 0.03$ fm
\cite{AARR.14}. Considering that the average experimental rms charge radius in
the nuclear chart is around 4.8 fm (see, for example, Fig. 23 in Ref.\ \cite{AARR.14} and
Figs. 2-4 in Ref.\ \cite{AM.13}), this amounts to high average precision of 0.625\% in the
prediction of charge radii. However, this information has to be taken with a grain of salt
because of the issues mentioned above with the measurements of absolute values of rms charge
radii and some reliance on interpolation/extrapolation procedures in the compilation
of Ref.\ \cite{AM.13}.

Thus, the differential mean-square (ms) charge radii (see Eq.\ (\ref{diff-radii}) below for definition), 
measured with high precision within the isotopic chains, become an important quantity.
The evolution of the charge radii within the isotopic chain with increasing neutron
number is defined by the pull on the proton states generated by neutrons gradually
added to the nuclear system. This is, in reality, a quite complicated and, in some cases,
contra-intuitive process. Here the strong nuclear symmetry energy acts to increase the
overlap between all the proton states and the overall nuclear density. This effect is
expected to be enhanced when the overlap between the wave functions is maximal.
The most investigated case here is the kink in charge radii at $N=126$ and the evolution
of charge radii above $N=126$ in the Pb isotopic chain. The pattern of these effects
critically depends on the occupation of the $2g_{9/2}$ and $1i_{11/2}$ orbitals, on
their relative energies, and on how close they are in energy \cite{RF.95,GSR.13}. Both in relativistic
and non-relativistic density functional theories (DFT), the single-particle rms neutron radius
of the $2g_{9/2}$ orbital  is larger than that of the $1i_{11/2}$ orbital (by $\approx 0.6$ fm
[see Table \ref{table-sp-radii} below] and  $\approx 1.0$ fm [see Ref.\ \cite{RF.95}],
respectively). Intuitively (for example, by using liquid drop model concepts), the  occupation
of the neutron $2g_{9/2}$ orbital would bring a larger proton radius as compared with the
occupation of the $1i_{11/2}$ orbital. However, the DFT calculations showed an opposite trend \cite{RF.95,GSR.13}
the deep microscopic origin of which has been found only in Ref.\ \cite{GSR.13}.
It is traced back to a nodal structure\footnote{Note that the nodal structure of the wavefunctions
plays an important role not only in the evolution of charge radii of spherical nuclei. In extremely
deformed nuclei, it also defines the necessary conditions for $\alpha$ clusterization in very light
nuclei, and its suppression with the increase of mass number \cite{AA.18}.  Moreover, the nodal
structure of the deformed wavefunctions allows us to understand the coexistence
of ellipsoidal mean-field-type structures and nuclear molecules
at similar excitation energies and the features of particle-hole
excitations connecting these two types of structures \cite{AA.18,A-epj-wc.18}.}
of these two orbitals ($n=1$ for $1i_{11/2}$ and $n=2$ for  $2g_{9/2}$, where $n$ stands for
principal quantum number; see also Fig. 6.2 in Ref.\ \cite{NilRag-book} for a pattern
of respective wavefunctions). The principal quantum
number of the neutron $1i_{11/2}$ orbital is the same as for the majority of occupied
proton orbitals (including deeply bound ones). This leads to a large overlap of their
wave functions and thus provides a large pull of these neutron states on proton
orbitals via the symmetry energy and allows the reproduction of the kink in charge
radii at $N=126$.

  A significant amount of experimental data on charge radii has been collected over
the years: the review \cite{AM.13} provides a compilation of such data measured by
early 2011\footnote{The experimental data shown in the present paper is based on
this compilation supplemented when available with more recent data.}.
In recent years an explosion of high-quality measurements of charge radii is observed
(see, for example, review \cite{CMP.16}).  They cover, for example, in recent years the
K \cite{K-radii.21},  Ca \cite{Ca-radii.2016,Ca-radii-prot-rich.19},
Cu \cite{Cu-radii.20}, Cd \cite{Cd-isiotopes-radii.18}, Sn
\cite{Gorges-Sn-radii.19,Sn-radii-Yordanov.20}, Hg \cite{Pb-Hg-charge-radii-PRL.21},
Bi \cite{Bi-radii.18},  At \cite{At-radii.19},   Ac \cite{Ac-radii.19} and No \cite{No-radii.20}
isotopic chains.   These experimental studies are
supplemented mostly by a theoretical analysis within non-relativistic Skyrme or Fayans
DFTs (see overview below) and occasionally by the analysis within Gogny
DFT (the Ca isotopes in Ref.\ \cite{Ca-radii.2016}),
 the CDFT (the Pb and Hg isotopes in Ref.\ \cite{Pb-Hg-charge-radii-PRL.21} and No isotopes
in Ref.\ \cite{ADFAA.20}) or  non-relativistic {\it ab-initio}
calculations (see Refs.\ \cite{Ca-radii.2016,K-radii.21}).

They reveal several interesting features.  The most familiar are the kinks in
charge radii at neutron numbers corresponding to shell closures and the odd-even staggering
(OES)  in charge radii.  In charge radii, a shell closure is observed as a sudden increase  in
the rate of the change of  charge radius of the isotopes just beyond magic shell closure; this
leads to the kinks in charge radii which are well known at $N=28, 50, 82$ and 126
\cite{AM.13,Ca-radii.2016,Gorges-Sn-radii.19,GV.20}. In addition,  the analysis of experimental
data presented in Ref.\ \cite{GV.20} reveals a puzzling feature related to similar slopes of
differential charge radii  $\delta \left< r^2 \right>$ as a function of neutron number above the
neutron shell closures for different isotopic chains (see Fig.\ 5 in Ref.\ \cite{GV.20} for even-
and odd-$Z$ isotopic chains and Figs.\ \ref{Pb-region-delta-r2}(a), \ref{Sn-region-delta-r2}(a),
\ref{Sr-region-delta-r2}(a) and \ref{Ca-region-delta-r2}(a) below for only even-$Z$ isotope chains).
On the contrary, all  these isotopic chains have different slopes of differential charge radii for
neutron numbers below the shell closures. To our knowledge, this observed feature,  which exists
in the Ca, Sr, Sn, and Pb regions, has not been addressed in a systematic theoretical analysis so
far.

Different theoretical approaches have been used with different degrees of success
to describe the evolution of charge radii in various isotopic chains. The initial focus of
such studies was the kink in differential charge radii of the Pb isotopes at $N=126$ but later
studies covered also other  isotopic chains.  The calculations in non-relativistic density functional
theories (NR-DFTs) based on conventional energy density functionals (EDFs) were unable to
reproduce the kink in the Pb isotopes \cite{TBFHW.93,Gorges-Sn-radii.19}. On the contrary,
this kink is quite successfully reproduced in covariant DFT (CDFT) for all employed covariant
energy density functionals (CEDFs) \cite{SLR.93,Sharma1995_PRL74-3744,Pb-Hg-charge-radii-PRL.21}.  
As discussed above the relative energies of the $1i_{11/2}$ and $2g_{9/2}$ neutron orbitals play a crucial
role in this difference between NR-DFT and CDFT results in the Pb isotopes. Two possible
ways of resolving the problem emerged in the NR-DFTs. The first one includes a modification
of spin-orbit interaction either by its fitting to CDFT results \cite{RF.95} or by introducing a
density dependence in the spin-orbit interaction \cite{Nakada.15}.
The second approach, introduced by 
Fayans et al.  \cite{FTTZ.94,FZ.96,FTTZ.00} includes an extension of non-relativistic functionals by 
adding gradient terms into the surface part and the pairing interaction. Although this approach has 
been reasonably successful  (especially after fitting the Fayans functionals to differential charge 
radii data in Ref.\ \cite{RN.17}) in the description of the evolution of charge radii
\cite{FZ.96,FTTZ.00,RN.17,Ca-radii.2016,Gorges-Sn-radii.19,Cd-isiotopes-radii.18,Cu-radii.20},
the microscopic origin of these gradient terms is not clear.  It was stated that the pairing
functional of the Fayans model is supposed to effectively account for the coupling to surface
vibrations \cite{RN.17}.  However, this contradicts to the observation that such a coupling is
quite small for the ground states in even-even nuclei in the Sn and Pb regions
(see discussion in Sec.\ \ref{sec-OES} below)  so that the charge radii of such nuclei are
not expected to be strongly modified by it.

 Alternative DFT approaches are based either on non-relativistic finite range Gogny
functionals,  a non-relativistic Yukawa interaction or on CEDFs. The Gogny DFT studies of 
differential charge radii are less frequent than those based either on the Skyrme or Fayans EDFs. 
For example, the isotopic evolution of charge radii in even-even and even-odd  Sr, Zr, and Mo 
isotopes with $N=47-68$ and the impact of triaxiality on charge radii of even-even Mo isotopes 
with $N=62, 64$, and 66 have been investigated in Ref.\ \cite{RSRP.10}. The differential charge 
radii of the $^{52,48}$Ca isotopes were studied at and beyond mean-field level with the Gogny 
functional D1S  in Ref.\ \cite{Ca-radii.2016}.
The non-relativistic HFB approach with a  finite-range Yukawa interaction \cite{Nakada.20} and 
density-dependent spin-orbit interaction has been successfully applied for the description of 
differential charge radii in spherical nuclei of the Ca, Ni, Sn and Pb isotopic chains 
\cite{Nakada.15,NI.15,Nakada.19}.  

The studies of differential charge radii are also rare within the CDFT\footnote{This is despite
the fact that global studies of charge radii of even-even nuclei located between the two-proton and
two-neutron drip lines with the assessment of systematic theoretical uncertainties have been performed
in Refs.\ \cite{AARR.14,AA.16} and tabulated values of charge radii are publicly available in
the supplemental material for Ref.\ \cite{AARR.14} (for the DD-PC1 functional) and at MassExplorer
\cite{Mass-Explorer} (for the DD-PC1, DD-ME2, DD-ME$\delta$ and NL3* CEDFs).}. The first-ever
[among any type of DFT] successful description of the kink in charge radii of the lead  isotopes has
been achieved in Ref.\ \cite{SLR.93} for CDFT with the NL-SH and NL1 functionals. This work
was followed by a number of the studies of differential charge radii in spherical even-even
nuclei in the Ca, Sn, and Pb isotopic chains in Refs.\ \cite{RF.95,UCP.16,DD-PCX}. The
odd-even staggering and  the kink in charge radii of the Pb and Hg isotopes has been
successfully described recently in Ref.\ \cite{Pb-Hg-charge-radii-PRL.21} using the DD-ME2 CEDF.
An ansatz for charge radii in CDFT has been suggested in Ref.\  \cite{AGZ.20}; it adds
the phenomenological term $a_0/\sqrt{A} |\sum\limits_{k>0}^n u_k v_k - \sum\limits_{k>0}^p u_k v_k|$
to the definition of charge radii without affecting the definitions of other physical observables.
Although it can describe the charge radii  and their OES in selected isotopic chains,
its physical meaning is not clear and it does not appear in conventional DFTs.

The goal of the present paper is to perform detailed studies of differential charge
radii within the CDFT framework in order to understand specific facts, such as the underlying
single-particle structure and the role of pairing, affecting the theoretical description of the
evolution of charge radii with the neutron number, the presence and magnitude of the kinks
and OES in the isotopic chains. Note that we employ conventional functionals which do
not use any new fit parameters nor the above mentioned modifications. The aim  is to understand to which extent they
can provide a satisfactory description of differential charge radii in the regions of the nuclear
chart in which available experimental data crosses shell closures at $N=28$, 50, 82
and 126. One should note that conventional relativistic functionals provide, amongst many other nuclear properties,
a reasonable description of rotational bands, which are sensitive to pairing, not only in even-even but also in
odd-$A$ nuclei (see Refs.\ \cite{AO.13,ZHA.20}). Thus, they provide access to OES in the
moments of inertia, the physical mechanism of which [blocking in odd-$A$ nuclei] is similar
to the one partially responsible for OES in charge radii and in binding energies. At present,
it is not clear whether the inclusion of gradient terms into the pairing functional will preserve this
feature. An additional goal of the present paper is to search for alternative physical mechanisms
affecting differential charge radii and OES.

    The paper is organized as follows.  A brief outline of the theory is given
in Sec.\ \ref{Theory}.  Charge radii and related indicators are discussed
in Sec.\ \ref{Indicators}.  The charge radii of the Pb isotopes are used in Sec.\
\ref{Pb-isotopes} as a testing ground to evaluate the importance of the underlying
single-particle structure and pairing in the evolution of differential charge radii
between the two-proton and two-neutron drip lines, in the appearance of the kinks
at shell closures and for the comparison of the results of calculations with
and without pairing. Charge radii, their evolution
with neutron number, the sources of the discrepancies between theory
and experiment, absolute and relative properties of differential charge radii
in different isotopic chains in the Pb, Sn/Gd, Sr and Ca regions are discussed
in detail in Secs.\ \ref{Pb-region-section}, \ref{Sn-region-section},
\ref{Sr-region-section} and \ref{Ca-region-section}, respectively.  Note that the
discussion in these sections is restricted to even-even nuclei. Sec.\ \ref{sec-OES}
is dedicated  to the analysis of odd-even staggering in charge radii and its origin.
A new mechanism of OES, relying on the fragmentation of the single-particle states
in odd-$N$ nuclei due to particle-vibration coupling, is suggested for the first time
in this section. Finally, Sec.\ \ref{Concl} summarizes the results of our paper.

\section{Theoretical framework}
\label{Theory}

In the present paper, the relativistic Hartree-Bogoliubov (RHB) approach is
used in the calculations.  The formalism of this approach is discussed in
detail in Refs.\ \cite{VALR.05,AARR.14}. The calculations are performed
with computer codes that preserve either spherical or axial symmetry.
The former code has been considerably modified to allow for fully self-consistent
calculations of the ground and excited states in odd-$A$ spherical nuclei;
in was applied for the first time in Ref.\ \cite{Pb-Hg-charge-radii-PRL.21}.
The axial code employed in Ref.\ \cite{AARR.14} has been used here for
calculations which include deformation.

To assess the dependence of the results on the underlying single-particle
structure,  several state-of-the-art covariant energy density functionals  (CEDFs)  such
as NL3* \cite{NL3*}, DD-PC1 \cite{DD-PC1}, DD-ME2 \cite{DD-ME2}, DD-ME$\delta$
\cite{DD-MEdelta} and PC-PK1 \cite{PC-PK1} are used in the present paper. They
represent the major classes of CEDFs and their global  performance in describing
ground state properties such as masses and charge radii of even-even nuclei has been
tested in Refs.\ \cite{AARR.14,AA.16,ZNLYM.14}. Note that many of the results
on charge radii and the deformations of the ground states employed in the present 
analysis are taken from Refs.\ \cite{AARR.14,Mass-Explorer}. This allows us to test the 
predictive power of the models with respect to the description of differential charge radii.

The separable pairing interaction of finite range, introduced as a simplification of the Gogny
pairing by Tian et al in Ref.\ \cite{TMR.09}, is used in the present paper.  Its matrix elements
in  $r$-space have the form
\begin{eqnarray}
\label{Eq:TMR}
V({\bm r}_1,{\bm r}_2,{\bm r}_1',{\bm r}_2') &=& \nonumber \\
= - f\,G \delta({\bm R}-&\bm{R'}&)P(r) P(r') \frac{1}{2}(1-P^{\sigma})
\end{eqnarray}
with ${\bm R}=({\bm r}_1+{\bm r}_2)/2$ and ${\bm r}={\bm r}_1-{\bm r}_2$
being the center of mass and relative coordinates.
The form factor $P(r)$ is of Gaussian shape
\begin{eqnarray}
P(r)=\frac{1}{(4 \pi a^2)^{3/2}}e^{-r^2/4a^2}
\end{eqnarray}
The parameters $G=728$ MeV fm$^3$ and $a=0.644$ fm of this interaction,
which  are the same for protons and neutrons,
have been derived by a mapping of the $^1$S$_0$ pairing gap of infinite
nuclear matter to that of the Gogny force D1S \cite{TMR.09} under the
condition that $f=1.0$. The particle number dependence of  the scaling
factor $f$  of the pairing force  is taken from Ref.\ \cite{AARR.14}.

  The proton quadrupole deformation $\beta_2$ is defined from proton 
quadrupole moment $Q_{20}$ as 
\begin{eqnarray}
\beta_2 = \frac{\sqrt{5 \pi} \,Q_{20}}{3ZR_0^2}
\label{beta2_def} 
\end{eqnarray}
where 
\begin{eqnarray}
Q_{20} &=& \int d^3r \rho({\bm r})\,(2z^2-r^2_\perp) \label{Q_20_def}
\end{eqnarray}
with $r^2_\perp=x^2+y^2$. Here $R_0=1.2 A^{1/3}$ and   $\rho({\bm r})$
stands for proton density.  Eq.\ (\ref{beta2_def}) is used also
in the extraction of experimental $\beta_2$ deformation parameters from
measured data \cite{RMMNS.87}. This justifies its application
despite the fact that this simple linear expression
ignores the contributions of higher power/multipolarity
deformations to the proton quadrupole moment. Including higher powers of
$\beta_2$, as in Ref.~\cite{NZ-def}, yields values of $\beta_2$
that are $\approx 10$\% lower.

\begin{figure}[htb]
\centering
\includegraphics[width=8.4cm]{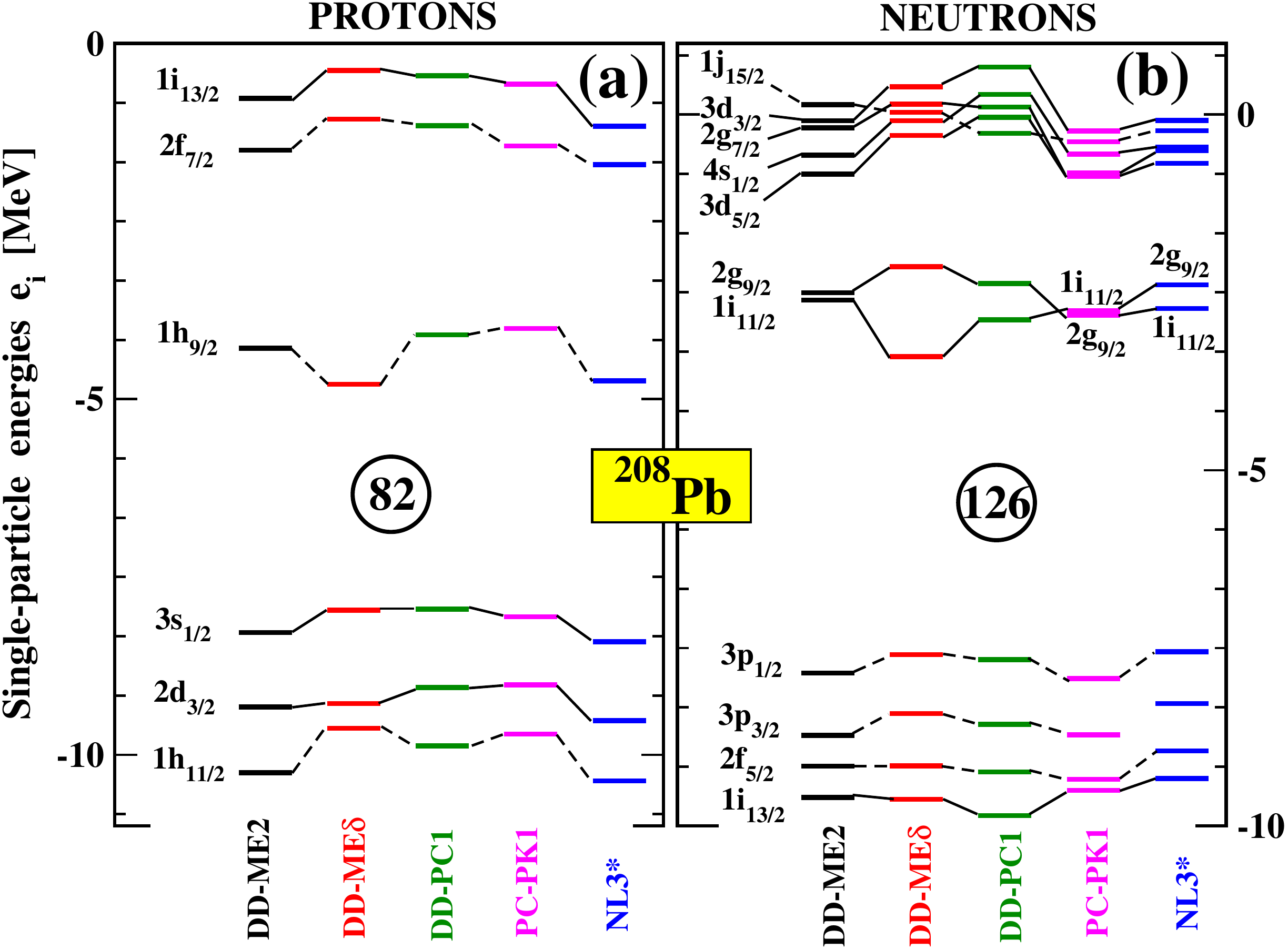}
\caption{Neutron and proton single-particle states at spherical shape in $^{208}$Pb
obtained in the calculations without pairing with the indicated CEDFs. Solid and dashed
connecting lines are used for positive- and negative-parity states. Spherical gaps are
indicated.
\label{208Pb-single-particle}
}
\end{figure}

\section{Charge radii and related indicators}
\label{Indicators}

\begin{figure}[htb]
\centering
\includegraphics[width=8.4cm]{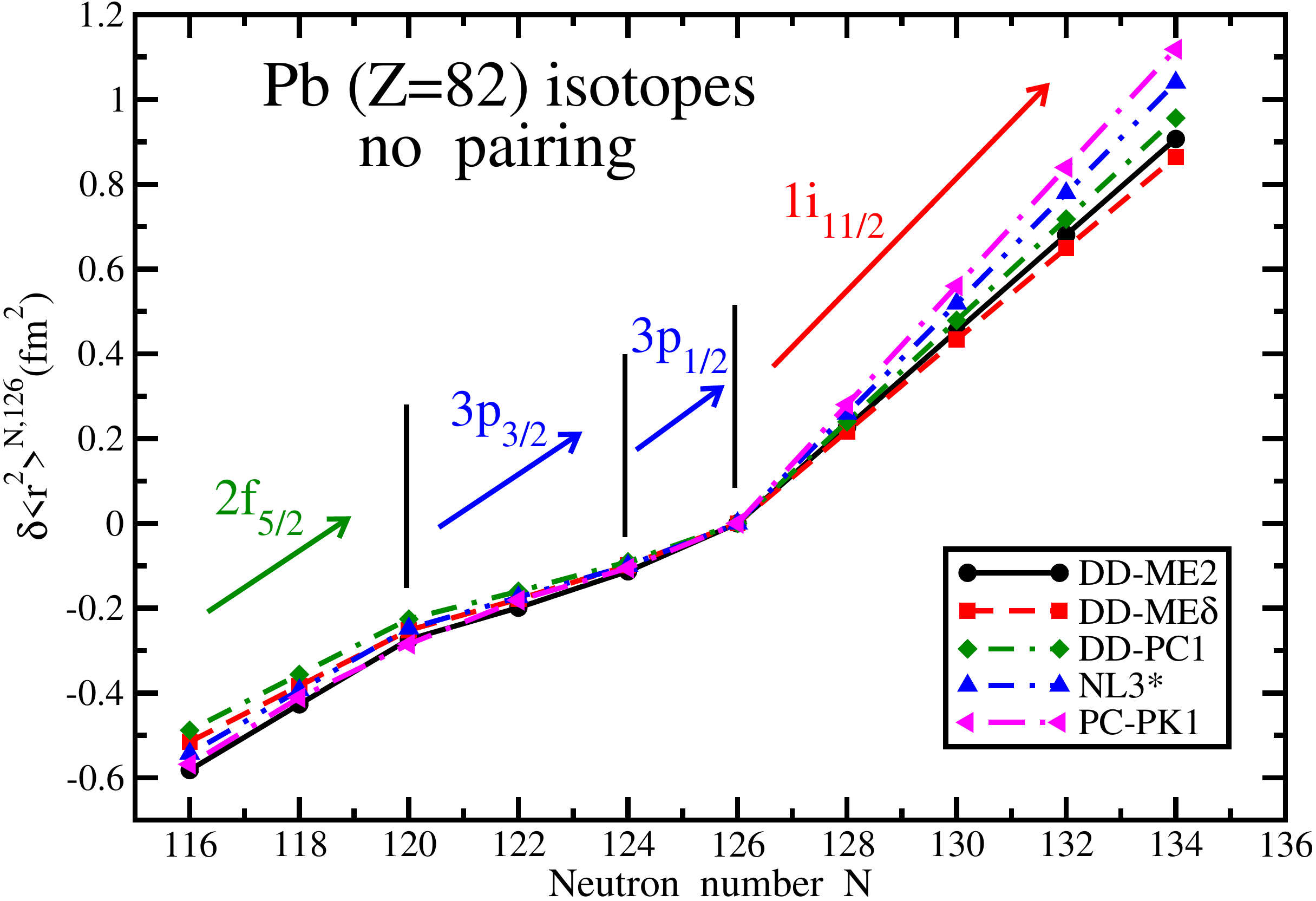}
\caption{The $\delta \left < r^2 \right>^{N,126}$ values of the Pb
isotopes relatively to $^{208}$Pb obtained in the calculations without
pairing with the indicated CEDFs. Vertical black lines indicate the neutron
numbers at which the slope of the $\delta \left < r^2 \right>^{N,N'}$
curves change. The spherical subshell labels indicate the orbitals
which are populated with increasing neutron number between these
vertical lines (see text for details). Note that the presentation is
restricted to the range of the nuclei in which experimental ground
states of the Pb isotopes are spherical \cite{Pb-Hg-charge-radii-PRL.21}.
\label{Delta-r2-Pb-even}
}
\end{figure}

The charge radii were calculated from the corresponding point proton
radii as
\begin{equation}
r_{ch} = \sqrt{<r^2>_p + 0.64}\,\,\,\, {\rm fm}
\label{r_charge}
\end{equation}
where $<r^2>_p$ stands for proton mean square point radius
and the factor 0.64 accounts for the finite-size effects of the
proton. Here we have neglected the small contributions to the charge
radius originating from the electric neutron form factor
and the electromagnetic spin-orbit coupling \cite{BFHN.72,NS.87}
as well as the corrections due to the center of mass motion.
Note that the functional DD-PC1~\cite{DD-PC1} has been adjusted 
only to nuclear binding energies.

  In addition, two differential indicators are commonly used  to
facilitate the quantitative comparison of the experimental results with those
from theoretical calculations. One of them is the differential mean-square
charge radius\footnote{This quantity is frequently written as a function of
mass number $A$. However, we prefer to define it as a function of neutron
number $N$ since this allows to see the behavior of the
$\delta \left < r^2 \right>_p^{N,N'}$ curves at neutron shell  closures.}
\begin{eqnarray}
\delta \left < r^2 \right>_p^{N,N'} &=& \left < r^2 \right>_p(N)
                                                          - \left < r^2 \right>_p(N') = \nonumber \\
                                                      && =r^2_{ch}(N) - r^2_{ch}(N')
\label{diff-radii}
\end{eqnarray}
where $N'$ is the neutron number of the reference nucleus. Another is the
three-point indicator
\begin{eqnarray}
\Delta \langle r^2 \rangle^{(3)}(N) && = \nonumber \\
= && \frac{1}{2} \left[ \langle r^2(N-1) \rangle +
\langle r^2(N+1) \rangle - 2  \langle r^2(N) \rangle \right] = \nonumber \\
= && \frac{1}{2} \left[ r_{ch}^2(N-1) + r_{ch}^2(N+1) - 2  r_{ch}^2(N) \right]
\label{Delta-3}
\end{eqnarray}
which quantifies OES in charge radii.

  In addition, the neutron skin thickness is commonly defined as the difference
of proton and neutron root-mean-square (rms) radii
\begin{eqnarray}
r_{\rm skin}=<r^2_n>^{1/2}-<r_p^2>^{1/2}.
\end{eqnarray}
The neutron skin thickness is an important indicator of isovector
properties.

\section{The Pb isotopes: from unpaired to paired results}
\label{Pb-isotopes}

For a better understanding of the physical features which affect the description
of charge radii it is very illustrative to start from the analysis of the results of the
calculations performed  without pairing but with different CEDFs representing
the major classes of the CDFT models.
They provide comparable global descriptions of the ground state properties
\cite{AARR.14,AANR.15,ZNLYM.14} but reveal visible differences in the
single-particle properties (see, for example, Fig.\ \ref{208Pb-single-particle}).
The addition of pairing will reveal how it affects the detailed properties of
differential charge radii.

The experimental absolute value of the charge radius of the nucleus $^{208}$Pb is well described
by the employed functionals (see Table \ref{table-ch-radius}), but there exist some
uncertainties in the prediction of the neutron skin in the model calculations and in
its experimental measurements. One can see that 
non-PREX\footnote{Different types of non-PREX experiments
are discussed in Refs.\ \cite{TSCD.12,RP.18} and references quoted therein. Note that the 
experimental data on the neutron skin are extracted in a model-dependent way in all these 
experiments. For instance, the neutron skin thicknesses $r_{skin} = 0.161 \pm 0.042$ fm 
\cite{KPVH.13} and $r_{skin} = 0.190 \pm 0.028$ fm \cite{Anti-an.2013} obtained from the energy
of the anti-analog giant dipole resonance rely on relativistic proton-neutron
quasiparticle random-phase approximation calculations based
on the RHB model. Another example is the value of the neutron
skin thickness of $r_{skin} = 0.15 \pm  0.03$(stat)$^{+0.01}_{-0.03}$(sys) fm
extracted from coherent pion photoproduction cross sections \cite{Pion-rskin}. 
In this case the extraction of information on the nucleon density distribution depends 
on the comparison of the measured $(\gamma,\pi^0)$ cross sections with model 
calculations. On the contrary, the electroweak probe (PREX types of experiment)
has the advantage over experiments using hadronic probes that it allows a nearly 
model-independent extraction of the neutron radius that is independent of most 
strong interaction uncertainties \cite{H.98}.}
 experiments provide neutron skins which are by $\approx 0.09$ fm smaller than the one obtained 
in the PREX-II experiment.

\begin{table}[h!]
\begin{center}
\caption{The charge radius $r_{ch}$ and the neutron skin $r_{skin}$ of the nucleus $^{208}$Pb
obtained in calculations with the indicated functionals. The experimental value of $r_{ch}$
is taken from Ref.\ \cite{AM.13}.  Two experimental values are  provided for $r_{skin}$: one
(approximate, labelled as non-PREX) obtained from  the experiments which do not employ
parity violating electron scattering on nuclei (PREX) (see discussion in
Sect. X of Ref.\ \cite{AARR.14} for more details) and another (labelled as PREX-II)
from the PREX-II  experiment \cite{PREX-II.21}.
\label{table-ch-radius}
}
\begin{tabular}{l|c|c} \hline \hline
  CEDF & $r_{ch}$ [fm]& $r_{skin}$ [fm] \\ \hline  			
DDME2             & 5.518 & 0.193 \\
DDME$\delta$  & 5.509 & 0.186 \\
DD-PC1             & 5.513 & 0.202 \\
NL3*                 & 5.509 & 0.288 \\
PCPK1             & 5.519 & 0.257 \\
exper.               &  $5.5012\pm 0.0013$  & $\approx 0.19$ [non-PREX]   \\
                         &                                     &  $0.283\pm0.071$ [PREX-II]  \\
			\hline	\hline
\end{tabular}
\end{center}
\end{table}

\begin{figure}[htb]
\centering
\includegraphics[width=8.4cm]{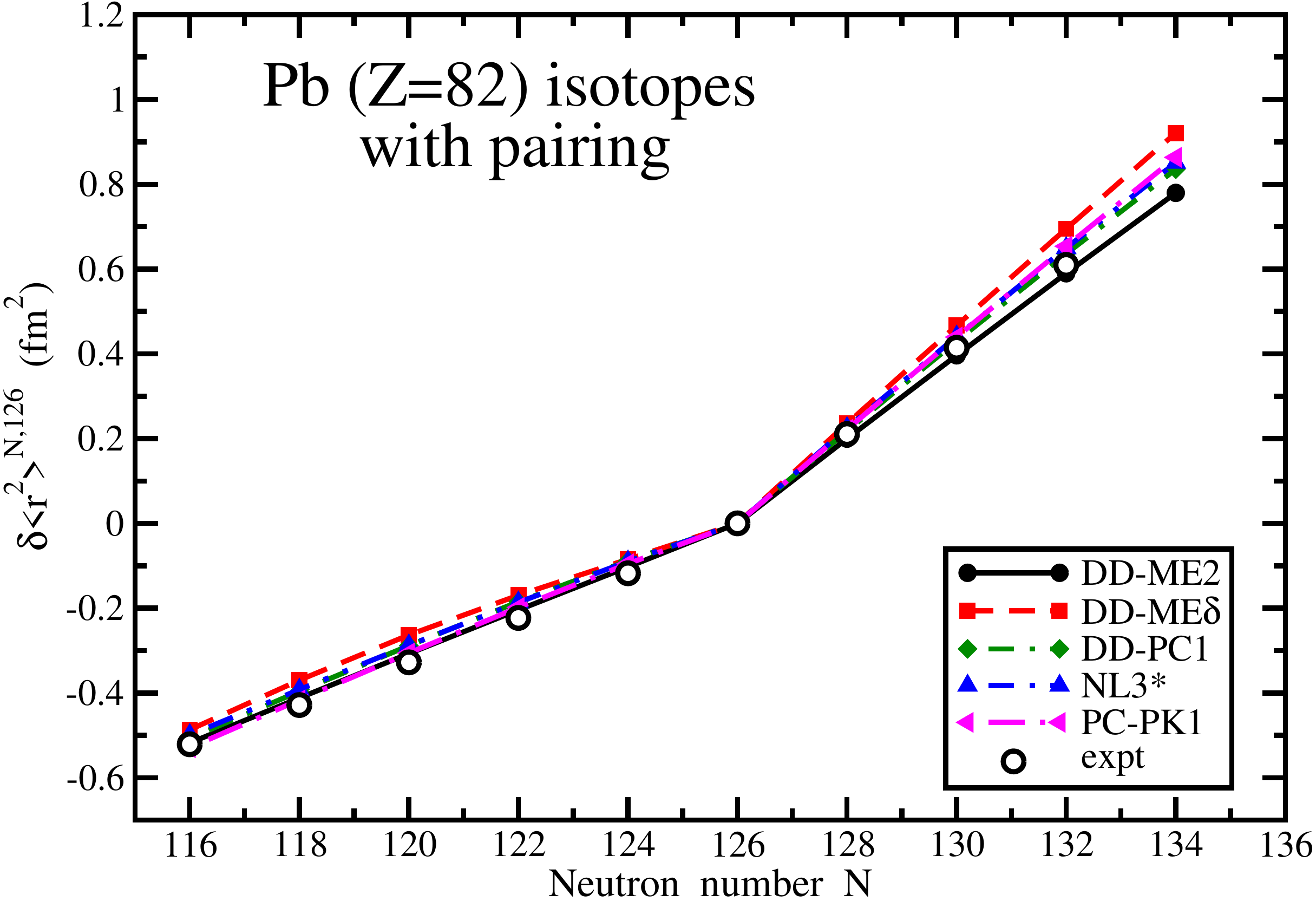}
\caption{The same as Fig.\ \ref{Delta-r2-Pb-even} but for the results of
calculations with pairing included. The experimental data are taken
from Ref.\ \cite{Pb-Hg-charge-radii-PRL.21}.}
\label{Delta-r2-Pb-even-pairing}
\end{figure}

\begin{figure}[htb]
\centering
\includegraphics[width=8.4cm]{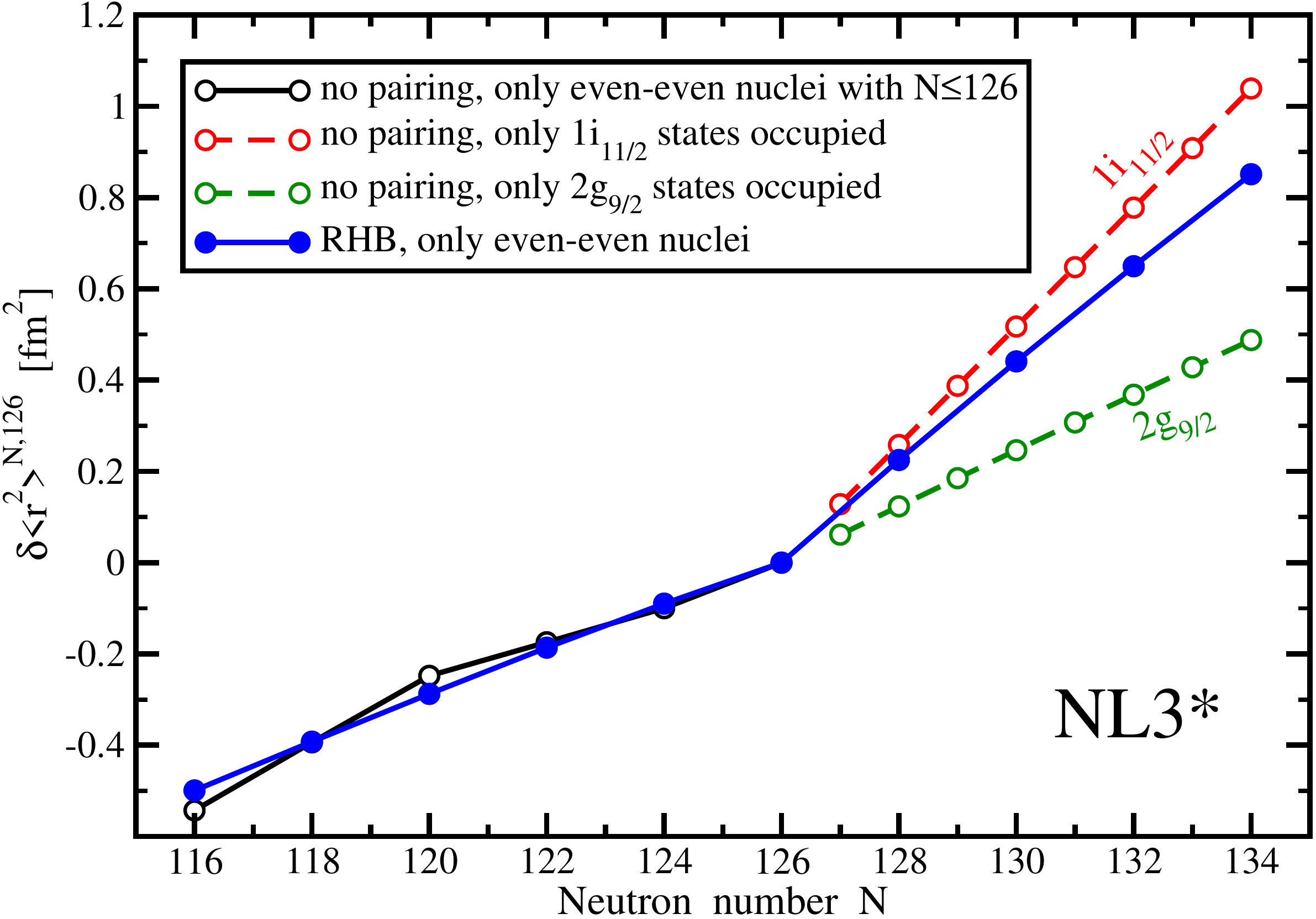}
\caption{The $\delta \left < r^2 \right>^{N,126}$ values of Pb isotopes
relatively to $^{208}$Pb obtained in calculations without pairing with
the NL3* CEDF (see text for details). The results of RHB calculations
with pairing for even-even nuclei are presented for comparison.}
\label{Delta-r2-Pb-no-pairing-NL3*}
\end{figure}

   The $\delta \left < r^2 \right>^{N,126}$ values of the Pb
isotopes obtained with various CEDFs in calculations without pairing
are shown in Fig.\ \ref{Delta-r2-Pb-even}. One can see that the
slope of this function (namely, the derivative $\delta \left < r^2 \right>^{N,126}/\delta N$)
changes at $N=120$, $N=124$  and $N=126$ in all functionals.  The changes in the
slope of $\delta \left < r^2 \right>^{N,126}$ as a function of $N$ are traced back to
the changes in the occupation of different spherical neutron subshells. The sequence
of the occupation of different spherical subshells with increasing neutron number is the
same for all functionals (see Figs.\ \ref{Delta-r2-Pb-even} and \ref{208Pb-single-particle}).
The $\nu 2f_{5/2}$ subshell is occupied for neutron numbers  $N=115 - 120$\footnote{In
the calculations without pairing, the occupation of either an odd neutron (in odd-$A$
nuclei) or a pair of neutrons (in even-even nuclei) from the same spherical subshell
leads  to the same slope of  the $\delta \left < r^2 \right>^{N,126}$ function. Thus, for
simplicity we consider only even-even nuclei in this part of the discussion.}. For higher
neutron numbers $N=121-124$, the $\nu 3p_{3/2}$ subshell gradually fills
up with increasing neutron number. This change of the occupation from the $\nu 2f_{5/2}$
to the $\nu 3p_{3/2}$ subshell leads to the change of the slope of the
$\delta \left < r^2 \right>^{N,126}$ function  at $N=120$ because these two subshells have
different density distributions and thus  different neutron radii. The next change
of the slope of this function takes place at $N=124$ at the transition from
the occupation  of the $\nu 3p_{3/2}$ to the $\nu 3p_{1/2}$  subshell.
Since these two subshells are spin-orbit partners, they have the same orbital
quantum number and, as a consequence, very similar spatial distributions
of the density. Minor differences in the latter are caused by the fact that
the  $\nu 3p_{1/2}$ subshell is located somewhat higher in energy than
the $\nu 3p_{3/2}$ one [see Fig.\ \ref{208Pb-single-particle}(b)] and thus in
the region of a somewhat larger radius of the nucleonic potential.

\begin{figure}[htb]
\centering
\includegraphics[width=8.4cm]{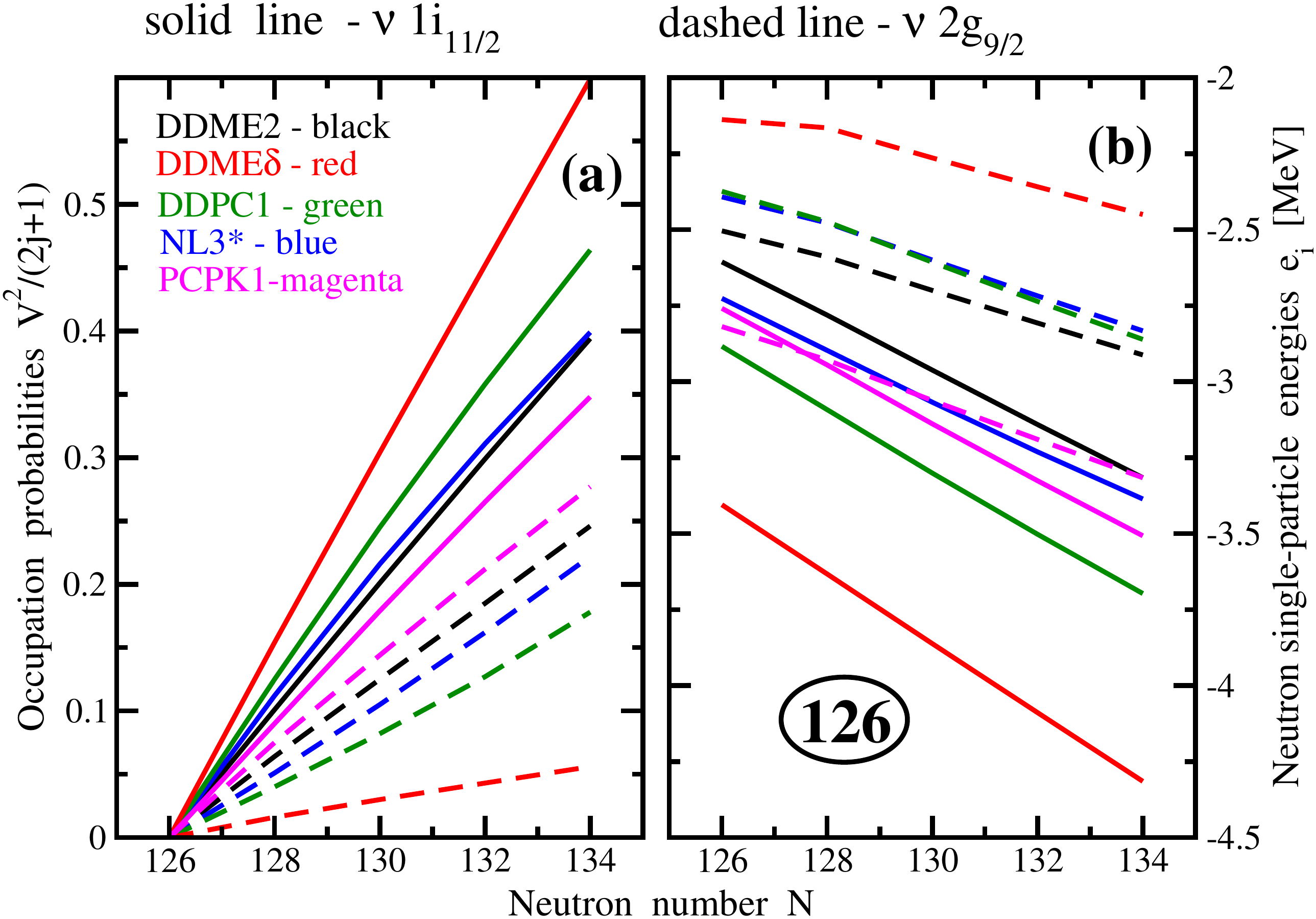}
\caption{(a) The evolution of occupation probabilities $v^2/(2j+1)$ of the
neutron $2g_{9/2}$ and $1i_{11/2}$ orbitals as a function of neutron number in
the $N>126$ nuclei. (b) The evolution of the energies of these
single-particle states as a function of neutron number.}
\label{Pb-vv2-sp-energies}
\end{figure}

 The inclusion of pairing modifies the results visibly as compared to those obtained in 
 the calculations without pairing (compare Fig.\ \ref{Delta-r2-Pb-even-pairing}
with Fig.\ \ref{Delta-r2-Pb-even}). First, the  changes of the slopes  of the
$\delta \left < r^2 \right>^{N,126}$ curves at $N=120$ and $N=124$, which are
present in the calculations without pairing, are almost washed out when pairing
is taken into account. This is because pairing modifies the occupation of different
subshells (see, for example, Fig.\ \ref{Pb-vv2-sp-energies}) and thus the evolution of
the $\delta \left < r^2 \right>^{N,126}$ values with neutron number
becomes more gradual at these particle numbers. Second, the
kink in the  $\delta \left < r^2 \right>^{N,126}$ values at $N=126$ still exists because
of the large shell closure at this particle number which leads to the collapse of pairing.
If (hypothetically) pairing would survive at $N=126$, the kink
would be less pronounced.  Third, the spreads and absolute magnitudes in the
predictions of the $\delta \left < r^2 \right>^{116,126}$ and
$\delta \left < r^2 \right>^{134,126}$ values (the values taken at the extremes
of the plots presented in Figs.\ \ref{Delta-r2-Pb-even} and
\ref{Delta-r2-Pb-even-pairing})  are reduced when the pairing is taken into
account.  These values are located in the ranges from $-0.58$ fm$^2$ to
$-0.48$ fm$^2$ (from $-0.51$ fm$^2$ to $-0.49$ fm$^2$)
 and from $0.86$ fm$^2$ to $1.12$ fm$^2$ (from $0.78$ fm$^2$ to
$0.92$ fm$^2$) in the calculations without (with) pairing, respectively.
Fourth, the relative order of the results obtained with different functionals
in Figs.\ \ref{Delta-r2-Pb-even} and  \ref{Delta-r2-Pb-even-pairing} at given
neutron number is different in the calculations with and without pairing.
This is best illustrated by considering the case of $N=134$ and
the sequence of functionals ordered according to the increase of
calculated $\delta \left < r^2 \right>^{N,126}$ values. The sequences of
functionals are DD-ME$\delta$, DD-ME2, DD-PC1, NL3* and PC-PK1
in the calculations without pairing (see Fig.\ \ref{Delta-r2-Pb-no-pairing-NL3*})
and  DD-ME2, DD-PC1, NL3*, PC-PK1 and DD-ME$\delta$ in the calculations
with pairing (see Fig.\ \ref{Delta-r2-Pb-even-pairing}).

    To better understand these features we analyze the results of
the calculations presented in Figs. \ref{Delta-r2-Pb-no-pairing-NL3*}  and
\ref{Pb-vv2-sp-energies}. The calculations without pairing clearly show that
the occupation of the $\nu 1i_{11/2}$ subshell for $N>126$ is  needed for
the formation of the experimentally observed kink at $N=126$ and that the
occupation of the $\nu 2g_{9/2}$ subshell above N=126 does not lead to
the formation of a kink at $N=126$ (see Fig.\ \ref{Delta-r2-Pb-no-pairing-NL3*}).
The inclusion of pairing leads to a partial occupation of these two subshells
(see Fig.\ \ref{Pb-vv2-sp-energies}) and thus to $\delta \left < r^2 \right>^{N,126}$
values located in between of those obtained in the calculations without pairing
for the occupation of these two subshells (see Fig.\ \ref{Delta-r2-Pb-no-pairing-NL3*}).

  Fig.\ \ref{Delta-r2-Pb-no-pairing-NL3*} also illustrates that the sequential
occupation of a given subshell (either $\nu 1i_{11/2}$ or $\nu 2g_{9/2}$) above
$N=126$ in odd-$A$ and even-even nuclei leads to  $\delta \left < r^2 \right>^{N,126}$
values that form a straight line as a function of neutron number. Thus, with this
occupation pattern, the OES of charge radii cannot be formed in the calculations
without pairing. However, as discussed in detail in Ref.\ \cite{Pb-Hg-charge-radii-PRL.21}
the scattering of  the occupation of the orbitals in these subshells will lead to the
formation of an OES in the charge radii which has a magnitude comparable to
experiment.

   Fig.\ \ref{Pb-vv2-sp-energies} allows to better understand the role
of pairing and the impact of the underlying single-particle structure on the magnitude
of the kink in the charge radii at $N=126$. This figure focuses on the occupation
pattern and the relative energies of the neutron $2g_{9/2}$ and $1i_{11/2}$ orbitals
located above $N=126$. Other states (such as $3d_{5/2}$, $4s_{1/2}$, $2g_{7/2}$
etc) do not play a significant role in the creation of the kink since they are separated
by a large energy gap from the pair of the states under study (see Fig.\
\ref{208Pb-single-particle}(b)) and their occupation in the presence of pairing
is small.  The occupation probability $v^2/(2j+1)$  of the respective subshell is defined in
such a way that it is equal to 1 or 0 when a given subshell of multiplicity $2j+1$ is either
fully occupied or empty.  This occupation probability grows almost linearly with
increasing neutron number [see Fig.\ \ref{Pb-vv2-sp-energies}(a)].

The largest energy gap between the neutron $2g_{9/2}$ and $1i_{11/2}$
subshells exists  in the DD-ME$\delta$ functional for all neutron numbers of interest (see Fig.\
\ref{Pb-vv2-sp-energies}(b)).  As a consequence, for a given neutron number the occupation
of the lowest (highest) in energy $1i_{11/2}$ ($2g_{9/2}$) subshell is large (small) but they
gradually raise with increasing neutron number (see  Fig.\ \ref{Pb-vv2-sp-energies}(a)).
This significant preference in the occupation of the $1i_{11/2}$ subshell leads to the largest
$\delta \left < r^2 \right>^{N,126}/\delta N$ values for $N>126$ isotopes among the considered
functionals which exceeds the experimental data (see Fig.\ \ref{Delta-r2-Pb-even-pairing}).
In all other functionals, the gap between  the $2g_{9/2}$ and $1i_{11/2}$ subshells is
smaller (see Fig.\ \ref{Pb-vv2-sp-energies}(b)) but still the occupation of the $1i_{11/2}$
subshell is favored\footnote{It is only in the PC-PK1 functional that the $2g_{9/2}$ subshell
is lower in energy than the $1i_{11/2}$ one for $N=126$ [see Fig.\ \ref{208Pb-single-particle}(b)].
However, already  at $N=128$ the $1i_{11/2}$ subshell dives below
$2g_{9/2}$ [see Fig.\ \ref{Pb-vv2-sp-energies}(b)].}. Thus, as compared with the DD-ME$\delta$
functional the difference in the occupation of these orbitals becomes smaller  [see Fig.\
\ref{Pb-vv2-sp-energies}(a)]. This leads to a reduction of the
$\delta \left < r^2 \right>^{N,126}/\delta N$ values for $N>126$ nuclei which now become
close to experiment (see Fig.\ \ref{Delta-r2-Pb-even-pairing}).

\begin{figure}[htb]
\centering
\includegraphics[width=8.4cm]{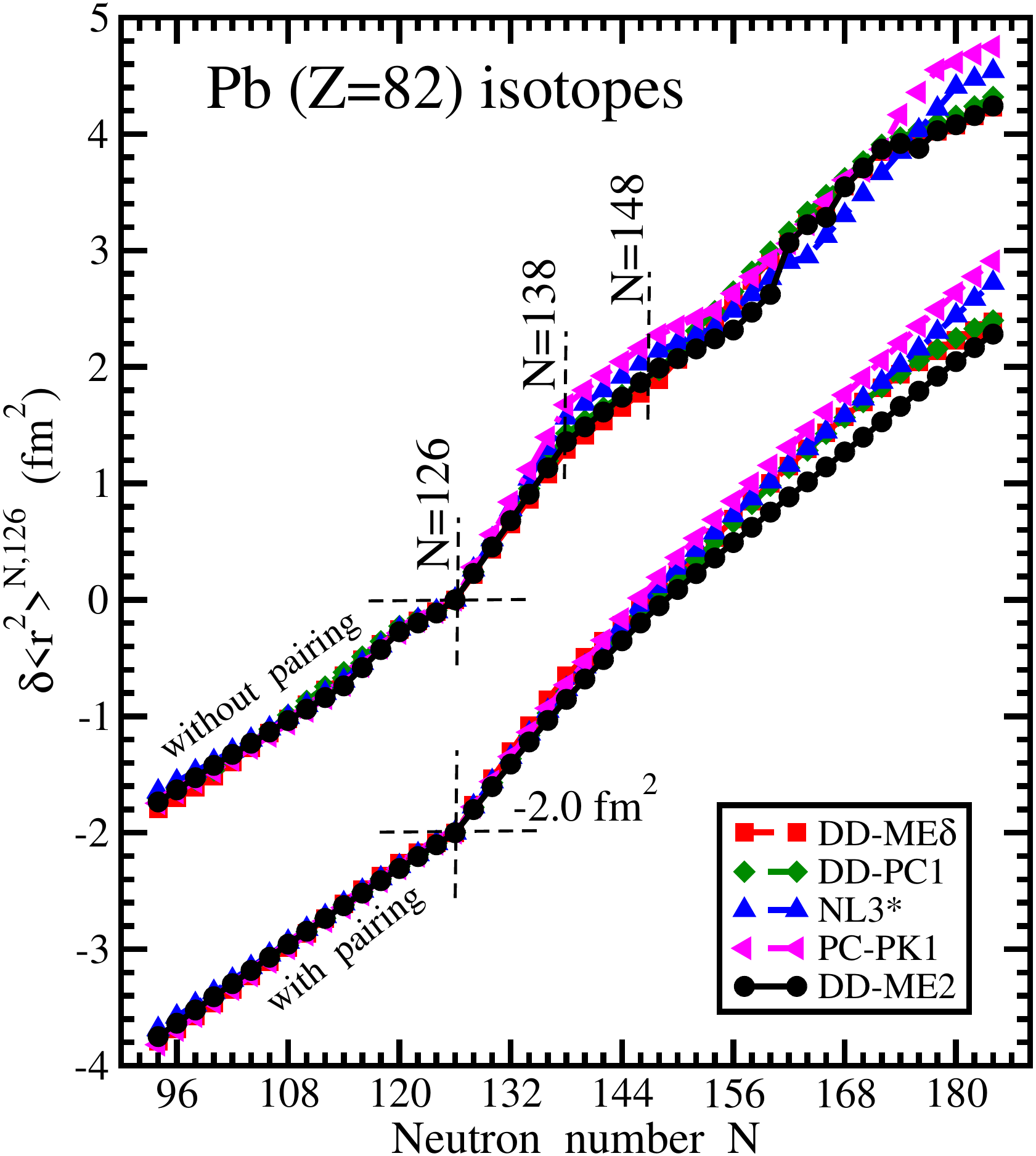}
\caption{The same as Figs.\ \ref{Delta-r2-Pb-even} and  \ref{Delta-r2-Pb-even-pairing},
but for all even-even Pb isotopes located between the two-proton and two-neutron drip lines.
Note that the results of the calculations with pairing are shifted down by $-2.0$ fm$^2$ in
order to compare the results of the calculations with and without pairing on the same panel.}
\label{Pb-full-chain}
\end{figure}

There are large similarities between the results obtained with the different
CEDFs presented in Fig.\ \ref{Delta-r2-Pb-even-pairing}. This is the consequence of
the fact that in all functionals (i) different single-particle subshells are well separated
in energy below $N=126$ (see Fig.\ \ref{208Pb-single-particle})  and (ii) the sequence
of the single-particle subshells  occupied with increasing neutron number is the same
(see Table \ref{table-filling-shells-Pb}). In order to see whether a similar situation persists
for higher neutron numbers, we performed calculations with and without pairing for
all even-even Pb isotopes located between the two-proton and the two-neutron drip lines (see
Fig.\ \ref{Pb-full-chain}). Note that we restrict the calculations to spherical shapes
to see the sources (not affected by the shape changes) of major differences between
the functionals. This is a somewhat hypothetical scenario since the calculations with deformation
included indicate the presence of deformation in the ground states of the Pb isotopes located in the
middle of the region between the magic neutron closures (see Fig. 19 of Ref.\
\cite{AARR.14}). The neutron single-particle rms radii of the single-particle orbitals are
shown in Table\ \ref{table-sp-radii}. One can see very large neutron rms radii of the
$3d_{5/2}$, $2g_{7/2}$, $3d_{3/2}$ and, especially, of the $4s_{1/2}$ subshells.
However, as discussed in the introduction and in Ref.\ \cite{GSR.13} the real impact of the 
occupation of these orbitals on the charge radii will be defined by the pull they exert on proton densities.

\begin{table}[htb]
\caption{Neutron single-particle rms radii $r_{neu}^{sp} = \sqrt{\left< r^2 \right>^{sp}}$  of the indicated single-particle orbitals 
obtained in  $^{208}$Pb in calculations without pairing with the CEDF DD-ME2. The order [from top to 
bottom] of the orbitals is the same as in Fig.\ \ref{208Pb-single-particle}(b).  The radii of the first two 
orbitals located above the $N=126$ spherical shell gap are shown in bold.}
\label{table-sp-radii}
\centering
\begin{tabular}{|c|c|} \hline \hline
s-p orbital        & $r_{neu}^{sp}$  [fm] \\ \hline
       1           &                 2              \\ \hline
$3d_{3/2}$       & 8.9411 \\
$2g_{7/2}$       & 7.4880 \\
$4s_{1/2}$       & 9.3128 \\
$3d_{5/2}$       & 8.2905 \\
$2g_{9/2}$       & {\bf 7.0227} \\
$1i_{11/2}$      & {\bf 6.4131} \\
$3p_{1/2}$       & 6.4775 \\
$3p_{3/2}$       & 6.3856 \\
$2f_{5/2}$       & 6.2215 \\
$1i_{13/2}$      & 6.4108 \\ \hline
\end{tabular}
\end{table}

\begin{table}[htb]
\caption{The sequence of the spherical subshells occupied in the Pb isotopes
with increasing neutron number in the calculations without pairing. Note that
these sequences are the same for the pairs of the NL3* and PC-PK1 as well
as DD-PC1 and DD-ME$\delta$ functionals. The numbers  in the brackets
$[N_1-N_2]$ provide the range of neutron numbers between $N_1$ and $N_2$
for which the occupation of a given subshell takes place. The sequences of the
subshells and  neutron number ranges are shown in the columns 2 and 3 only
starting from the point at which the difference with the column 1 emerges.
\label{table-filling-shells-Pb}
}
\begin{tabular}{|c|c|c|} \hline \hline
NL3*/PC-PK1 & DD-PC1/DD-ME$\delta$ & DD-ME2 \\ \hline
       1              &                 2                      &       3        \\ \hline
$1i_{13/2}$ ~~ $[101-114]$  & & \\
$2f_{5/2}$   ~~ $[115-120]$  & & \\
$3p_{3/2}$  ~~ $[121-124]$  && \\
$3p_{1/2}$  ~~ $[125-126]$  && \\
$1i_{11/2}$  ~~ $[127-138]$  &&   \\
$2g_{9/2}$  ~~ $[139-148]$  && \\
$3d_{5/2}$  ~~  $[149-154]$ & $1j_{15/2}$~~ $[149-164]$ &  $3d_{5/2}$  ~~  $[149-154]$ \\
$2g_{7/2}$  ~~  $[155-162]$  & $2g_{7/2}$~~ $[165-172]$ &  $4s_{1/2}$  ~~  $[155-156]$ \\
$4s_{1/2}$  ~~  $[163-164]$  & $3d_{5/2}$~~ $[173-178]$ &   comp.~~ $[157-172]$ \\
$1j_{15/2}$ ~~  $[165-180]$  & $4s_{1/2}$~~ $[179-180]$ & $3d_{5/2}$~~ $[173-178]$ \\
$3d_{3/2}$  ~~ $[181-184]$   & $3d_{3/2}$~~ $[181-184]$ & $4s_{1/2}$~~ $[179-180]$ \\
                                              &                                             & $3d_{3/2}$~~ $[181-184]$ \\ \hline
\end{tabular}
\end{table}

   We start from the analysis of calculations performed without pairing (see
upper curves in Fig.\  \ref{Pb-full-chain}).  The significant (comparable to that seen at
$N=126$) changes of the slope of the  $\delta \left < r^2 \right>^{N,126}$ curves are
observed at $N=138$: at this neutron number the $1i_{11/2}$ subshell is completely filled and
the $2g_{9/2}$ subshell is filled at higher neutron number (up to $N=148$). The
slopes of the $\delta \left < r^2 \right>^{N,126}$ curves for different CEDFs as a function
of the neutron number are similar in the neutron range $N=126-148$, and these curves
do not cross.  However, the situation starts to change above $N=148$ because of
the different sequences of the occupation of the single-particle subshells (see Table
\ref{table-filling-shells-Pb}) caused by the fact that five different subshells, clustered into
an energy window which is slightly larger than 1 MeV (see top of Fig.\ \ref{208Pb-single-particle}),
have somewhat different relative energies for the different functionals. The NL3* and PC-PK1
functionals have the same sequences of filling of spherical subshells with increasing neutron
number (see column 1 in Table  \ref{table-filling-shells-Pb}).  As a consequence, the
$\delta \left < r^2 \right>^{N,126}$ curves for these two functionals have comparable evolutions
as a function of neutron number with minor changes of the slope at neutron numbers at
which the transition from filling of one subshell to another one takes place.  A similar
situation also exists in the pair of functionals DD-PC1 and DD-ME$\delta$ (see Fig.\
\ref{Pb-full-chain}) for which the sequences of the occupation of the single-particle subshells
with increasing neutron number are the same (see column 2 in Table  \ref{table-filling-shells-Pb}).
Note that the $\delta \left < r^2 \right>^{N,126}$ values calculated with these two functionals
are extremely close to each other.

   The situation is more complex in the case of the DD-ME2 functional. Calculations
with this functional reveal a complicated interplay of the occupation of different spherical subshells with
increasing neutron number.  The neutron $3d_{5/2}$ and
the $4s_{1/2}$ subshells are gradually occupied for $N=149-156$ (see Table \ref{table-filling-shells-Pb}).
However, in the neutron number range $N=157-172$  a complicated interplay of the occupation
of the $2g_{7/2}$ and $1j_{15/2}$ subshells and deoccupation of the $3d_{5/2}$ and $4s_{1/2}$
subshells with increasing neutron number take place [this region is labeled as "comp." in Table
\ref{table-filling-shells-Pb}].  It leads to substantial irregularities in the $\delta \left < r^2 \right>^{N,126}$
curve at these particle numbers (see Fig.\ \ref{Pb-full-chain}). At $N=172$ the $2g_{7/2}$ and
$1j_{15/2}$ subshells are fully occupied, and the repetitive occupation of the $3d_{5/2}$ and
$4s_{1/2}$ subshells as well as the occupation of the $3d_{3/2}$ subshell takes place at higher
neutron numbers.

   The inclusion of pairing leads to a redistribution of the occupation of the
single-particle states located in the vicinity of the neutron Fermi level and basically removes
all the changes in the slopes of the $\delta \left < r^2 \right>^{N,126}$ curves  seen
in the calculations  without pairing for $N>126$ (compare the results of the calculations with and without
pairing shown in Fig.\ \ref{Pb-full-chain}).  However, important differences between the
functionals still exist. This is best illustrated by comparing the results obtained with the DD*
functionals. For these functionals $\delta \left < r^2 \right>^{184,126}\approx 4.3$ fm$^2$ both
in the calculations with and without pairing (see Fig.\ \ref{Pb-full-chain}). At $N=184$, neutron
pairing collapses (see, for example,  Fig. 2 in Ref.\ \cite{AARR.15}) but proton pairing is present
due to the reduced size of the $Z=82$ shell closure as compared with the one for the N=126 isotope.
It is weakest for the DD-ME2 functional and strongest for the DD-ME$\delta$ one.  This feature
explains the slightly larger spread $\Delta(\delta \left < r^2 \right>^{184,126})=0.117$  fm$^2$ of the
$\delta \left < r^2 \right>^{184,126}$ values obtained with the DD* functionals in the calculations
with pairing as compared without pairing.

Thus, the results of the DD* functionals are the same at $N=126$ (by normalization) and almost
the same at $N=184$.  However, the difference between the DD-ME2 and
DD-PC1/DD-ME$\delta$ functionals\footnote{Note  that  the $\delta \left < r^2 \right>^{N,126}$
curves obtained with the CEDFs DD-PC1 and DD-ME$\delta$ are almost the same for
all neutron numbers. This is the
consequence of (i) the same sequences of the filling of spherical subshells with increasing
neutron number (see Table \ref{table-filling-shells-Pb}) and (ii) similar isovector properties of
these two functionals (see Table III in Ref.\ \cite{AA.16}).} is increasing on moving
away from the shell closures, and it is maximized at $N=170$  where it reaches 0.302 fm$^2$
(see bottom curves of Fig.\ \ref{Pb-full-chain}). This is due to different sequences of the filling
of spherical subshells in these two groups of the functionals (see Table
\ref{table-filling-shells-Pb}).

  Thus, the relative energies of the single-particle states and the patterns of their occupation
with increasing neutron number are still important even in the calculations with pairing. They
can lead to different predictions in different functionals and to discrepancies with experiment. It
is important to remember that the group of the single-particle subshells discussed above is located
in a very narrow energy range ($\approx 1.0$ MeV, see Fig.\ \ref{208Pb-single-particle}).
Thus, the correct description of the
sequence of the occupation of the single-particle states with increasing neutron number requires an
enormous accuracy (within approximately 200 keV) for the description of the energies of the single-particle
states. Such an accuracy is unachievable in the present generation of energy density
functionals (both covariant and non-relativistic ones). This is because the structure of the
experimental ground states in odd-$A$ nuclei is reproduced globally only in approximately
40\% of the nuclei in the DFTs, and there are substantial differences between experimental
and calculated single-particle spectra \cite{BQM.07,AS.11,DABRS.15}. The  inclusion of  particle-vibrational
coupling increases the accuracy  of the description of the single-particle configurations in odd-$A$
nuclei but such studies are, so far, limited to spherical nuclei (see Refs.\ \cite{CSB.10,LA.11,AL.15}).

The results of the calculations with PC-PK1 and NL3* functionals show the same trends in the
$\delta \left < r^2 \right>^{N,126}$ curves (see Fig.\ \ref{Pb-full-chain}) reflecting the same sequence
of the occupations of the spherical subshells (see Table \ref{table-filling-shells-Pb}). At $N=184$, the
$\delta \left < r^2 \right>^{184,126}$ values obtained with these two functionals are higher than
those obtained with the DD* ones. This is most likely the consequence of the different isovector
properties of the compared functionals (see Ref.\ \cite{AA.16}).

\begin{figure*}[htb]
\centering
\includegraphics[width=16.0cm]{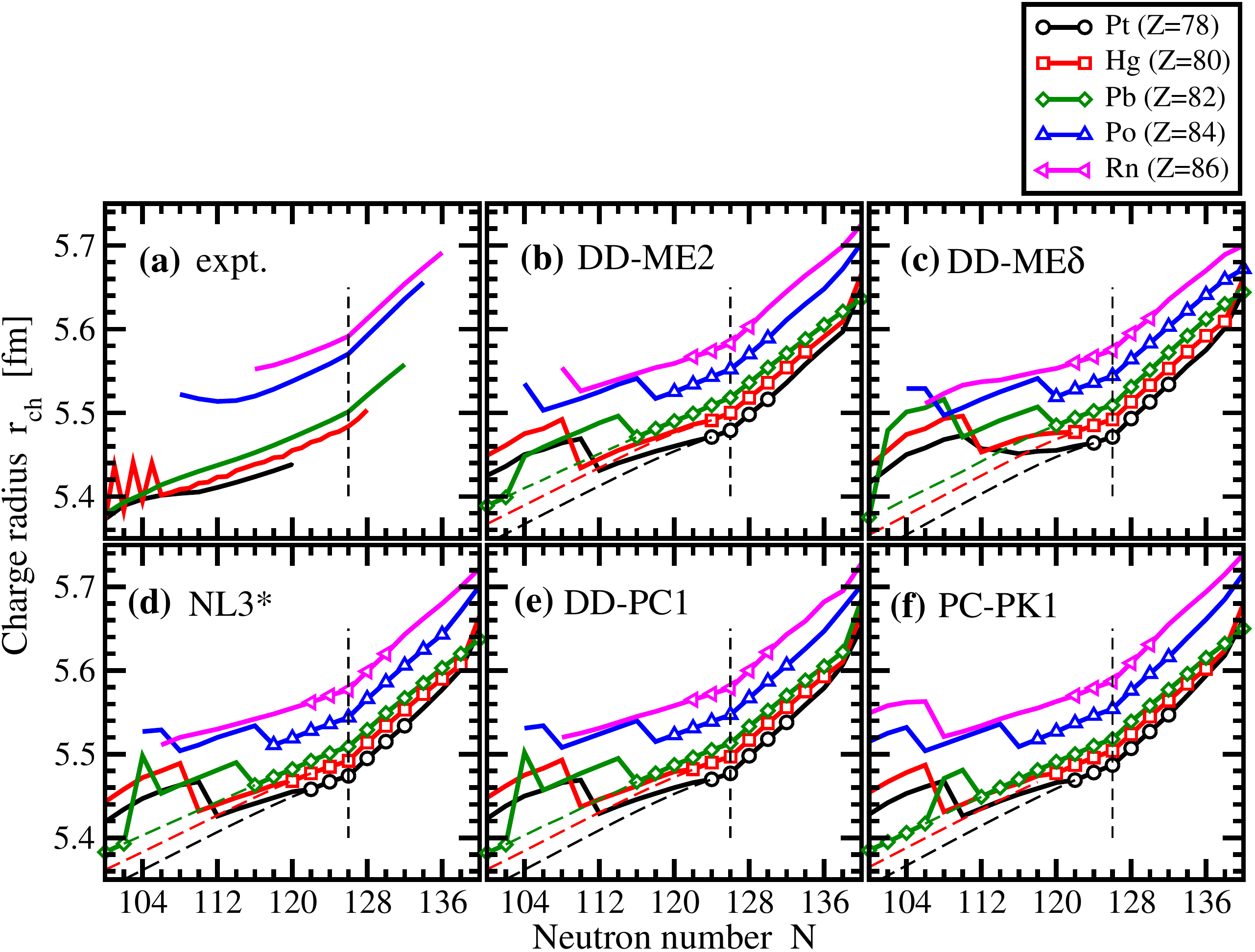}
\caption{The charge radii $r_{ch}$ of the Pt ($Z=78$), Hg ($Z=80$),  Pb ($Z=82$), Po ($Z=84$)  
and  Rn ($Z=86$) isotopes as a function of neutron number. In panel (a), experimental data 
are shown only for even-even nuclei. The only exception is the Hg isotopes for which experimental
data for odd-$N$ isotopes is included in order to illustrate a typical magnitude of OES in charge
radii induced by shape coexistence (the $N=101-106$ region, see Sec.\ \ref{sec-OES-shape} for a 
detailed  discussion)  or by other effects when neighboring even and odd-$N$ isotopes have 
comparable shapes (the rest of the Hg curve).  In panels (b)-(f),  thick  solid lines 
show  the  $r_{ch}$ values   obtained in the calculations for the lowest in energy
solutions in each isotopes. Open symbols show the isotopes for which these
solutions are either spherical or quasi-spherical ($|\beta_2|\leq 0.05$).
This  "line-symbol" convention is used in all figures below. Thin dashed lines show the charge 
radii of spherical solutions in neutron-poor Pt, Hg and Pb isotopes. Vertical black dashed line 
indicates $N=126$.
\label{Pb-region-radii}
}
\end{figure*}

\section{Charge radii in isotopic chains of the Pb region}
\label{Pb-region-section}

\begin{figure*}[htb]
\centering
\includegraphics[width=16.0cm]{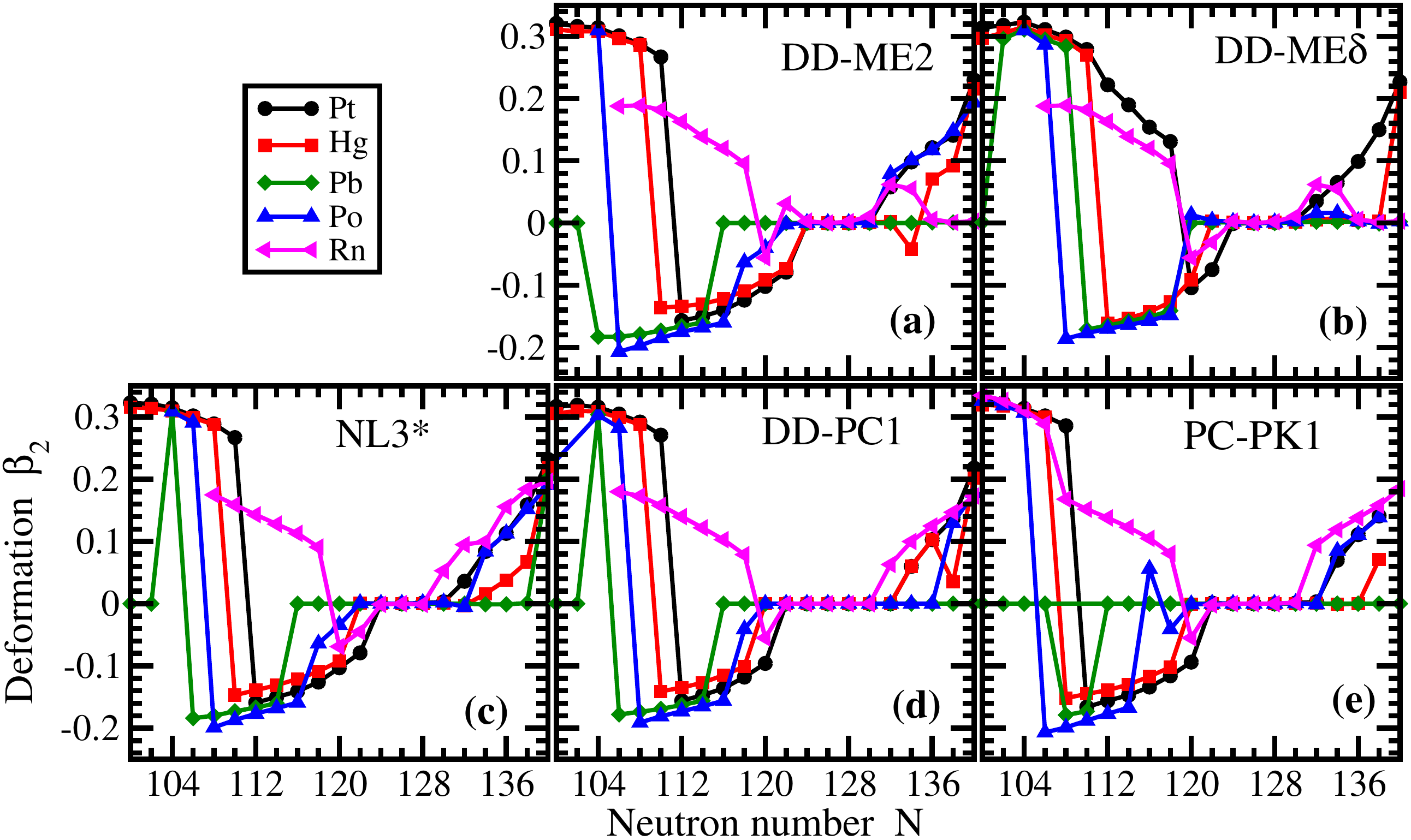}
\caption{Quadrupole deformations $\beta_2$ of the lowest in energy solutions
of the Pt ($Z=78$), Hg ($Z=80$),  Pb ($Z=82$), Po ($Z=84$) and  Rn ($Z=86$)
isotopes obtained in the calculations with the indicated functionals.
\label{Pb-region-deformation}
}
\end{figure*}

 The absolute values of  experimental and calculated charge radii of the Pt, Hg, Pb, Po, and Rn isotopes
 are compared in Fig.\  \ref{Pb-region-radii}, while a similar comparison for differential charge radii is
 presented in Fig.\ \ref{Pb-region-delta-r2} below. Note that in these and in other figures, which cover
 the nuclei over the specific region, we consider only experimental even-$Z$ and calculated even-even
 nuclei.  This is done in order to focus on general global features and avoid the discussion of odd-$A$ nuclei and
 related OES in charge radii which will be separately considered in Sec.\ \ref{sec-OES} below. The calculated
 quadrupole  deformations of the lowest in energy solutions are presented in Fig.\ \ref{Pb-region-deformation}.

   The relative charge radii of the Pt, Hg and Pb isotopes are reasonably well reproduced in the region
of neutron numbers around $N\approx 120$ (see Fig.\  \ref{Pb-region-radii}).  The increase of proton number above
$Z=82$  leads to a gradual increase of charge radii in model calculations with all employed CEDFs (see Figs.\
\ref{Pb-region-radii}(b), (c), (d), (e) and (f)). In contrast, there is a substantial gap in experimental charge radii
of the $Z=82$ and $Z=84$ isotopes (see Fig.\  \ref{Pb-region-radii}(a)) which is larger than that predicted
in the calculations. The increase of charge radii in going on from the Po to Rn isotopes is somewhat smaller in
experiment as compared with theoretical results.   Note that at this point  it is not clear whether these differences
are due to the deficiencies of the model predictions or experimental evaluations of absolute values of charge radii
(see discussion in the Introduction).

  The evolution of experimental charge radii in the Rn isotopes are rather well reproduced in model calculations
(see Fig.\ \ref{Pb-region-radii}).  Note that only the isotopes in the vicinity of the neutron shell  closure 
at $N=126$ are spherical in the lowest in energy solutions (see Fig.\ \ref{Pb-region-deformation}). Some 
moderate deformation $|\beta \approx 0.10|$ appears in the calculations for experimentally known nuclei 
at $N=116-120$, but their potential energy curves (PEC) are very soft (see Fig.\ 1 in the supplemental material) 
so they are expected to be transitional. The shift from spherical to transitional nuclei does not trigger visible 
changes  in charge radii in experiment since beyond mean-field effects are expected to smooth out this 
transition.

\begin{figure*}[htb]
\centering
\includegraphics[width=16.0cm]{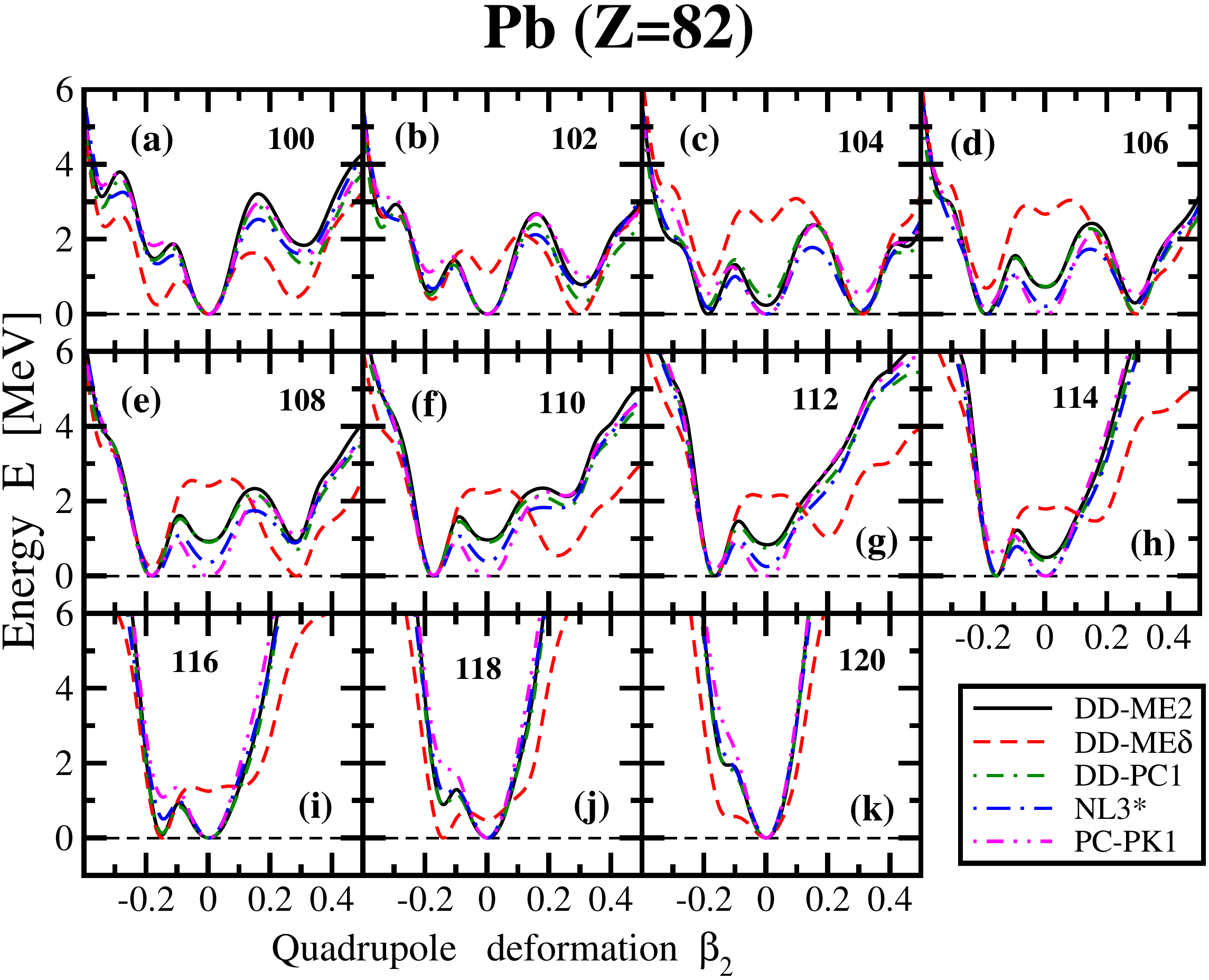}
\caption{Potential energy curves (PECs) of the Pb isotopes with neutron numbers $N=100-120$ as a function of
quadrupole deformation $\beta_2$ obtained in the RHB calculations with the indicated functionals. Note that their
energies are normalized to zero at the global minimum.}
\label{Pes-Pb}
\end{figure*}

  The evolution of charge radii in the Po isotopes is reproduced rather well from the isotope with
the highest neutron number accessible in experiment down to $N=120$ in DD-ME$\delta$, $N=118$ in DD-ME2,
NL3* and DD-PC1 and $N=116$ in PCPK1 CEDFs (see Fig.\ \ref{Pb-region-radii}).  At lower neutron numbers, the
experimental charge radii gradually bend up so that at $N=108$ experimental $r_{ch}$ exceed the value defined from the
trend of charge radii defined at $N=118-126$ by $\approx 0.04$ fm. However this process is more
abrupt in the calculations, since the calculated radii for $N=108-116$ exceed the above mentioned trend 
by $\approx 0.04$ fm. In the calculations,  this abrupt shift in calculated charge radii at $N\approx 118$ is caused 
by the transition from spherical to oblate
shapes with $\beta_2 \approx -0.18$ (see Fig.\ \ref{Pb-region-deformation}). These facts suggest that
the $r_{ch}$ values of the $N=108$  and $N\approx 118$ [depends on functional] isotopes are rather well
reproduced in the calculations, but the mean-field calculations fail to reproduce the gradual transition in $r_{ch}$
seen between $N=106$ and $N\approx 118$ isotopes. This gradual transition is most likely due to beyond
mean-field effects  since the nuclei in this neutron number range have soft PECs (see Fig. 2 in the supplemental 
material).  In addition, the triaxiality could play a role in this gradual transition.

   The charge radii of the Pb isotopes in the $N=116-132$ range are well described in the model calculations
of Sec.\ \ref{Pb-isotopes}. Here we focus on more neutron-poor Pb isotopes in the range of $N=100-116$. Experimental
charge radii in this neutron range continue the trend seen at $N=116-126$ [see Fig.\ \ref{Pb-region-radii}(a)]. This
suggests that the shapes of the nuclei in the measured states are either spherical or near-spherical.  Indeed, if
we consider spherical solutions in these nuclei (see green dashed lines in panels (b)-(f) of Fig.\ \ref{Pb-region-radii}),
then the experimental data are rather well reproduced.  However, the calculations predict either oblate or prolate
shapes for the ground states of the $N=104-114$ isotopes in CEDFs DD-ME2,  NL3* and  DD-PC1, of the $N=102-118$
isotopes in DD-ME$\delta$ and of the  $N=108-110$ isotopes in PC-PK1 (see Fig.\ \ref{Pb-region-deformation}).
Despite that, spherical minima, located either close in energy to the ground states or at some excitation energy,
exist in all isotopes in all functionals with the exception of DD-ME$\delta$ (see Fig.\ \ref{Pes-Pb}).  Note that PECs of 
these nuclei are rather soft in quadrupole deformation (see Fig.\ \ref{Pes-Pb} ). Thus, the correlations beyond
mean-field can play an important role in these nuclei.

\begin{figure}[htb]
\centering
\includegraphics[width=8.5cm]{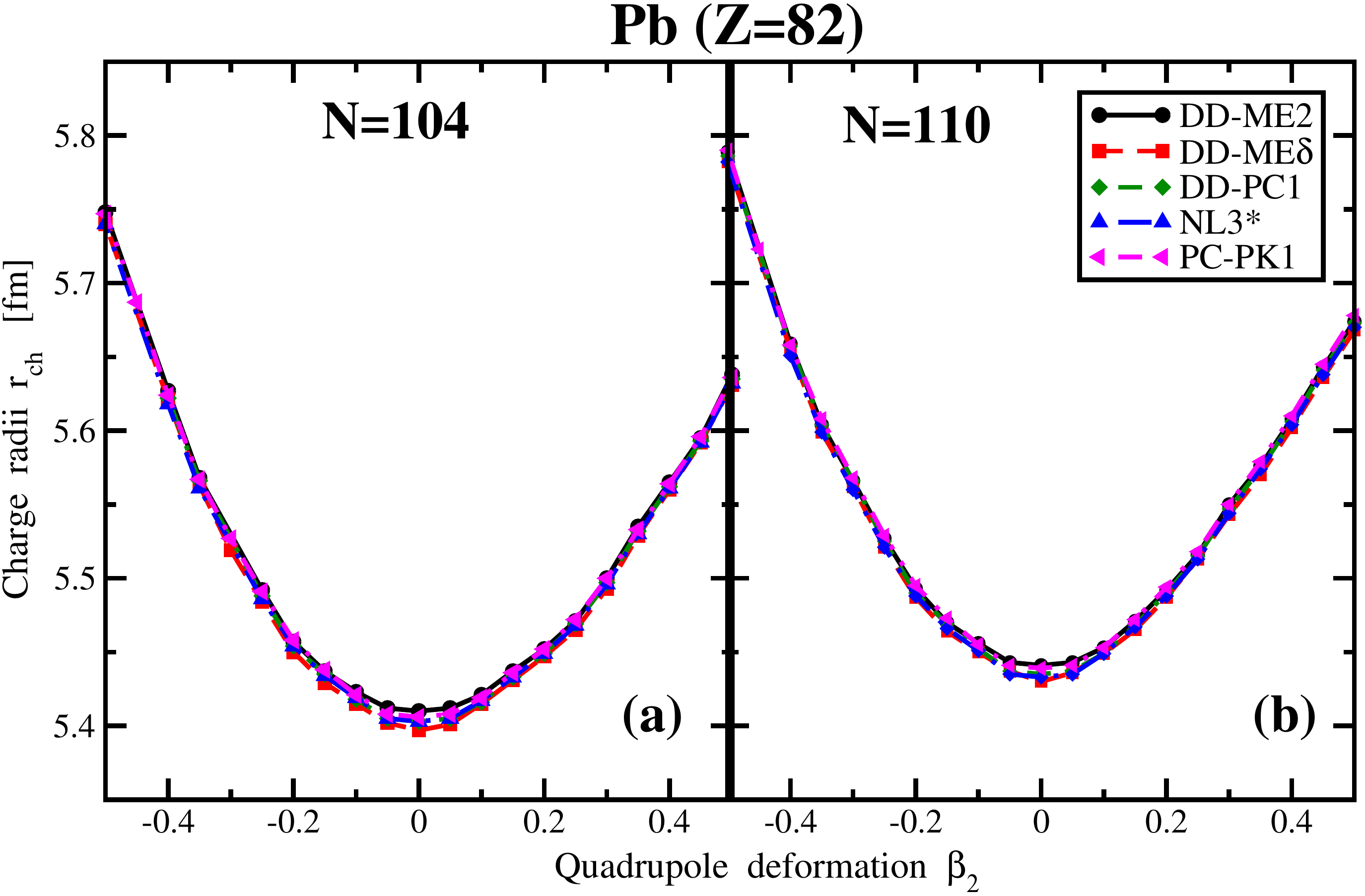}
\caption{The variations of charge radii $r_{ch}$ in $^{186}$Pb
and $^{192}$Pb nuclei as a function of the quadrupole deformation obtained
in RHB calculations with the indicated  functionals.
\label{Radii-func-beta}
}
\end{figure}

According to the droplet model (DM) \cite{Otten.89} the following relation
\begin{eqnarray}
\left< r^2 \right>_{DM} = \left< r^2 \right>^{spher}_{DM} (1 +\frac{5}{4\pi} \beta_2^2)
\label{Radii-def}
\end{eqnarray}
exists between the predictions of charge radii $\left< r^2 \right>_{DM}$ and  $\left< r^2 \right>^{spher}_{DM}$
at quadrupole deformation $\beta_2$ and at spherical shape, respectively.  
This relation is frequently used in the experimental analysis of the data for the extraction of 
quadrupole deformations. This equation tells us that
the charge radii form a parabolic function of $\beta_2$ with the minimum at spherical shape;  this function 
is symmetric with respect to a sign change of the deformation.  Realistic calculations presented  in Fig.\ \ref{Radii-func-beta}
confirm this parabolic-like dependence of charge radii on the quadrupole deformation. However, it is somewhat
asymmetric with respect of the change of the sign of the deformation\footnote{Note that Ref.\ \cite{MS.83} 
provides even higher order expansion of charge radii in terms of multipole deformations within the droplet 
model. However, the asymmetry of  $\left< r^2 \right>_{DM}$ as a function of   $\beta_2$ is opposite to that 
seen in the RHB calculations in Fig.\ \ref{Radii-func-beta}
because of cubic term in $\beta_2$ .
}. This difference between Eq.\ (\ref{Radii-def})
and the results of Fig.\ \ref{Radii-func-beta} are most likely due to neglecting higher-order deformations
(such as $\beta_4$ etc.) in Eq.\ (\ref{Radii-def}). Note that the functional dependence of the charge radii on 
deformation $\beta_2$ almost does not depend on the functional (see Fig.\ \ref{Radii-def}).  This fact is quite useful in the selection
of the most probable scenario when comparing the experimental situation in the Pb region with the 
results of the calculations that provide several local closely lying minima.

\begin{figure*}[htb]
\centering
\includegraphics[width=16.0cm]{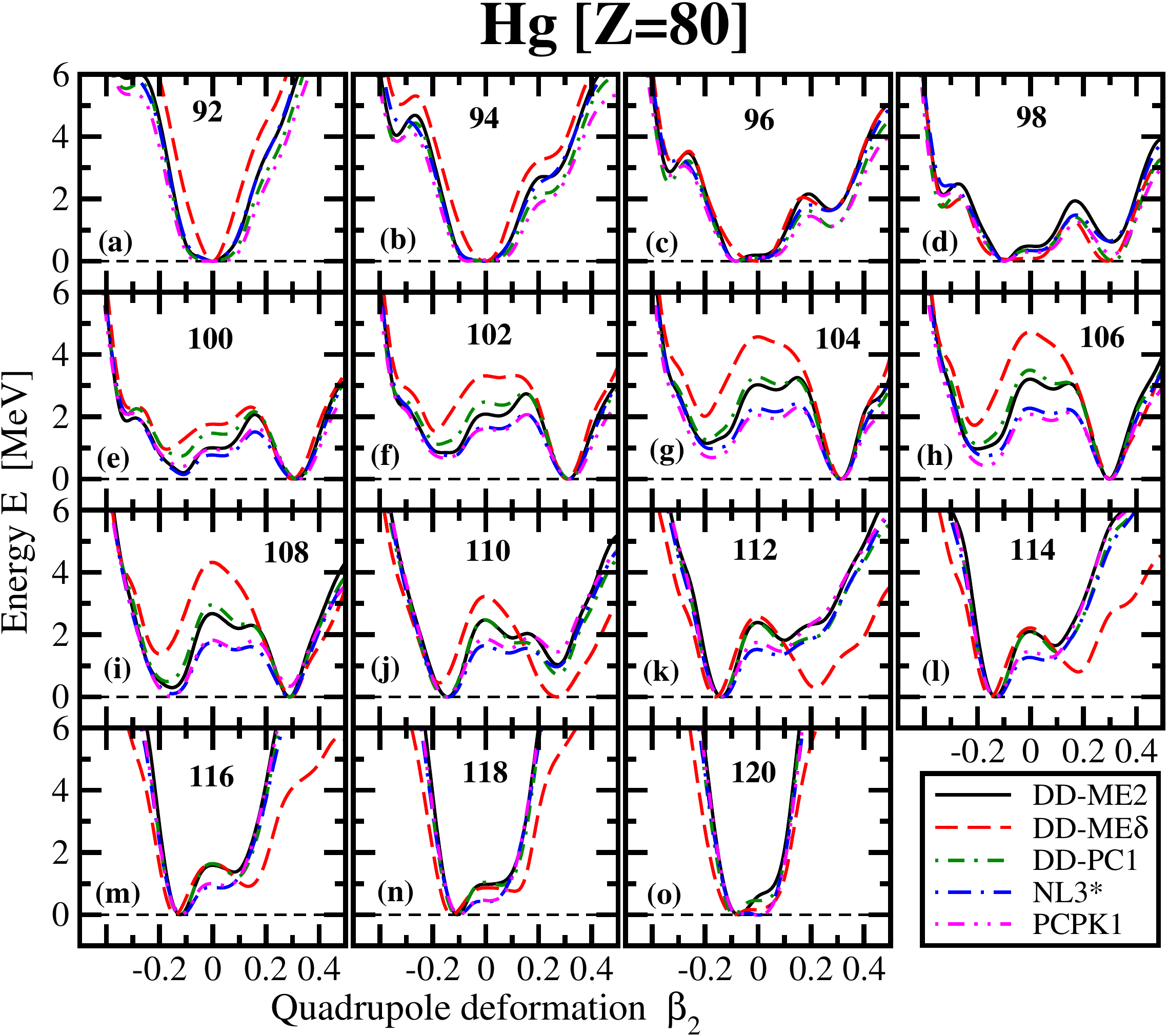}
\caption{The same as Fig.\ \ref{Pes-Pb} but for the Hg isotopes with $N=100-120$.
}
\label{PEC-Hg}
\end{figure*}

  Detailed investigations within the CDFT framework of charge radii in Hg isotopes with neutron 
numbers $N=121-128$  are presented in Refs.\ \cite{Pb-Hg-charge-radii-PRL.21,Pb-Hg-radii-PRC.21}. 
Good agreement between theory and experiment is obtained. The charge radii are also well described for neutron
numbers $N=110-120$ in calculations with DD-ME2, NL3* and DD-PC1, for $N=112-120$  with DD-ME$\delta$, and
for $N=108-120$ in PC-PK1 (see Fig.\ \ref{Pb-region-radii}). However, they also suggest that these nuclei are oblate
with $\beta_2 \approx -0.15$  in their ground states (see Fig.\ \ref{Pb-region-deformation} and Fig.\ \ref{PEC-Hg}) 
which leads to a slight increase of $r_{ch}$ as compared with the ones for the spherical solution (compare dashed 
red lines with solid red lines with red  squares in panels (b)-(f) of Fig.\ \ref{Pb-region-radii}). A significant odd-even 
staggering in the Hg charge radii exists for $N=100-106$ [see Fig.\ \ref{Pb-region-radii}(a)] the origin of which is 
discussed in Sec.\ \ref{sec-OES-shape}.

\begin{figure*}[htb]
\centering
\includegraphics[width=16.0cm]{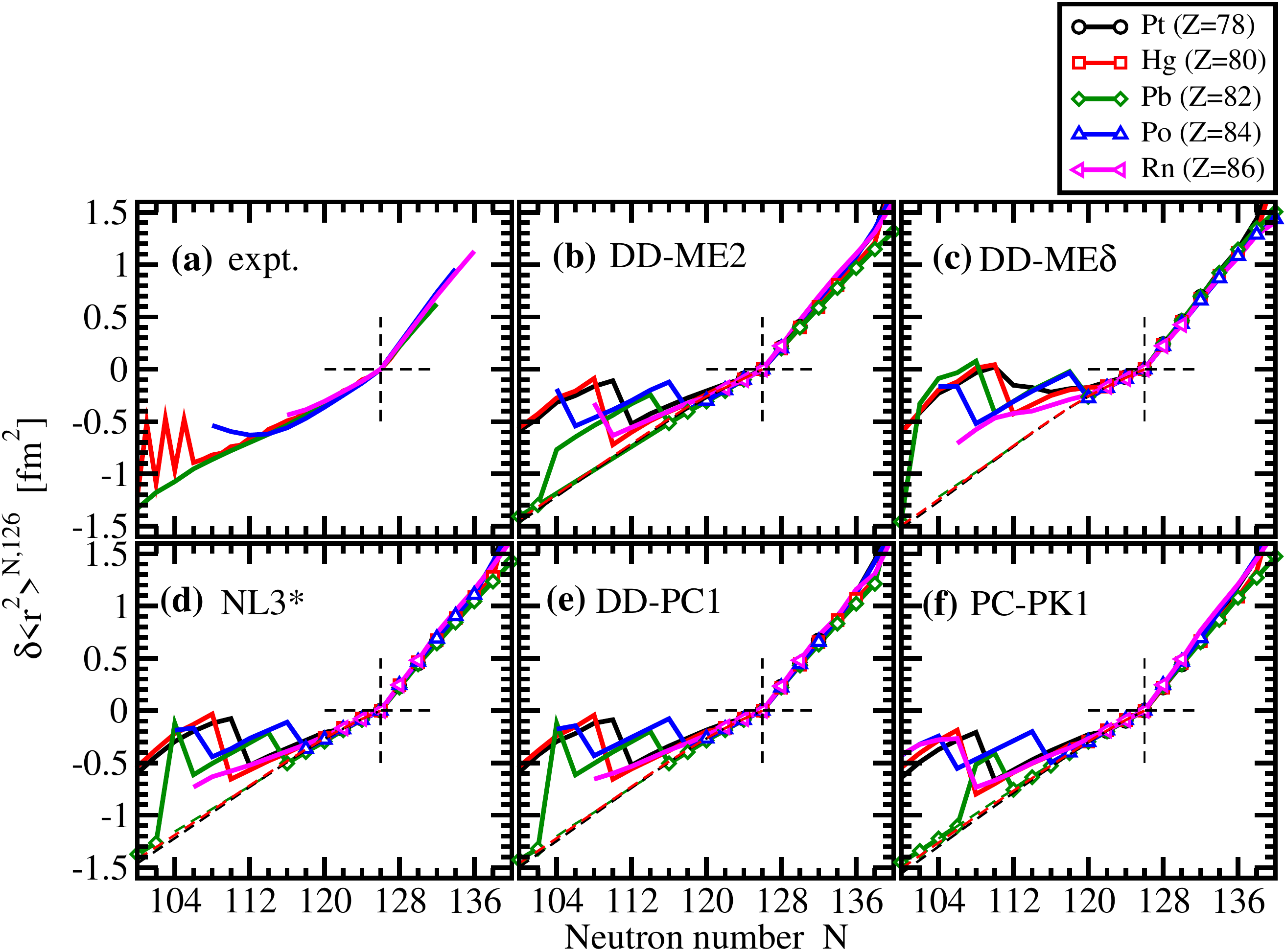}
\caption{The $\delta \left < r^2 \right>^{N,126}$ values of the Pt ($Z=78$), Hg ($Z=80$),
Pb ($Z=82$), Po ($Z=84$) and  Rn ($Z=86$) isotopes as a function of neutron number.
Thick lines (with open symbols for spherical of quasi-spherical solutions)
show the  $\delta \left < r^2 \right>^{N,126}$ values obtained for the lowest in energy 
solutions in each isotopes. Thin dashed lines show the charge radii of spherical solutions 
in neutron-poor Pt, Hg and Pb isotopes. Vertical black dashed line indicates $N=126$ 
and horizontal dashed line $\delta \left < r^2 \right>^{N,126}=0$.
\label{Pb-region-delta-r2}
}
\end{figure*}

    Except for calculations with the DD-ME$\delta$ functional, the
evolution of experimental data in the Pt isotopes is rather well reproduced for the $N=112-120$
nuclei.  The results of the calculations with DD-ME2, NL3*, DD-PC1 and PC-PK1 suggest
that the ground states of these nuclei have weakly deformed oblate shapes with
$\beta_2 \approx -0.13$ (see Fig. 3 in the supplemental material).  However, the competing
prolate minimum exists in all these isotopes.  With decreasing neutron number, this
prolate minimum with $\beta_2 \approx 0.3$ becomes the lowest in energy at $N=110$
(see Fig.\ \ref{Pb-region-deformation} and Fig. 3 in the supplemental material). This leads  to
a sharp increase in charge radii which overshoots the experimental data by roughly 0.04 fm
(see Fig.\ \ref{Pb-region-radii}). However,  if we would associate excited oblate states with
deformation $\beta_2 \approx -0.2$ in the $N=100-110$ nuclei with the observed ground
states, then the experimental data on charge radii would be much better described since
they are characterized by lower charge radii (see discussion of Fig.\ \ref{Radii-func-beta}
above). The change of the slope of experimental charge radii at $N\approx 106$ is 
possibly  an indicator of such a transition from oblate shapes with nearly constant $\beta_2
\approx -0.2$ for $N=100-106$ to oblate shapes where the deformation decreases in
absolute value with increasing neutron number above $N=106$ (see Fig.\
\ref{Pb-region-deformation} and Fig. 3 in the supplemental material).

  Fig.\ \ref{Pb-region-delta-r2}  shows the evolution of the  $\delta \left< r^2 \right>^{N,126}$
values in the Pt, Hg, Pb, Po, and Rn isotopes. These curves are similar [clustered] for
different isotopic  chains for $N=112-136$, but their slopes change at $N=126$.
Below $N=112$ this feature is disturbed in the Po isotopes  due to the gradual transition to prolate
shapes,  but still, it is present for the isotopic chains of  Pb, Hg (excluding odd nuclei), and Pt.
This clustering of the $\delta \left< r^2 \right>^{N,126}$ values for different isotopic chains is
well reproduced in all functionals above $N=126$. The situation is
somewhat different below $N=126$. The DD-ME2, NL3*, DD-PC1 and PC-PK1 functionals
reproduce this clustering of the $\delta \left< r^2 \right>^{N,126}$ values down to $N\approx 116$
and for the Pt, Hg, and Rn isotopic chains even to lower neutron numbers. However, as discussed above
in the low-$N$ region the calculations do not reproduce correctly the lowest in energy minimum leading to
 the discrepancies between theory and experiment.  Note that among the functionals under consideration
DD-ME$\delta$ provides the worst description of experimental data in the nuclei with $N<126$.

\begin{figure}[htb]
\centering
\includegraphics[width=8.4cm]{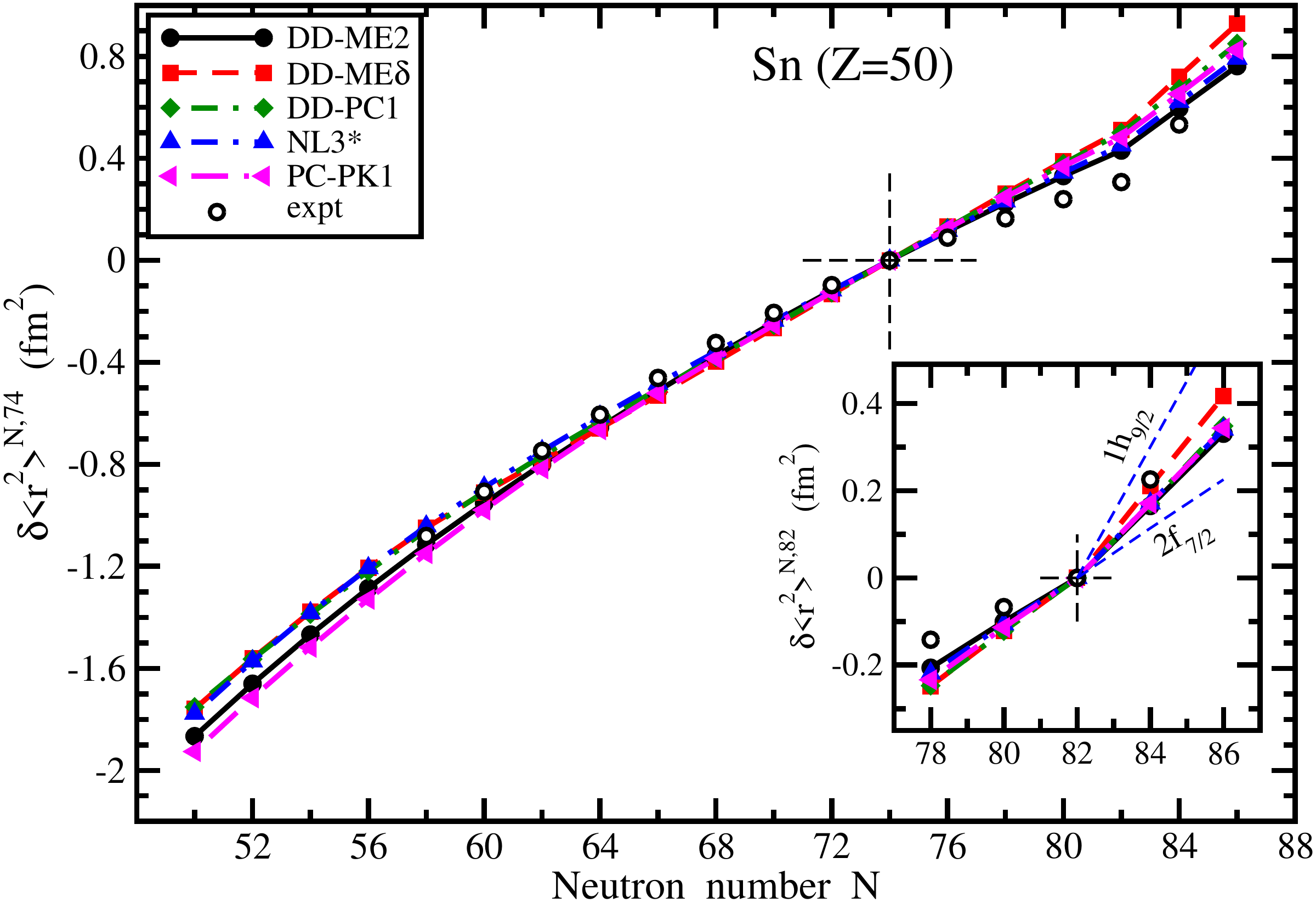}
\caption{The same as Fig.\ \ref{Delta-r2-Pb-even-pairing} but for the Sn
isotopes. The experimental data are taken from Ref.\ \cite{Gorges-Sn-radii.19}.
The $N=74$ and $N=82$ nuclei are used as reference nuclei in the full figure and in the
inset, respectively. The latter is done in order to show the kink at $N=82$ in more
detail. The dashed lines in the inset show the results of the calculations without
pairing and with the CEDF NL3* in which either only neutron $1h_{9/2}$ or only
neutron $2f_{7/2}$ orbitals are occupied in the $N>82$ nuclei.
\label{Sn-isotopes-delta-r2}
}
\end{figure}

\section{Charge radii in isotopic chains of the Sn/Gd region}
\label{Sn-region-section}

  The calculated and experimental differential charge radii for the Sn
isotopes are compared in Fig.\ \ref{Sn-isotopes-delta-r2}. The experimental
$\delta \left<r^2\right>^{N,74}$ curve is well described in the $N=58-74$ range
in model calculations with the best reproduction provided  by the NL3*,
DD-ME$\delta$ and DD-PC1 functionals. However, the slope
of the experimental $\delta \left<r^2\right>^{N,74}$ curve is somewhat
overestimated in the $N=74-82$ range by all functionals. The experimental
data shows a kink at $N=82$, the magnitude of which is underestimated
by the NL3*, PC-PK1, DD-ME2 and DD-PC1 CEDFs. Only the
DD-ME$\delta$ functional reproduces it reasonably well with the slope of the
experimental $\delta \left<r^2\right>^{N,82}$ curve for $N>82$ described
almost perfectly (see inset in Fig.\ \ref{Sn-isotopes-delta-r2}).

\begin{figure}[htb]
\centering
\includegraphics[width=8.4cm]{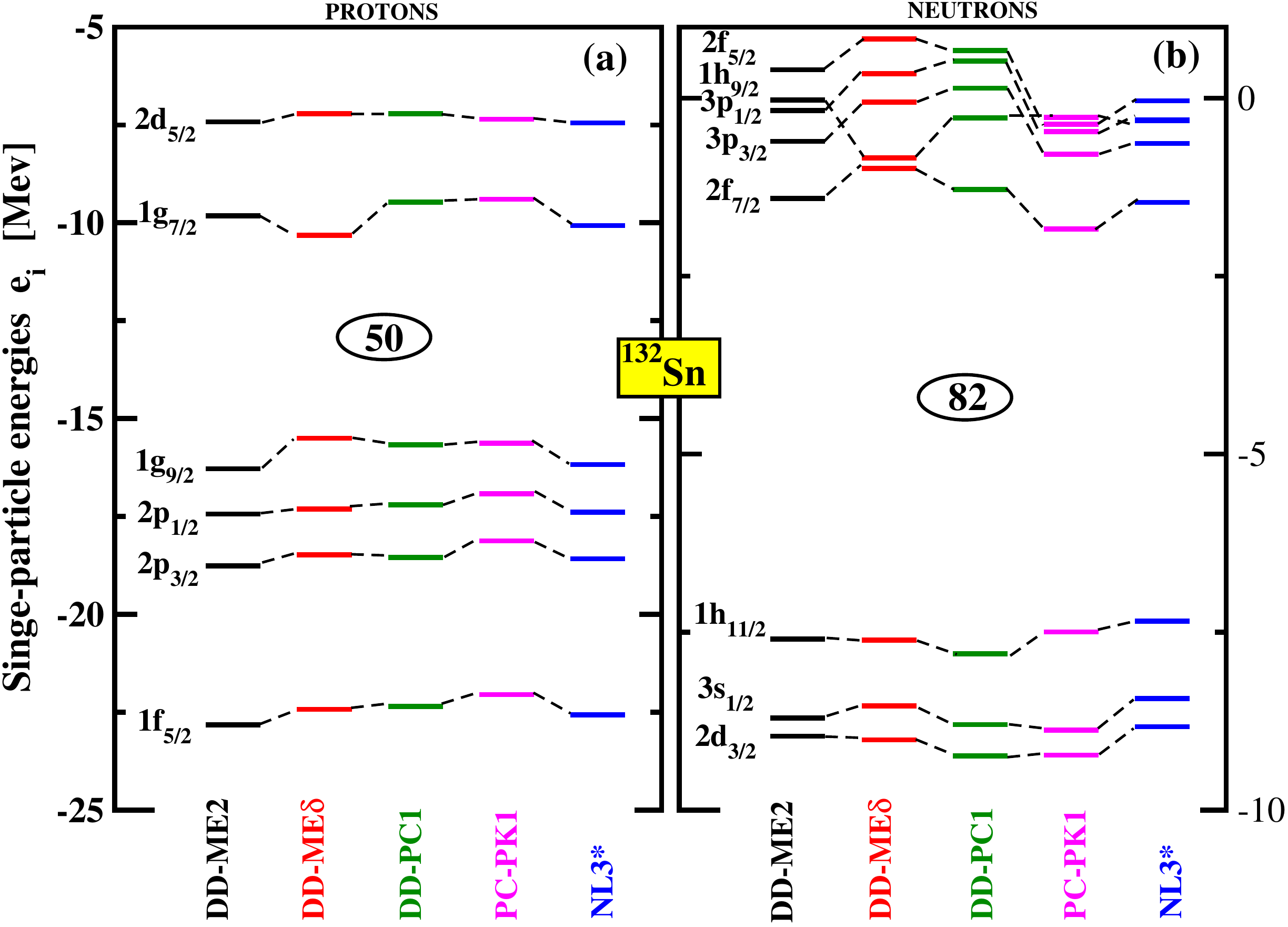}
\caption{The same as Fig.\ref{208Pb-single-particle}  but for $^{132}$Sn.
\label{Sn132-sp-spectra}
}
\end{figure}

  The origin of this feature can be traced back to the occupation pattern
of the neutron $1h_{9/2}$ and $2f_{7/2}$ subshells located above the $N=82$
shell closure (see Fig.\ \ref{Sn132-sp-spectra}(a)).  The calculations without pairing
show that the occupation of the $2f_{7/2}$ orbital, which is the lowest in energy
subshell above $N=82$ (see Fig.\ \ref{Sn132-sp-spectra}(a)), does not create
a kink  at $N=82$ (see inset in Fig.\ \ref{Sn-isotopes-delta-r2}). A similar situation
exists for the occupations of other orbitals such as the $n=2$ orbital $2f_{5/2}$ and
the $n=3$ orbitals $3p_{3/2}$, and  the $3p_{1/2}$ orbitals located above $N=82$.
It is only the occupation of the $n=1$ $1h_{9/2}$ orbital which drives the $N=84$
isotope to a visibly larger charge radius and creates a kink at $N=82$ (see inset in
Fig.\ \ref{Sn-isotopes-delta-r2}). This situation is very similar to the one seen in the
Pb isotopes in which only the occupation of the $n=1$ $1i_{11/2}$ subshell above
$N=126$ creates a kink in charge radii at this particle number (see Sec.\
\ref{Pb-isotopes}).

\begin{figure}[htb]
\centering
\includegraphics[width=8.4cm]{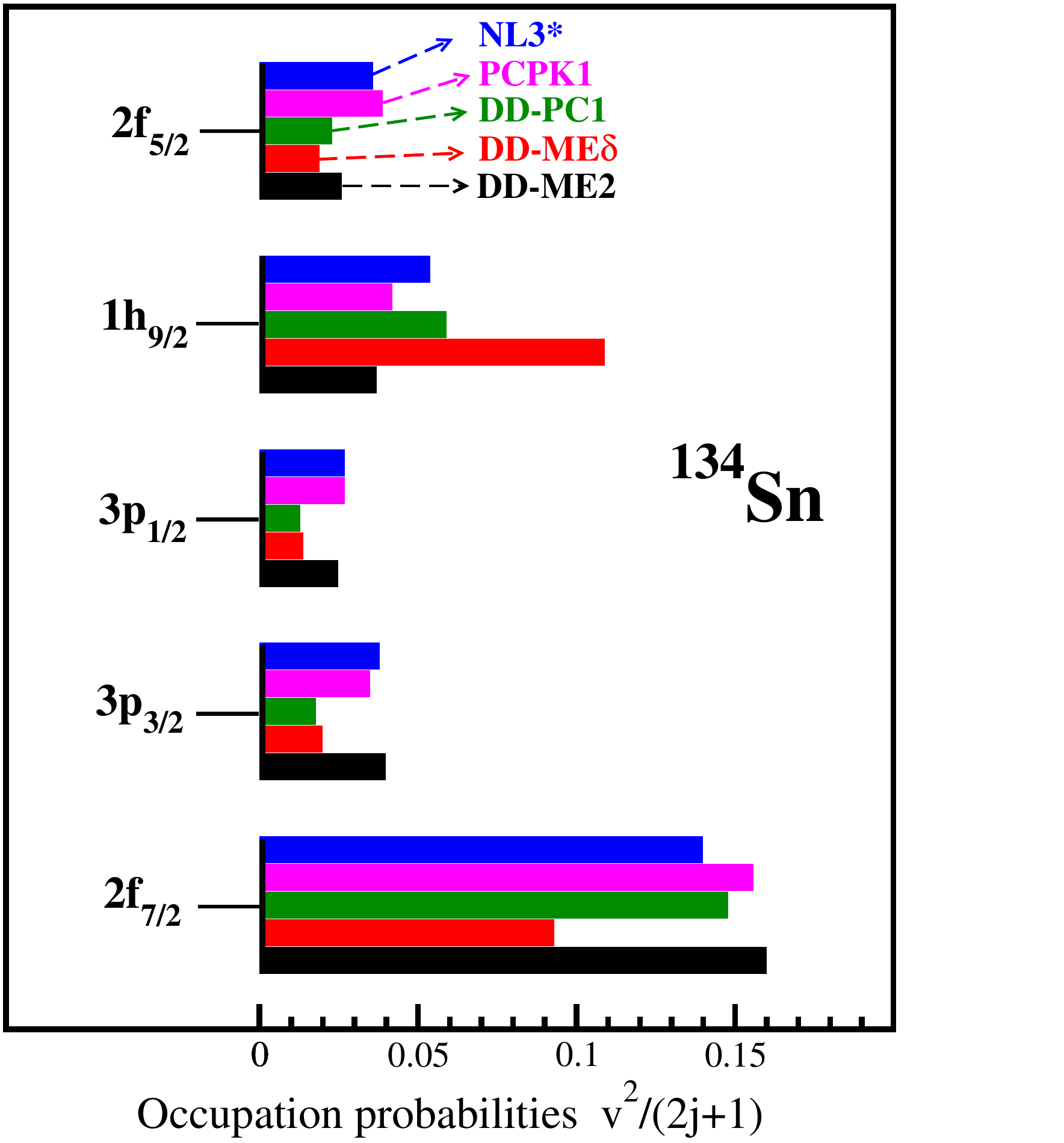}
\caption{The occupation probabilities $v^2/(2j+1)$ of the neutron orbitals
of $^{134}$Sn located above the $N=82$ shell closure.
\label{Sn132-134-v2}
}
\end{figure}

   Pairing leads to a redistribution of the occupation of different single-particle
orbitals in the $N=84$ isotope of Sn (see Fig.\ \ref{Sn132-134-v2}). However, it
is only for the DD-ME$\delta$ functional that the $1h_{9/2}$ orbital is strongly occupied
in the RHB calculations [stronger than the $2f_{7/2}$ one] because of the closeness of 
the $2f_{7/2}$ and $1h_{9/2}$ orbitals in energy (see Fig.\ \ref{Sn132-sp-spectra}(b)). 
And this balance of the occupation of these orbitals leads to an almost perfect reproduction 
of the slope of differential charge in the $N>82$ Sn isotopes (see  inset in Fig.\ 
\ref{Sn-isotopes-delta-r2}). In contrast, in other functionals the  $1h_{9/2}$ orbital is higher 
in energy by approximately 2 MeV than the $2f_{7/2}$ one (see Fig.\ \ref{Sn132-sp-spectra}(b))
and, as a result, its occupation probability is by a factor three or four [dependent on CEDF] 
smaller  than that for $2f_{7/2}$ (see Fig.\ \ref{Sn132-134-v2}). As a consequence,
the slope of differential charge radii in the $N>82$ isotopes as well as the magnitude
of the kink at $N=82$ is underestimated by these functionals (see inset in Fig.\ 
\ref{Sn-isotopes-delta-r2}).

\begin{figure*}[htb]
\centering
\includegraphics[width=17.0cm]{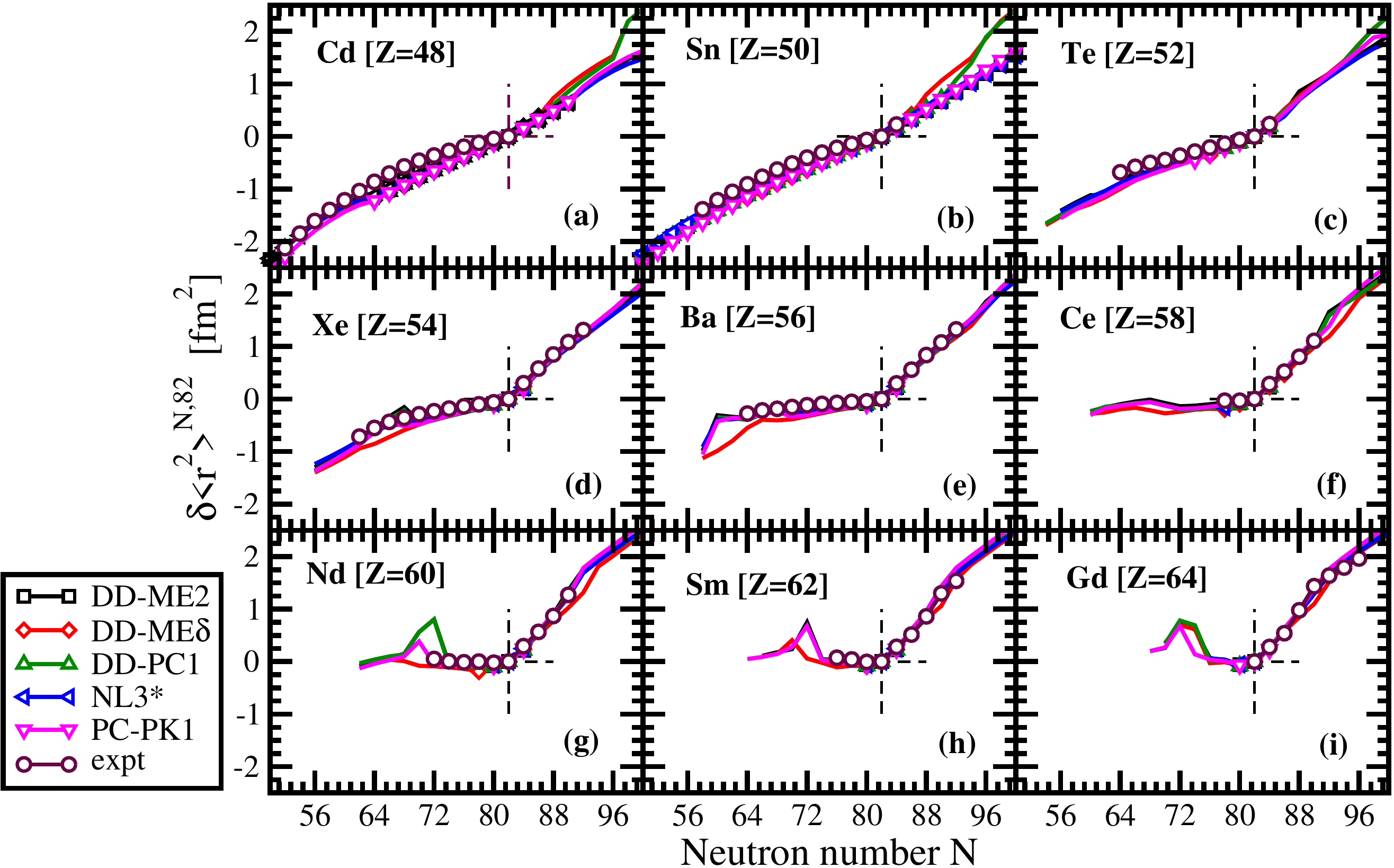}
\caption{The $\delta \left < r^2 \right>^{N,82}$ values of the isotopic chains in the Sn/Gd
region. The experimental data are taken from Ref.\ \cite{AM.13}.
\label{Sn-region-delta-r2-individual}
}
\end{figure*}

\begin{figure*}[htb]
\centering
\includegraphics[width=16.0cm]{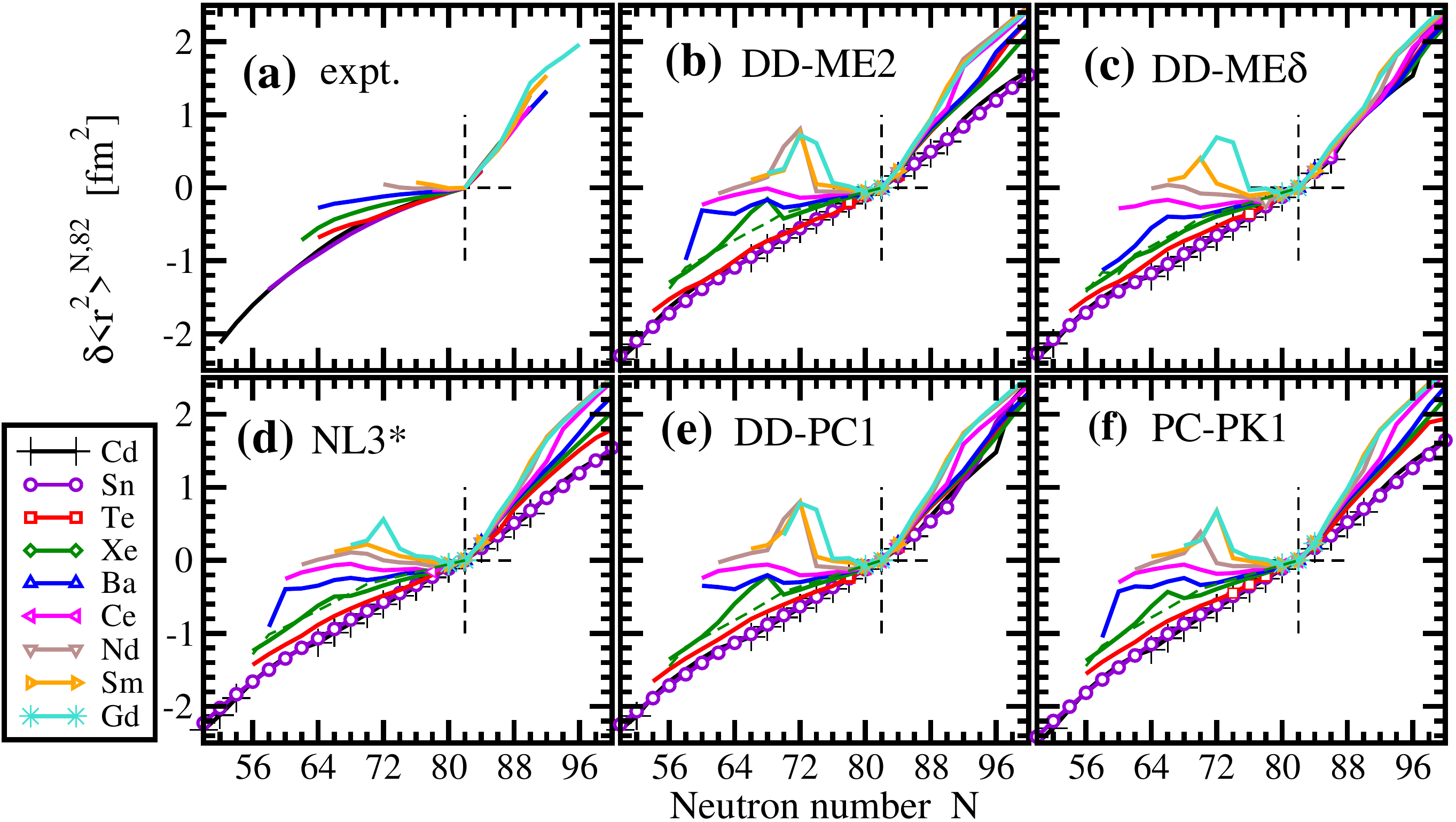}
\caption{The same as Fig.\ \ref{Pb-region-delta-r2} but for the $\delta \left < r^2 \right>^{N,82}$
values of the Sn ($Z=50$), Te ($Z=52$), Xe ($Z=54$), Ba ($Z=56$), Ce ($Z=58$), Nd ($Z=60$),
Sm ($Z=52$), and  Gd ($Z=54$) isotopes.  The experimental data are
taken from Ref.\ \cite{AM.13}.
\label{Sn-region-delta-r2}
}
\end{figure*}

   The experimental and calculated differential charge radii in the Cd, Sn, Te, Xe,
Ba, Ce, Nd, Sm, and Gd isotopic chains are compared in Figs.\ \ref{Sn-region-delta-r2-individual}
and \ref{Sn-region-delta-r2}. One can see that, on average, the experimental data are well
reproduced.

  For the nuclei with $N<82$, the experimental $\delta \left<r^2\right>^{N,82}$ curves
diverge away from each other with decreasing neutron number (see Fig.\ \ref{Sn-region-delta-r2}).
The only exceptions
from this trend are the Cd ($Z=48$) and Sn ($Z=50)$ isotopic chains, for which the experimental differential
charge radii are almost the same. These features are best reproduced by the NL3* functional
[compare panels (a) and (d) of  Fig.\ \ref{Sn-region-delta-r2}].  The similarity of differential
charge radii in the Cd and Sn isotopic chains is reproduced in all functionals. Note that the
ground states of all Sn isotopes are predicted to be spherical (see Figs. 4 and 12 in the supplemental
material). On the contrary, the $N=56-62$ Cd isotopes are predicted to be slightly prolate with
$\beta_2 \approx 0.15$ (see Fig. 4 in the supplemental material), but the PECs of many of them are soft
in quadrupole deformation (see Fig. 13 in the supplemental material) so that the effects beyond 
mean-field could play a role in the definition of the exact ground state deformation. The relative properties
of differential charge radii of the Te ($Z=52$)  and Sn isotopes are reproduced rather well in all
functionals. Note that the calculations predict that many of the Te nuclei are soft in their ground
states (see Fig. 11 in the supplemental material).

  The evolution of differential charge radii in the Xe $(Z=54)$  isotopes and the kink
at $N=82$ are reasonably well described by all functionals [see Fig.\
\ref{Sn-region-delta-r2-individual}(d)]. However, some functionals slightly underestimate
the experimental data in the $N<82$ nuclei. The relative properties of differential
radii of the Xe and Te isotopic chains are somewhat better described by the
DD-ME$\delta$ functional (compare panels (c) and (a) in Fig.\ \ref{Sn-region-delta-r2}).
Note that in this functional the $N=64-68$
isotopes are predicted to be oblate with $\beta_2 \approx -0.2$, while other isotopes are prolate
with $\beta_2 \approx 0.2$ (see Fig.\ 4(b) in the supplemental material). 
In contrast, in other functionals the calculated
ground state deformation is prolate at $N=64-68$, and it reaches a maximum value at $N\approx 66$
 (see Figs.\ 4(a), (c), (d) and (e) in the supplemental material). As a result, the differential
radii show a small peak at these neutron numbers (see  Figs.\ \ref{Sn-region-delta-r2}(b), (d), (e)
and (f)).  However, the excited oblate minimum in those functionals is only slightly higher in energy
than the prolate one (see Fig. 10 in the supplemental material). If one associates this oblate minimum
with the ground state in the $N=64-68$ nuclei, then the experimental differential charge radii will be
well reproduced by model calculations [see dashed green lines in Figs.\ \ref{Sn-region-delta-r2}(b),
(d), (e) and (f)].

   Experimental data on differential charge radii of the Ba $(Z=56)$ isotopes extends down to
$N=64$ with $\delta \left< r^2 \right>^{N,82}$ values being nearly constant but slightly decreasing
with decreasing neutron number [see Fig.\ \ref{Sn-region-delta-r2-individual}(e)]. Similar trends are observed in
the calculations with all CEDFs which reasonably well reproduce the evolution of experimental
$\delta \left< r^2 \right>^{N,82}$ curve as well as the kink at $N=82$. However, the calculated curves
are disturbed by a small peak at $N=68$ in the calculations with DD-ME2, NL3*, DD-PC1, and PC-PK1
[see Figs.\ \ref{Sn-region-delta-r2}(b), (d), (e) and (f)] and a substantial downslope of the
$\delta \left< r^2 \right>^{N,82}$ curve with decreasing neutron number  which starts at $N=64$ in the
calculations with DD-ME$\delta$ [see Figs.\ \ref{Sn-region-delta-r2-individual}(e) and Fig.\
\ref{Sn-region-delta-r2}(c)]. This peak takes place at neutron numbers where the rate of the increase
of prolate deformation with decreasing neutron number is enhanced (see Fig.\ 4 in the supplemental
material).

  In the $N<82$ nuclei, the experimental data for even-even Ce $(Z=58)$ isotopes are
available only for $N=78$ and 80. It is reasonably well described in all model calculations [see
Fig.\ \ref{Sn-region-delta-r2-individual}(f)].  The experimental $\delta \left< r^2 \right>^{N,82}$
values of the Nd $(Z=60)$ isotopes are close to zero for $N=72-82$ and this feature is described
by all functionals [see Fig.\ \ref{Sn-region-delta-r2-individual}(g)].  The only exception is the $N=72$
isotope in the calculations with the CEDFs DD-ME2 and DD-PC1 for which a substantial increase of
differential charge radii is predicted [see Figs.\ \ref{Sn-region-delta-r2-individual}(g) and
 \ref{Sn-region-delta-r2}(b), (e)].  This is caused by the drift of the prolate minimum from
 $\beta_2 \approx 0.2$ to $\beta_2\approx 0.4$ (see Fig.\ 7 in the supplemental material).
The experimental  $\delta \left< r^2 \right>^{N,82}$ values of the Sm $(Z=62)$ isotopes are slightly
higher than those of the Nd ones and are decreasing with increasing neutron number in the
$N=76-80$ range [see \ref{Sn-region-delta-r2-individual}(h)]. These two features are reasonably
well described in the model calculations. The only exception is the DD-ME$\delta$ functional which
does not predict this decreasing trend [see Figs.\ \ref{Sn-region-delta-r2}(c)].

\begin{figure*}[htb]
\centering
\includegraphics[width=17.0cm]{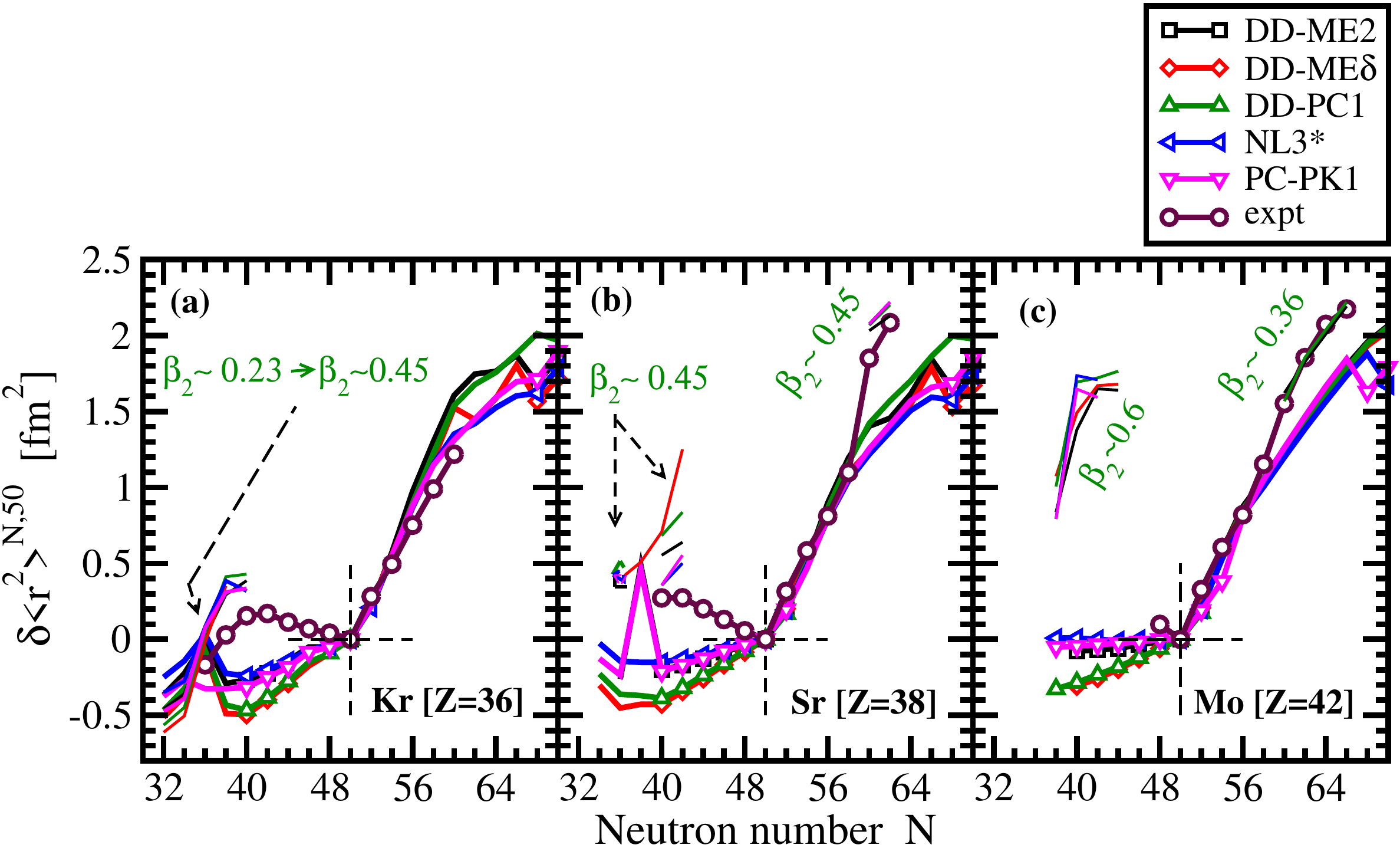}
\caption{The $\delta \left < r^2 \right>^{N,50}$ values of the isotopic chains in the Sr
region.  
The calculated charge radii in excited prolate minima are shown by the  thin lines
without symbols. The typical deformation in these minima are indicated. 
\label{Sr-region-delta-r2-individual}
}
\end{figure*}

\begin{figure*}[htb]
\centering
\includegraphics[width=16.0cm]{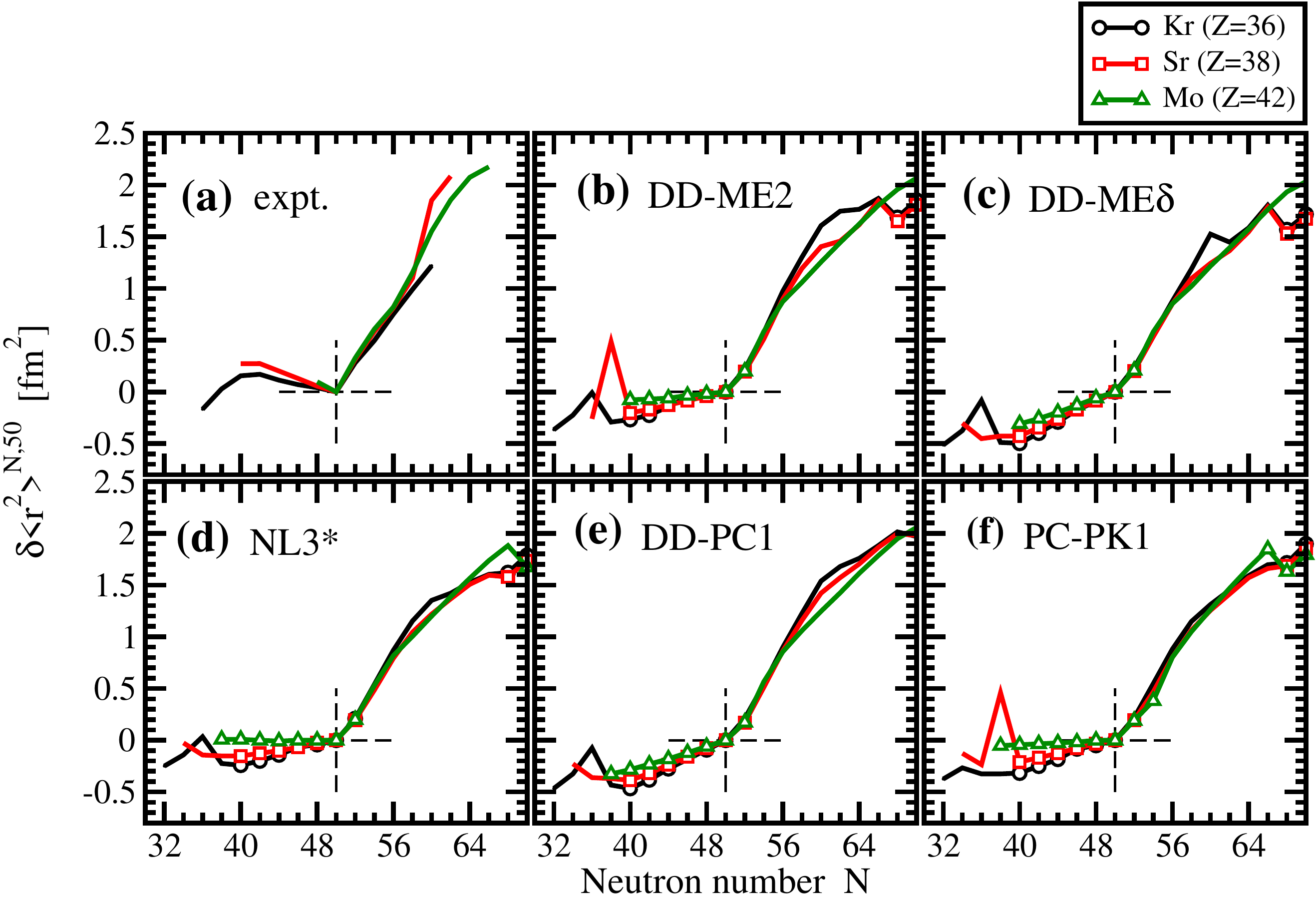}
\caption{The same as Fig.\ \ref{Pb-region-delta-r2} but for the $\delta \left < r^2 \right>^{N,50}$
values of the Kr ($Z=36$), Sr ($Z=38$),  and  Mo ($Z=42$) isotopic chains. 
\label{Sr-region-delta-r2}
}
\end{figure*}

   Let us consider the evolution of differential charge radii in the $N>82$ nuclei.
Note that experimental data for the Cd isotopes stops at $N=82$ and the one for the Sn
and Te isotopic chains at $N=84$. The isotopic Ce and Nd chains extend up to $N=90$,
those of  Xe, Ba, and Sm up to $N=92$, and the Gd isotopic chain up to $N=96$. The
substantial kink in charge radii at $N=82$ is present in all these isotopic chains; the
$\delta \left< r^2 \right>^{84,82}$ values range from 0.226 fm$^2$ in Sn up to 0.297 fm$^2$
in Gd. These kinks are reasonably well described in the majority of the calculations
(see Fig.\ \ref{Sn-region-delta-r2-individual}). Note also that the calculations
reasonably well reproduce the $N>82$ branches of differential charge radii.

  The experimental differential charge radii of the indicated isotopic chains cluster in the $N>82$
nuclei [see Fig.\ \ref{Sn-region-delta-r2}(a)] and this feature is reproduced well only in the calculations
with DD-ME$\delta$ (compare panels (c) and (a) in Fig.\ \ref{Sn-region-delta-r2}).
The spread of the $\delta \left< r^2 \right>^{90,82}$ values  obtained in the calculations with
DD-ME2, NL3*, DD-PC1, and PC-PK1 for the isotopic chains under study is larger by a factor of
approximately two than that seen in the experiment (compare panels (b), (d), (e), and (f) with panel
(a) in Fig.\ \ref{Sn-region-delta-r2}). Note that theoretical results for the Cd, Sn,  and Te isotopic
chains have to be ignored in such a comparison since experimental data in these chains extends
only up to either $N=82$ or 84.

 The calculations indicate that only the $N=82$ and 84 Xe, Ba, Ce, Nd, Sm, and Gd isotopes
are spherical in their ground states, while higher $N$ isotopes are prolate [with a pair of
exceptions for DD-ME$\delta$] (see Fig. 4 in the supplemental
material). This suggests that the clustering of differential charge radii of these isotopic chains
has to be, in part, traced back to the similarity of calculated deformations. Indeed, in the neutron 
number range of $82-92$, the smallest spread of calculated  deformations is seen in the calculations
with DD-ME$\delta$ (see Fig. 4(b) in the supplemental material), and this functional provides the best
description of the clustering (see Fig.\ \ref{Sn-region-delta-r2}). However, this is probably not the complete
picture since the spread of calculated deformations for DD-PC1 is only slightly higher
than in the calculations with DD-ME$\delta$ (compare panels (b) and (d) in Fig. 4 of the supplemental
material), but it does not produce a good description of clustering.  As a result, alternative sources of
the clustering of differential charge radii for the nuclei above shell closures may be possible. One possibility
is that it is related to the lowering of the energy of the $1h_{9/2}$ neutron subshell to the vicinity of
the $N=82$ shell closure, which is present only for the DD-ME$\delta$ functional
(see Fig.\ \ref{Sn132-sp-spectra}).

   It is interesting to compare the situation with the clustering of differential charge
radii  above the neutron shell closure in the Pb and Sn regions. This clustering in the Pb region
for $N>126$ is defined by only three isotopic chains, namely, Pb, Po, and Rn
(see Fig.\ \ref{Pb-region-delta-r2}). Moreover, the calculations suggest that the nuclei in the 
$N=126-130$ range are spherical, and that only a weak deformation  $(|\beta_2\approx 0.1|$) 
develops for $N=132, 134$, and 136
(see Fig.\ \ref{Pb-region-deformation}), the highest neutron numbers in the experimentally observed
isotopic chains of Pb, Po, and Rn, respectively. So, the latter nuclei remain quasi-spherical. All these
factors explain why it is easier to reproduce the clustering of differential charge radii in the
Pb region as compared with the Sn/Gd one.

\begin{figure}[htb]
\centering
\includegraphics[width=8.4cm]{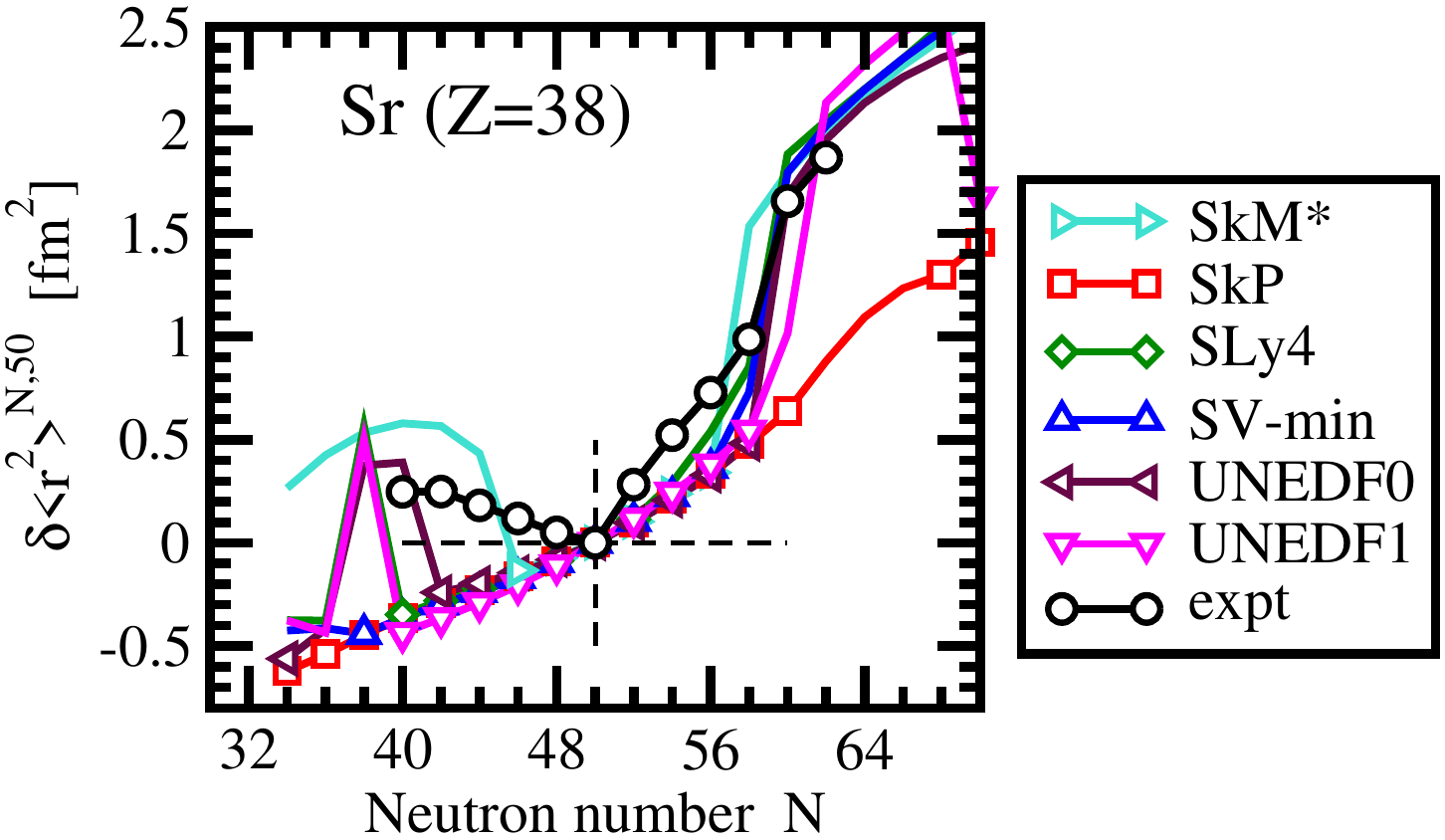}
\caption{The $\delta \left < r^2 \right>^{N,50}$ values for the
Sr ($Z=38$) isotopes.
The experimental data are taken from Ref.\ \cite{AM.13} and the results of
the Skyrme DFT calculations from Mass Explorer \cite{Mass-Explorer}.
\label{Sr-isotopes-Skyrme}
}
\end{figure}

\section{The charge radii in the isotopic chains of the Sr region}
\label{Sr-region-section}

   The isotopic chains of Kr, Sr, and Mo allow us to test how the differential radii
are modified when the $N=50$  shell closure is crossed. In Fig.\
\ref{Sr-region-delta-r2-individual} the experimental
data for these chains are compared with the results of the calculations.

   Let us first consider the differential radii for the $N>50$ isotopes.
In the Kr isotopes, the calculations reproduce rather well the experimental data for
$N=52$ and 54, but start to overestimate it for $N=56$, 58 and
60  with the biggest overestimate given by the DD* functionals [see Fig.\
\ref{Sr-region-delta-r2-individual}(a)]. All functionals rather well reproduce the experimental differential radii
of the Sr isotopes for $N=50-58$ [see Fig.\ \ref{Sr-region-delta-r2-individual}(b)].
Note that in both isotopic chains, the ground states have moderately oblate deformation with
$\beta_2 \approx -0.18$ above $N=52$ (see Figs. 14 and 16  in the supplemental
material). Except for PC-PK1, the differential charge radii
of the Mo isotopes are also rather well reproduced for $N=50-58$
[see Fig.\ \ref{Sr-region-delta-r2-individual}(c)].  Note that the calculations
show a competition of prolate and oblate shapes in this neutron number
range (see Fig. 14 in the supplemental material).

  The differential charge radii drastically increase for $N=60$ and 62 in the Sr
isotopes and $N=60, 62$, and 64 in the Mo isotopes [see Figs.\
\ref{Sr-region-delta-r2-individual}(b) and (c)].  This increase is related to the
transition of the ground states to a prolate minimum with $\beta_2 \approx 0.4$.
Such a minimum is excited by 100-200 keV in the Sr $N=60$ and 62 isotopes
in the calculations with DD-ME2, DD-PC1, and NL3*
(see Figs. 16 (n) and (o) in the supplemental material), and by
approximately 1 MeV as compared with the oblate
ground  state minimum in the $N=60, 62$, and 64 Mo isotopes (see Figs. 15 (l), (m)
and (n) in the supplemental material). Note that for some functionals and some
neutron numbers such a minimum is either non-existent [appears as shoulder in the PEC] or
separated by a very small barrier from the minima with lower deformation. Figs.\ \ref{Sr-region-delta-r2-individual}(b) 
and (c) illustrates that experimental $\delta\left< r^2 \right>^{N,50}$ values for these neutron
numbers are rather well reproduced if the calculated charge radii in this minimum are used for comparison.

 It is interesting to compare the performance of relativistic and non-relativistic functionals 
in the description of differential charge radii in the $N>50$ nuclei of the Sr region. The results of the
calculations with non-relativistic Skyrme functionals for the Sr isotopes are presented in Fig.\
\ref{Sr-isotopes-Skyrme}. The best reproduction of experimental data in the
$N=50-58$ range is provided by the SLy4 functional, but in general, the Skyrme
functionals provide a less accurate description of  the $\delta\left< r^2 \right>^{N,50}$
values in this neutron number range as compared with the CEDFs.  At higher neutron number,
except for the SkP functional, the prolate minimum  with $\beta_2 \approx 0.37$ is
the lowest in energy in the Skyrme DFT calculations, and this allows explaining the experimental data
at $N=60$ and 62.

Similar results  have also been obtained in Ref.\ \cite{RSRP.10} in calculations with Gogny D1S for the 
Sr, Mo, and Zr isotopic chains.  They underestimate experimental
$\delta\left< r^2 \right>^{N,50}$ values for $N=52-58$ but correctly predict the existence of a
highly deformed prolate minimum above $N=58$.  This minimum is the lowest in energy
in the Sr isotopes but an excited configuration in Mo. Moreover, similar to our results
for these shapes, they overestimate $\delta\left< r^2 \right>^{N,50}$ values for the Sr isotopes,
but correctly reproduce them for Mo. Despite all these differences,
both relativistic and non-relativistic functionals predict a similar trend of the evolution of charge
radii with increasing neutron numbers above $N=50$ provided that the correct minimum is associated
with the experimental data.

   Both covariant and non-relativistic DFTs fail to reproduce the evolution
of experimental $\delta\left< r^2 \right>^{N,50}$ curve in the $N=40-50$ Kr and Sr
isotopes (see  Figs.\ \ref{Sr-region-delta-r2-individual} and \ref{Sr-isotopes-Skyrme} in
the present paper and Fig.\ 7 in Ref.\ \cite{DGLGHPPB.10}). This is, because in the absolute
majority of the cases, these models predict spherical ground states for these isotopes
(see Figs. 14, 15, 16 and 17 in the supplemental material for CDFT results, Fig.\ \ref{Sr-isotopes-Skyrme}
and  Mass Explorer at FRIB \cite{Mass-Explorer} for Skyrme DFT results and Ref.\
\cite{Gogny-compilation} for Gogny DFT results).  However,
the PECs obtained in the CDFT calculations indicate the presence of a prolate minimum with
$\beta_2 \approx 0.5$ in the $N=36-42$ Sr isotopes (see Figs. 16 (b), (c), (d) and (e) in the
supplemental material) which [with the exception of the calculations with a few functionals
in the $N=38$ isotope]  is the excited one. The calculated charge radii in this minimum
somewhat overestimate experimental data [see the lines without symbols in Fig.\
\ref{Sr-region-delta-r2-individual}(b)].  The existence of similar prolate minima is seen
also in the Skyrme DFT calculations: it becomes the lowest in energy in the calculations
with SkM*  for $N=34-44$, with UNEDF1 and SLy4 at $N=38$ and  with UNEDF0
at $N=38-40$ (see Fig.\ \ref{Sr-isotopes-Skyrme}).  Similar to our results, the Skyrme
calculations for this minimum somewhat overestimate experimental
$\delta\left< r^2 \right>^{N,50}$  values. An excited prolate minimum exists also in PECs of the Kr
isotopes in the CDFT calculations: its deformation drifts from $\beta_2 \approx 0.45$ for $N=40$ and 42 down to
$\beta_2 \approx 0.25$ for $N=34$ and 36 (see Fig. 17 in the supplemental material). This drift
explains the experimentally observed decrease of  $\delta\left< r^2 \right>^{N,50}$ on going
from $N=40$ to $N=36$ [see \ref{Sr-region-delta-r2-individual}(b)].

  Note that the PECs obtained for the Kr and Sr isotopes in the Gogny DFT calculations
with the  D1S force (see Ref.\ \cite{Gogny-compilation}) are very similar to those obtained in our
calculations.  The inclusion of the correlations beyond mean-field within the framework
of a five-dimensional collective quadrupole Hamiltonian based on the Gogny DFT allows to
improve the description  of charge radii in the $N<50$ Sr isotopes (see the discussion of
Fig. 7 in Ref.\ \cite{DGLGHPPB.10}).  Considering the similarity of mean-field PECs
obtained in the CDFT and Gogny DFT calculations, it is reasonable to expect that the
inclusion of the correlations beyond mean-field will also improve the description of
charge radii in the $N<50$ nuclei of the Sr region in the CDFT framework.

   Finally, Fig.\ \ref{Sr-region-delta-r2} compares the relative properties of
differential charge radii of the Kr, Sr, and Mo isotopic chains obtained in
the calculations with the employed functionals. One can see that the clustering
of these radii seen for $N=50-58$ in the experiment is reasonably well reproduced
in the model calculations.

\begin{table}[h!]
\begin{center}
\caption{The experimental and calculated $\delta \left < r^2 \right>^{28,20}$ values [in fm$^2$] of
the Ca ($Z=20$) isotopes and their connection to the nuclear matter properties (such as symmetry
energy $J$ and its slope $L_0$) for the employed CEDFs. "N/A" means that the data are either
not applicable or not available. The functionals are arranged in such a way that the calculated
$\delta \left < r^2 \right>^{28,20}$ values decrease. References for either the
functionals or for related results are shown in the first column.
\label{table-Ca-4840-radii+NMP}
}
\begin{tabular}{l|c|c|c|} \hline \hline
    CEDF            & $\delta \left < r^2 \right>^{28,20}$ [fm$^2$]    &  $J$ [MeV]   & $L_0$ [MeV]  \\ \hline
exper.  \cite{Ca-radii.2016}    &-0.001  &  N/A          &     N/A     \\
NL-IT  \cite{RF.95}             & $\approx 0.06$    &  39.4        &        N/A         \\
NL-SH \cite{NLSH}               & 0.040  &  36.13   & 113.68                \\
NL5(E) \cite{AAT.19}            & 0.031  &  38.93   & 124.96   \\
NL5(D) \cite{AAT.19}            & 0.003  &  38.87   & 123.98   \\
NL1  \cite{NL1}                 &-0.006  &  43.46   & 140.07    \\
NL-I   \cite{RF.95}             & $\approx -0.01$   &  39.7        &        N/A         \\
NL3  \cite{NL3}                 &-0.014  &  37.40   &    118.53             \\
NL-Z \cite{RF.95}               &-0.015  &  41.72   &   133.91              \\
NL3*  \cite{NL3*}               &-0.028  &  38.68   & 122.60          \\
DD-MEX \cite{TAAR.20}           &-0.056  &  32.87   &  47.81                       \\
NL5(A) \cite{AAT.19}            &-0.088  &  34.92   & 108.85     \\
NL5(C) \cite{AAT.19}            &-0.092  &  35.925  & 112.31  \\
NL5(B) \cite{AAT.19}            &-0.094  &  34.92   & 108.33   \\
PCPK1  \cite{PC-PK1}            &-0.098  &  35.60   & 113.00         \\
DDME2  \cite{DD-ME2}            &-0.111  &  32.40   & 49.40         \\
DD-PCX  \cite{DD-PCX}           &-0.178  &  31.12 &  46.32  \\
DD-PC1 \cite{DD-PC1}            &-0.229  &  33.00   & 68.40        \\
DDME$\delta$ \cite{DD-MEdelta}  &-0.296  &  32.35   & 52.90          \\ \hline
\end{tabular}
\end{center}
\end{table}

\begin{figure}[htb]
\centering
\includegraphics[width=8.0cm]{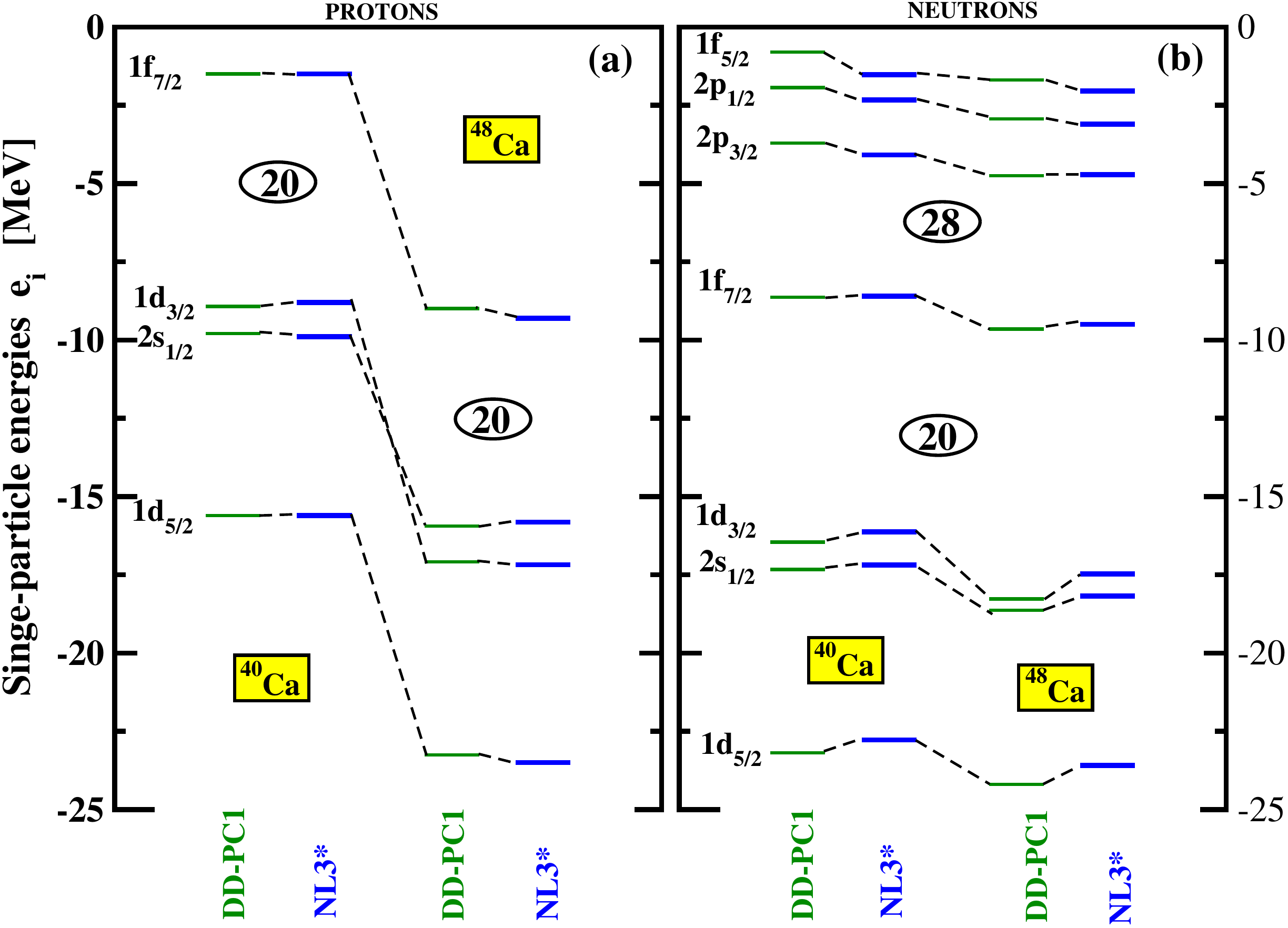}
\caption{The same as Fig.\ \ref{208Pb-single-particle} but for the $^{40,48}$Ca isotopes.
}
\label{Ca40-48-spe}
\end{figure}

\section{The charge radii in the isotopic chains of the Ca region}
\label{Ca-region-section}

\begin{figure}[htb]
\centering
\includegraphics[width=8.4cm]{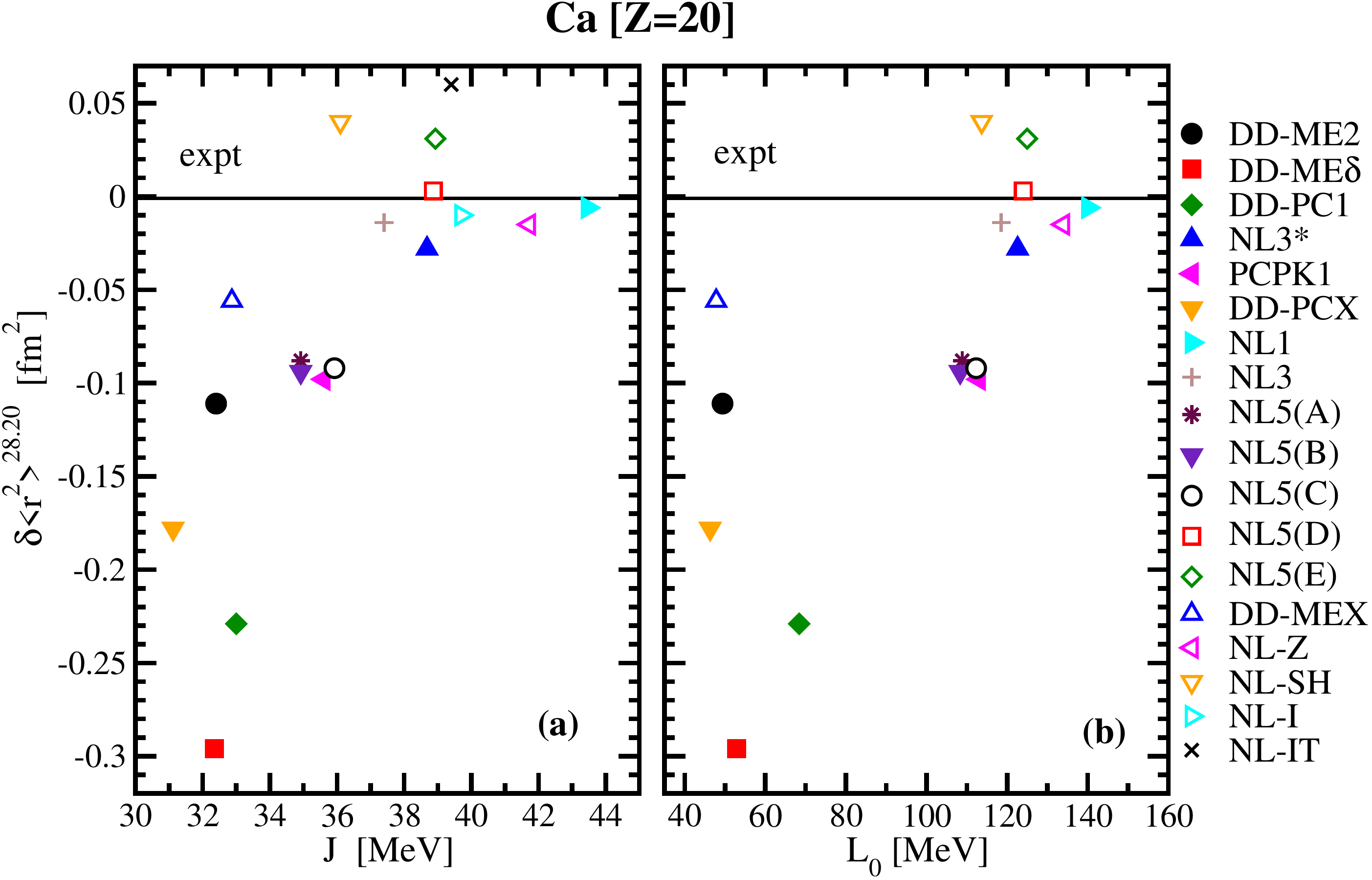}
\caption{The relations between the calculated $\delta \left< r^2 \right>^{28,20}$ values
in the Ca isotopes and the symmetry energy $J$ and its slope $L_0$ of employed functionals.
See Table \ref{table-Ca-4840-radii+NMP} for additional details.
\label{JL-rad}
}
\end{figure}

   The Ca isotopes have been in the focus of extensive experimental
and theoretical studies over the years. In the context of the studies of charge radii,
there are two puzzling features of these isotopes, namely, (i) almost  exactly the
same  charge radii of the $^{40,48}$Ca isotopes and (ii) a large and unexpected increase
of charge radii in neutron-rich beyond $N=28$ nuclei (see Ref.\ \cite{Ca-radii.2016}).

  First, let us consider the similarity of the charge radii in the $^{40,48}$Ca nuclei.
These two nuclei are doubly magic with a proton shell closure at $Z=20$ and the neutron shell
closures at $N=20$ and 28, respectively (see Fig.\ \ref{Ca40-48-spe}).  As a consequence,
proton and neutron pairings are expected either to collapse or to be extremely weak and
thus, these two nuclei are ideal candidates  for testing of the particle-hole channel of DFTs, underlying
EDFs and their isovector dependencies. This is because the theoretical results will not be
polluted by the uncertainties in the treatment of pairing. In addition, the PECs presented
in Figs.\ 22(e) and (i) of the supplemental material indicate extreme localization of the ground
states of these two nuclei at spherical shape with little or no expected impact from beyond
mean-field effects; these features also do not depend on the CEDF.

  Table \ref{table-Ca-4840-radii+NMP} and Fig.\ \ref{Ca-region-delta-r2} present 
the summary of published and newly calculated  $\delta \left < r^2 \right>^{20,28}$ values for 
the Ca isotopes and their connections with nuclear matter properties of the functionals. One
can see that  $\delta \left < r^2 \right>^{28,20} \approx 0$ is produced by the functionals (such
as NL3) characterized by large values of the symmetry energy $J \approx 40$ MeV and its slope 
$L_0 \approx 110$ MeV while the large negative values of  $\delta \left < r^2 \right>^{20,28}$ 
are produced by the CEDFs with low values of $J$ and $L_0$. The latter feature is also seen 
for traditional non-relativistic Skyrme EDFs such as SkM*, SkP, SLy4, SV-min,  UNEDF0 and
UNEDF1 which are characterized by low values of the symmetry energy  $J \approx 32$ MeV
and its slope $L_0 \approx 50$ MeV (see Fig.\ \ref{Ca-isotopes-Skyrme})\footnote{The
Fayans functionals of Ref.\ \cite{RN.17} are also able to describe
$\delta \left < r^2 \right>^{28,20} \approx 0$ but they are specifically designed for that
by the use of experimental value of $\delta \left < r^2 \right>^{28,20}$ in the fitting protocol.}.
It is not likely that the fact that $\delta \left < r^2 \right>^{28,20} \approx 0$ in the Ca
isotopes can be related to the details of the single-particle spectra since they are very
similar (especially for the proton subsystem) for the functionals which provide different predictions
for $\delta \left < r^2 \right>^{28,20}$ (see Fig.\ \ref{Ca40-48-spe}).

\begin{figure*}[htb]
\centering
\includegraphics[width=17.0cm]{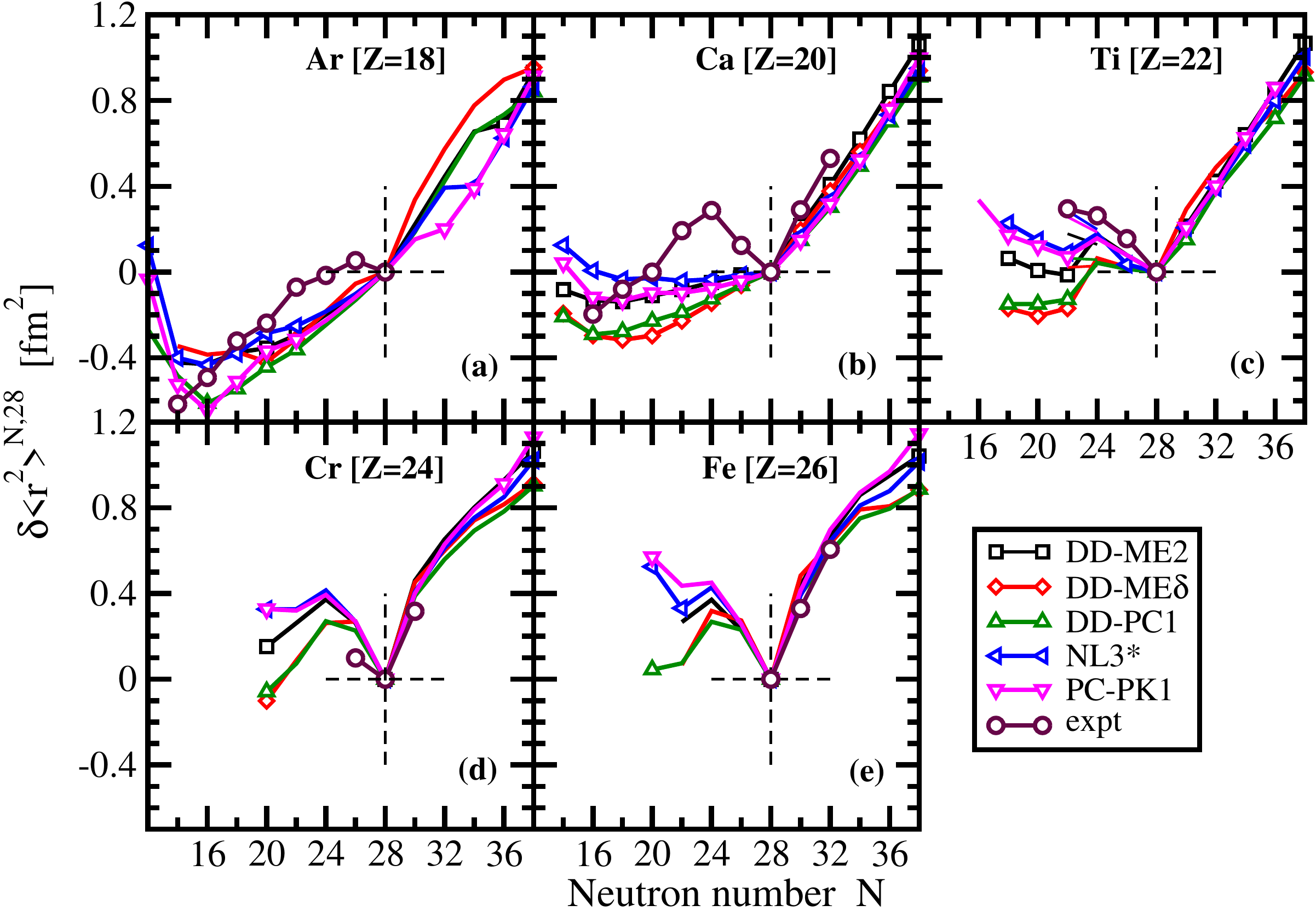}
\caption{The $\delta \left < r^2 \right>^{N,28}$ values of the isotopic chains in the Ca
region. The experimental data for the Ar, Ti, Cr and Fe isotopes are taken from Ref.\ \cite{AM.13}.
The experimental data for the Ca isotopes are mostly taken from Ref.\ \cite{Ca-radii.2016} while that
for the $^{39,41}$Ca isotopes from Ref.\ \cite{39-41Ca-radii.96} and for $^{36,37,38}$Ca from Ref.\
\cite{Ca-radii-prot-rich.19}.
\label{Ca-region-delta-r2-individual}
}
\end{figure*}

\begin{table}[h!]
\begin{center}
\caption{The experimental and calculated $\delta \left < r^2 \right>^{32,28}$ values [in fm$^2$] of
the  Ca ($Z=20$) and Fe ($Z=26$) isotopes.
\label{table-N32-N28-Ca-Fe}
}
\begin{tabular}{l|c|c} \hline \hline
    CEDF      & Ca      & Fe      \\ \hline
exper.                 &  0.530    &   0.606    \\
DDME2              &   0.407         &   0.670        \\
DDME$\delta$   &   0.377         &   0.633        \\
DD-PC1             &   0.302         &  0.591         \\
NL3*                  &   0.336         &   0.645         \\
PCPK1              &    0.317         &    0.697        \\ \hline	
\end{tabular}
\end{center}
\end{table}

\begin{figure}[htb]
\centering
\includegraphics[width=8.4cm]{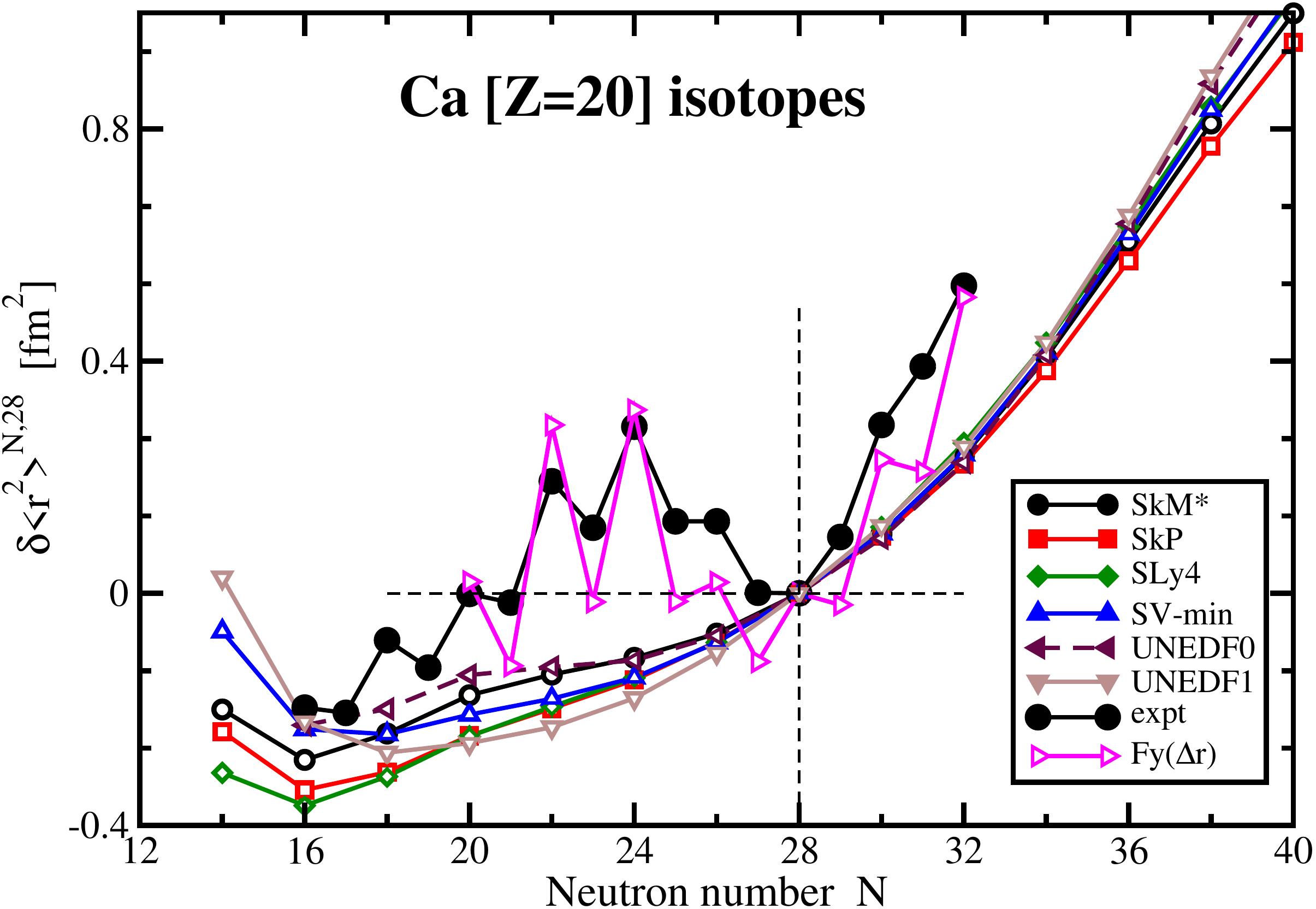}
\caption{The $\delta \left < r^2 \right>^{N,28}$ values of the Ca ($Z=20$) isotopes
relative to the $N=28$ isotope. The experimental  data for the Ca isotopes are mostly
taken from Ref.\ \cite{Ca-radii.2016} while that for the $^{39,41}$Ca isotopes from Ref.\
\cite{39-41Ca-radii.96} and for $^{36,37,38}$Ca from Ref.\ \cite{Ca-radii-prot-rich.19}. 
The results of the Skyrme DFT calculations are taken from Mass Explorer at FRIB
\cite{Mass-Explorer}. The results of the Fayans Fy($\Delta$r, HFB) functional are 
taken from Fig.\ 4 of Ref.\ \cite{RN.17}.
\label{Ca-isotopes-Skyrme}
}
\end{figure}

  Although there are the correlations between nuclear matter properties and  the  
$\delta \left < r^2 \right>^{20,28}$ values in the Ca isotopes, other factors contribute 
to experimental $\delta \left < r^2 \right>^{20,28} \approx 0$ value.

  First,  non-relativistic HFB  calculations of Ref.\ \cite{Nakada.15} with the semi-realistic 
M3Y-P6a interaction can reproduce $\delta \left < r^2 \right>^{28,20}$ and $\delta \left 
< r^2 \right>^{32,28}$ values in the Ca isotopes: this is attributed to the density 
dependence of the three-nucleon spin-orbit interaction and its impact on the density
distributions of specific single-particle orbitals.  However, these calculations fail to 
reproduce the inverted parabola-like behavior of differential charge radii for $N=22-26$.

  Second,  it is necessary to recognize that the functionals used in the present paper have 
been fitted in a time when the importance of the fine structure in the charge radii and alternative 
mechanisms contributing to them have not been completely recognized. As a result, an 
approximate Eq.\ (\ref{r_charge}) has been used for charge radii in the absolute majority of the
publications within the CDFT framework. However, a more general expression
for a charge radius in the CDFT is  given by \cite{HP.12,Kurasawa.19} 
\begin{eqnarray}
r^2_{ch} = \left< r^2_p \right>  + r_p^2  + \left< r^2_p \right>_{SO} 
                  + \frac{N}{Z}  \left( r^2_n +  \left< r^2_n \right>_{SO} \right)
\label{r_charge_general}
\end{eqnarray} 
where $<r_p^2>$ stands for proton mean square point radius, $r_p$ and $r_n$ 
for single proton and neutron radii, respectively, and $\left< r^2_p \right>_{SO}$ 
and $\left< r^2_n \right>_{SO}$ for proton and neutron spin-orbit contributions 
to the charge radius. So, Eq.\ (\ref{r_charge}) takes into account only the first two
terms of this general expression. 

  It turns out that Eq.\ (\ref{r_charge}) is a quite reasonable approximation to  Eq.\ 
(\ref{r_charge_general}) for medium and heavy mass nuclei especially for differential
charge radii. This is because the spin-orbit contribution to charge radii 
decreases with increasing  the mass of nuclei and it almost does not depend on the 
CEDF. These features are  illustrated in Table II of Ref.\ \cite{HP.12}. Since the 
calculations of this reference are restricted to spherical  shape and neglect pairing 
correlations,  the values quoted for spin-orbit  contribution to charge radii  in this table 
for non-doubly magic nuclei  have to be considered as an upper limit. This 
is because deformation and pairing give rise to the fragmentation of the spin-orbit 
strength which results in  a smoothing of the spin-orbit correction to charge radii as a 
function of particle  number \cite{RN.21}.  In addition, for existing experimental data the 
range of the variation of the $\frac{N}{Z}  r^2_n$ term is significantly smaller  in medium 
and heavy mass nuclei as compared with light nuclei and its contribution to the differential
charge radii is cancelled to a large degree.
 
  However, the last two terms of Eq.\ (\ref{r_charge_general}) are important
in light nuclei (see Refs.\ \cite{HP.12,Kurasawa.19}).  For example, when these terms
are taken into account the differential charge radius $\delta \left < r^2 \right>^{20,28}$ 
for the Ca isotopes changes from $-0.013$ fm$^{2}$ to $+0.164$ fm$^{2}$ in the NL3 
CEDF and from $-0.07$ fm$^{2}$ to $+0.108$ fm$^{2}$ in the NLSH one (see Table 
1 in Ref.\  \cite{Kurasawa.19}) moving away from experimental value of  $-0.001$ fm$^{2}$. 
Let us assume a similar range of corrections by these two terms of Eq.\ (\ref{r_charge_general})
for other functionals. Then the accuracy of the description of experimental 
$\delta \left < r^2 \right>^{20,28} \approx 0$ value after inclusion of these corrections
would degrade for the functionals  listed in upper part (down to NL3*) of Table  
\ref{table-Ca-4840-radii+NMP}, would remain similar for DD-MEX  and would improve 
for the functionals located in the bottom part of Table \ref{table-Ca-4840-radii+NMP}.

   An analysis of the contributions of spin-orbit densities and other terms to charge
radii has also been performed in the non-relativistic framework (see Refs.\ 
\cite{FN.75,RN.21,NCLX.21}).  There are some differences  between relativistic and 
non-relativistic treatments of these terms (see detailed discussion in Ref.\ 
\cite{Kurasawa.19}), however, in general  a comparable  modification of charge radii 
is generated by such terms in non-relativistic DFTs. For example,  these contributions 
change $\delta \left < r^2 \right>^{20,28}$ of the Ca isotopes
from $-0.198$ fm$^2$ to $-0.048$ fm$^2$ 
(see Table 1 in Ref.\  \cite{Kurasawa.19} and Fig.  1 in Ref.\ \cite{NCLX.21} in the 
Skyrme DFT calculations with the SLy4 functional bringing them closer to experimental
data. Similar results are obtained for the Skyrme SV-bas and Fayans Fy($\Delta r$, HFB) 
functionals in Ref.\ \cite{RN.21} with  spin-orbit density and $\frac{N}{Z}  r^2_n$ term 
providing approximately 0.1 fm$^2$ and 0.04 fm$^2$ contributions to the  $\delta \left < r^2 \right>^{20,28}$ 
value, respectively. Assuming that similar corrections appear for other non-relativistic functionals 
shown in Fig.\ \ref{Ca-isotopes-Skyrme}, it is clear that their addition will improve the description 
of experimental $\delta \left < r^2 \right>^{20,28}$ value in these functionals. However, their 
addition do not allow either to  reproduce the inverted parabola-like behavior  of differential charge 
radii for $N=22-26$ or improve the description of the large $\delta \left < r^2 \right>^{32,28}$ experimental 
value in the Ca isotopes (see Fig.\ 1 in Ref.\ \cite{NCLX.21}). The latter is because spin-orbit 
densities do not modify substantially the calculated $\delta \left < r^2 \right>^{32,28}$ value.
 
   Since the experimental data on charge radii is included in the fitting protocols
(which rely on Eq.\ (\ref{r_charge}) for the definition of charge radii) of the CEDFs 
employed in the present paper, they partially include the corrections provided by
additional terms of Eq.\ (\ref{r_charge_general}). Thus, to avoid double counting
of these corrections, new fits of CEDFs with charge radii defined by Eq.\ 
(\ref{r_charge_general})  are needed. They will provide a more consistent 
(and, hopefully, more accurate) description of charge radii.  However, in the context of 
the present study with the employed functionals one can conclude that for a given isotopic 
chain the relative properties of charge radii and differential charge radii provided by two
CEDFs should not be very much affected by these corrections. This is 
because existing studies indicate their weak dependence  on the functional.
On the other hand, the calculated absolute values of these radii are expected to be
partially affected by these corrections and this fact is taken into account in further
discussion. However, this effect is expected to be reasonably small for the calculated
differential charge radii of isotopic chains in the experimentally available range in 
neutron number if this range is rather short (see Ref.\ \cite{RN.21}). These are Ti, Cr 
and Fe isotopic chains (see, for example, Fig.\ \ref{Ca-region-delta-r2-individual}).

  These new functionals are also needed to answer the question on whether
$\delta \left < r^2 \right>^{20,28}\approx 0$ of the Ca isotopes can provide a 
meaningful constraint on nuclear matter properties. The discussion provided 
above on  that issue is not conclusive. The analysis of proton density 
distributions determined by electron scattering and the neutron density distributions
determined by proton elastic scattering in the $^{40,48}$Ca nuclei supplemented
by the studies within Skyrme DFT and CDFT support low  values of $J\approx 28$ MeV and 
$L=28-50$ MeV \cite{ZUSY.21}.  In contrast, the recent PREX-II measurements of 
the neutron skin in $^{208}$Pb \cite{PREX-II.21} imply a stiff equation of state with 
large  $J=38.1 \pm 4.7$ and $L_0= 106\pm 37$ MeV values \cite{RFHP.21}.

 The calculated and experimental differential charge radii are compared in Figs.\
 \ref{Ca-region-delta-r2-individual} and \ref{Ca-region-delta-r2}.
One can see that all employed CEDFs fail to describe the evolution of differential
charge radii in the Ca isotopes in the neutron number range $N=16-28$ and especially
the peak at $N=24$ [see Fig.\ \ref{Ca-region-delta-r2-individual}(b)]. However, the
same problem exists also in all non-relativistic Skyrme EDFs (see Fig.\
\ref{Ca-isotopes-Skyrme}).  As discussed above, the accounting 
of spin-orbit densities and the $\frac{N}{Z} \left< r_n^2 \right>$ term in Eq.\ 
(\ref{r_charge_general}) can somewhat modify this situation but in no way will it
correct the problem with the description of the peak at $N=24$. The analysis of the occupation 
probabilities indicates that mostly neutron $1f_{7/2}$ states are occupied in the transition 
from $^{40}$Ca to $^{48}$Ca, and this leads either to a linear increase or nearly constant 
differential charge radii in conventional functionals. This figure also indicates that the inverted
parabola-like behavior of differential charge radii in the $N=20-28$ isotopes is reproduced
on average only in the Fayans Fy($\Delta$r) functional which includes gradients both in
surface and pairing terms and was fitted to experimental (absolute and relative) data on
charge radii in $^{40,44,48}$Ca in Ref.\ \cite{RN.17}.

The rise in the differential charge radii above the $N=28$ shell closure is underestimated in 
the model calculations (see Fig.\ \ref{Ca-region-delta-r2-individual}(b) and Table
\ref{table-N32-N28-Ca-Fe}).  The non-relativistic results of Ref.\ \cite{NCLX.21}
indicate that the  $\delta \left < r^2 \right>^{32,28}$ value is only moderately affected
by spin-orbit densities. If that result holds also in the CDFT, then the best reproduction of experimental
$\delta \left < r^2 \right>^{28,32}$ values is achieved by the DD-ME2 functional
which underestimates it only by  23\% (see Table \ref{table-N32-N28-Ca-Fe}).
Note that the problem with the description of the rise in charge radii in the Ca isotopes above
$N=28$ also exist in non-relativistic Skyrme and Gogny calculations and in non-relativistic ab initio
 calculations (see, for example,  Fig.\ \ref{Ca-isotopes-Skyrme} in the present
paper and Fig. 3 in Ref.\ \cite{Ca-radii.2016}).

   The slope of the experimental $\delta \left < r^2 \right>^{N,28}$ curve for the Ar
($Z=18$) isotopes is best  reproduced in the calculations with the DD-PC1 and PC-PK1
functionals [see Fig.\ \ref{Ca-region-delta-r2-individual}(a)].  However, even these functionals
underestimate the experimental data by $\approx 0.2$ fm$^2$.  Based on
available estimates in this mass region (see Refs.\ \cite{HP.12,Kurasawa.19,RN.21,NCLX.21}), 
the contribution of spin-orbit densities to the differential charge is below this value.
 Note that in the calculations, many of the Ar isotopes are oblate in their ground 
states (see Fig. 18 in the supplemental material) and that the PECs of almost all Ar isotopes 
are extremely soft in quadrupole deformation (see Fig. 23 in the supplemental material). In 
such a situation, the correlations beyond mean-field are expected to play an important role.

  The differential charge radii in the Ti ($Z=22$) isotopes are gradually decreasing from
$\delta \left < r^2 \right>^{N,28} \approx 0.3$ fm$^2$ at $N=22$ down to zero at
$N=28$  (see Fig.\ \ref{Ca-region-delta-r2-individual}(c)). This trend is reasonably
well reproduced for $N=24-28$ by  the NL3*, PC-PK1, and DD-ME2 functionals. This is due
to two facts. First, the $N=24$ and 26 isotopes are prolate in the calculations
with $\beta_2 \approx 0.15$ (see Fig. 18 in the supplemental material), so their charge radii are
larger than those at the spherical shape. Second, these are two functionals that reproduce
reasonably well the near equality of charge radii in $^{40,48}$Ca (see Table
\ref{table-Ca-4840-radii+NMP}).  Other functionals (DD-PC1 and DD-ME$\delta$), which 
cannot reproduce this feature, also fail to describe the evolution of charge radii in the Ti
isotopes. Note that in the calculations, the PEC of the N=22, 24, and 26 isotopes are soft in
quadrupole deformation, which suggests that beyond mean-field effects could play an important
role in the reproduction of experimental data. Assuming that these effects could lead to the creation
of a prolate minimum with $\beta_2\approx 0.2$ in the $N=22$ isotope, this would also allow explaining
 the charge radius of this isotope in the calculations with PC-PK1, DD-ME2, and NL3* (see
results presented by thin solid magenta, black and blue lines in Fig.\
\ref{Ca-region-delta-r2-individual}(c)).

   The experimental differential charge radii of the Cr ($Z=24$) isotopes
show asymmetric parabola-like features at $N=26-30$ with the minimum at $N=28$
[see Fig.\ \ref{Ca-region-delta-r2-individual}(d)].  The experimental $\delta \left < r^2 \right>^{26,28}$
value is overestimated by a factor of approximately 1.5 in all calculations.
Note that in the model calculations, the $N=26$ and $N=30$ isotopes are prolate with
$\beta_2 \approx 0.22$, while the $N=28$ isotope is spherical (see Fig. 18 in the supplemental
material).  The PECs of these isotopes are soft in quadrupole deformation (see Fig. 19
in the supplemental material), so the ground-state properties of these nuclei could be
affected somewhat by beyond mean-field correlations.

    The experimental differential charge radii of the Fe $(Z=26)$ isotopes are available
for $N=28-32$ [see Fig.\ \ref{Ca-region-delta-r2-individual}(e)]. The large increase in
$\delta \left < r^2 \right>^{N,28}$ observed in the experiment above $N=28$ is rather well reproduced in
model calculations (see above mentioned figure and Table \ref{table-N32-N28-Ca-Fe}).  The
$N=28$ isotope is spherical or quasi-spherical in model calculations, while the $N=30$ and 32
isotopes are prolate with a quadrupole deformation of $\beta_2 \approx 0.23$ (see Figs. 18 and
19 in the supplemental material).

   It is clear that the description of differential charge radii in the Ca region represents a case of the
"hit-or-miss" situation when some data are rather well described while others are difficult to reproduce.
This is especially the case for the relative properties of differential charge radii of different isotopic chains.
Figure \ref{Ca-region-delta-r2}  shows that it is more difficult to reproduce them in the Ca region as compared 
with other regions.  It is expected that this conclusion will not be affected by the inclusion of the 
spin-orbit densities or the $\frac{N}{Z} \left< r_n^2 \right >$ term.

  It is interesting that the $N=22-28$ Ca and Ti isotopes show very similar trends in differential charge radii
[see Fig.\ \ref{Ca-region-delta-r2}(a)]. It is also likely that the Cr isotopes will  follow the same
trend  if experimental data are extended to lower neutron numbers.  The challenge here lies in the fact that
in the model calculations, the $N=22-26$ isotopes are prolate and spherical in the Ti and Ca isotopic chains,
respectively (see Fig. 18 in the supplemental material).  In  contrast to the experiment, this suggests a quite
substantial difference in charge radii. Indeed, the model calculations with some functionals can reproduce 
the data in the Ti isotopes but fail to do that in the Ca ones.

 One possible exit from this
contradiction is the possibility that the $N=22-26$ Ca isotopes are much softer in quadrupole deformation
as compared with the suggestion provided by PECs obtained at the mean-field level (see Fig. 22 of the
supplemental material).  The  low excitation energy of the superdeformed (SD) band in
$^{40}$Ca (see Ref.\ \cite{Ca40-PRL.01}), which is the lowest amongst all SD bands in the nuclear chart,
may point to a substantial softness of the Ca isotopes. To our knowledge, mean-field calculations
substantially overestimate the excitation energy of this SD band and thus, quite likely, predict stiffer Ca
isotopes.   The enhanced influence of the correlations beyond mean-field on the charge radii of the Ca
isotopes as compared with the Sn and Pb isotopes has already been pointed in Refs.\ \cite{RF.95,FTTZ.00}.
However, its magnitude and neutron number dependence are not sufficient to reproduce the peak in the differential
charge radii at $N=24$.  If the mean-field PECs are softened (especially for the $N\neq 20,28$
isotopes) then the correlations beyond mean-field could be enhanced and maybe the experimental peak
in  $\delta \left < r^2 \right>^{N,20}$ at $N=24$ could be reproduced.  In contrast, the analysis of 
Ref.\ \cite{BB.85}  partially based on the QRPA calculations  indicates that the ground state correlations associated 
with the surface modes of the Ca isotopes are important and that they qualitatively explain the observed inverted
parabola-like behavior  of differential charge radii with neutron number for $N=20-28$. However, these 
calculations rely on a number assumptions which have to be verified in more microscopic calculations. 
All these calculations point to  potential limitations of the mean field approximation in the description of the 
ground state properties of light nuclei such as the Ca isotopes and the need to include correlations
beyond mean field. 

\begin{figure*}[htb]
\centering
\includegraphics[width=18.0cm]{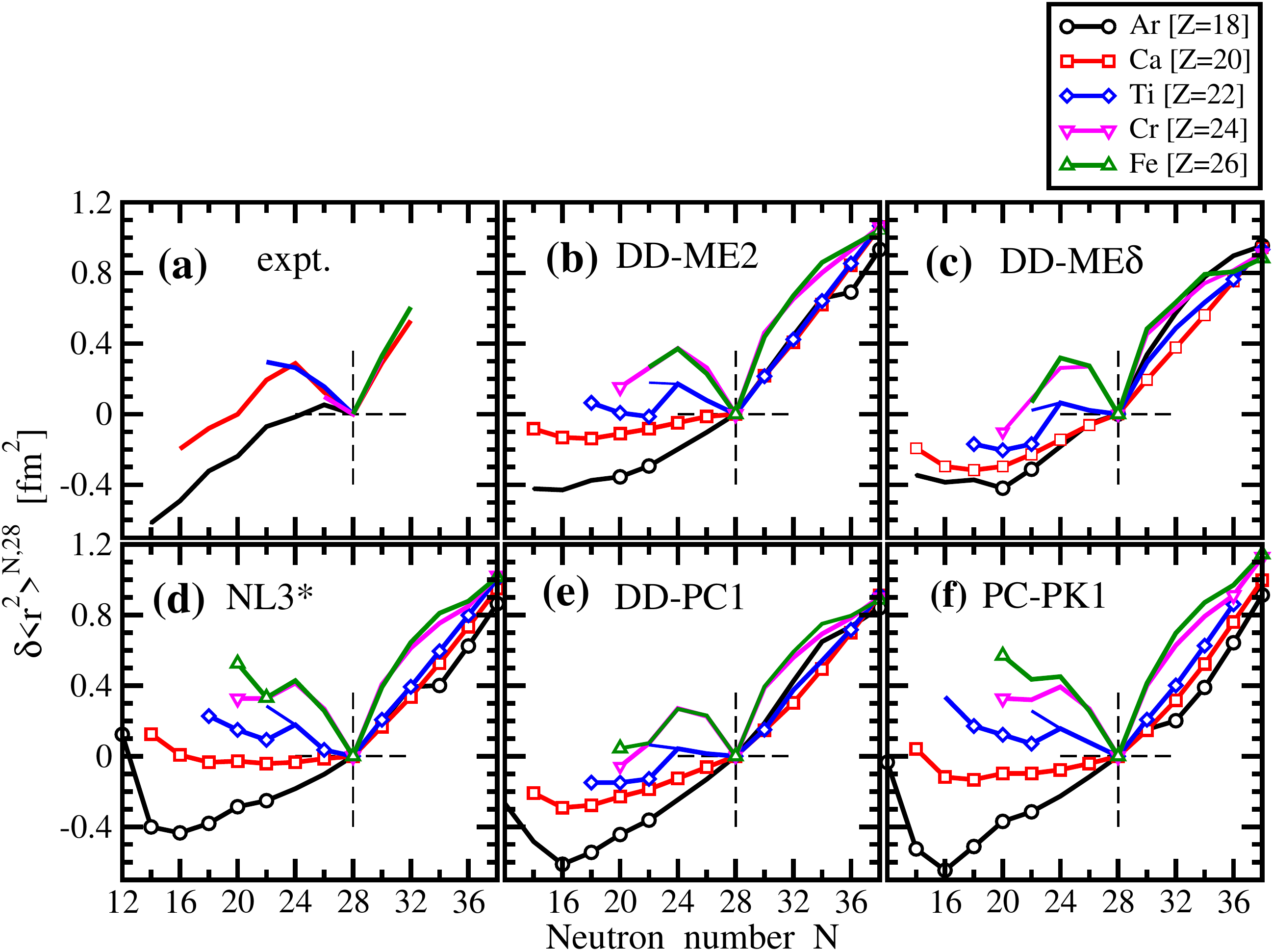}
\caption{The same as Fig.\ \ref{Pb-region-delta-r2} but for the $\delta \left < r^2 \right>^{N,28}$
values of the Ar ($Z=18$), Ca ($Z=20$),   Ti ($Z=22$) and  Fe ($Z=26$) isotopic chains. Thin
blue solid line connecting $N=22$ and $N=24$ illustrates the situation which would emerge if
the $N=22$ Ti nucleus would be deformed with $\beta_2=0.2$ instead of  being spherical.
}
\label{Ca-region-delta-r2}
\end{figure*}

\begin{figure*}[htb]
\centering
\includegraphics[width=4.62cm]{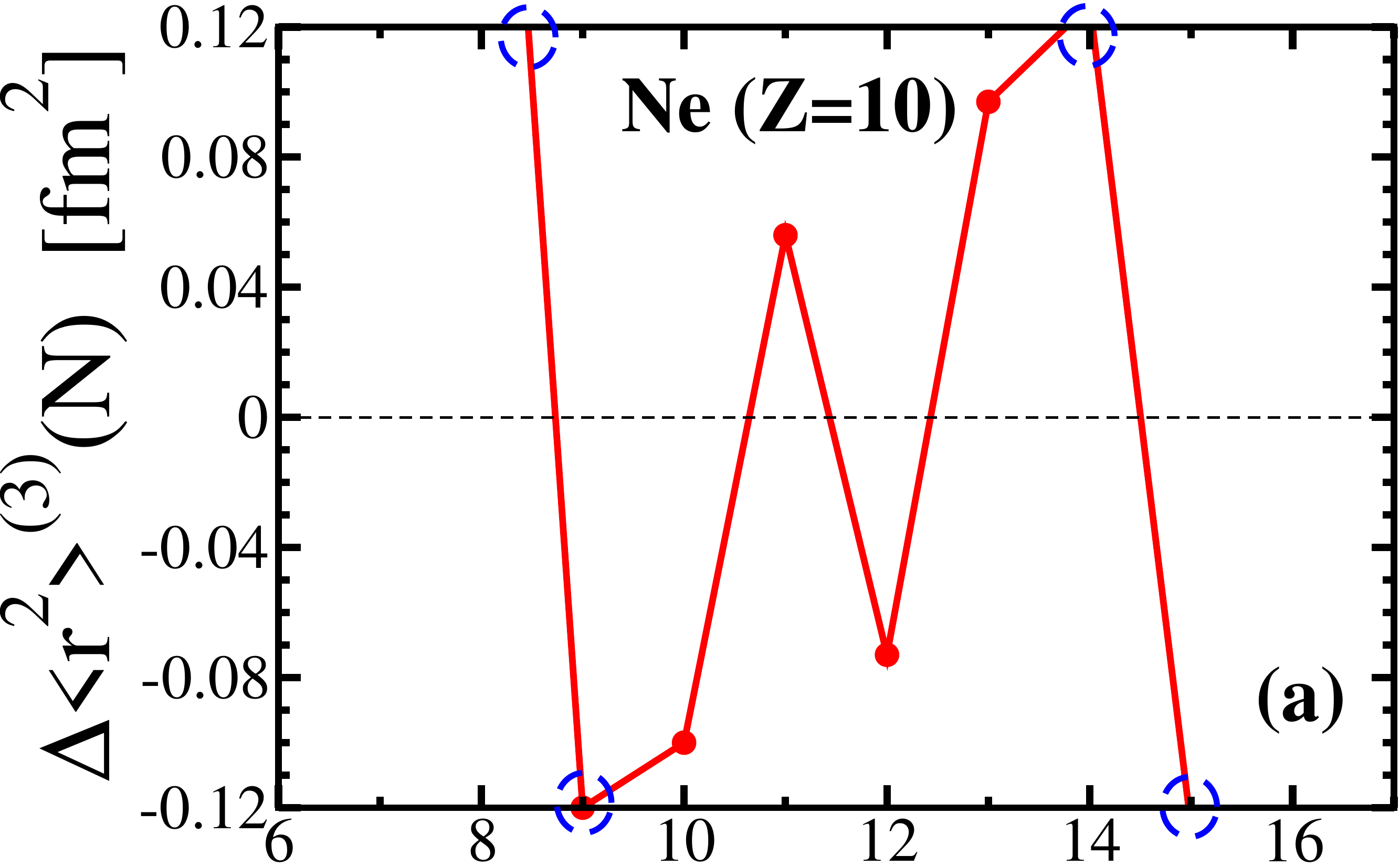}
\includegraphics[width=4.15cm]{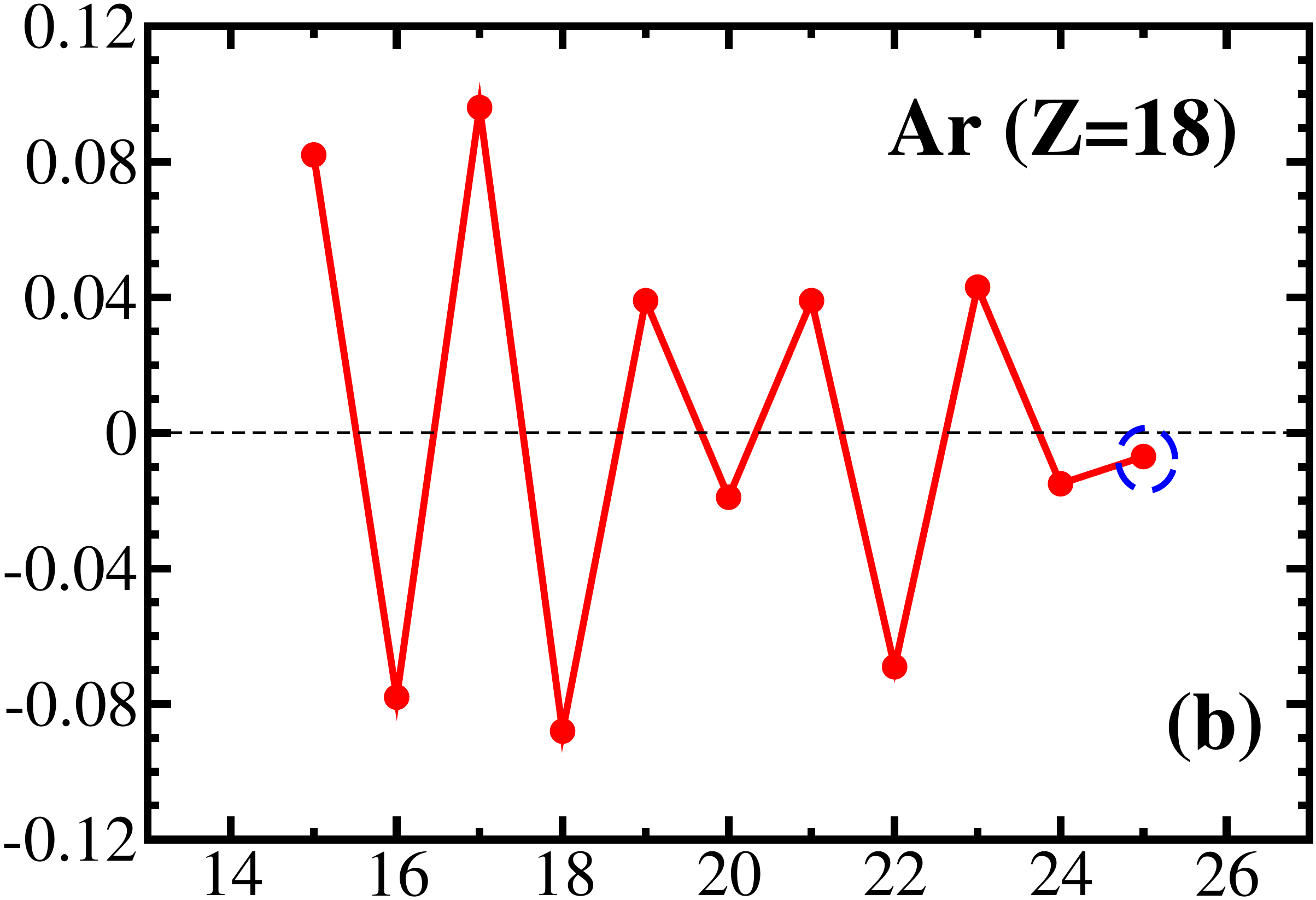}
\includegraphics[width=4.18cm]{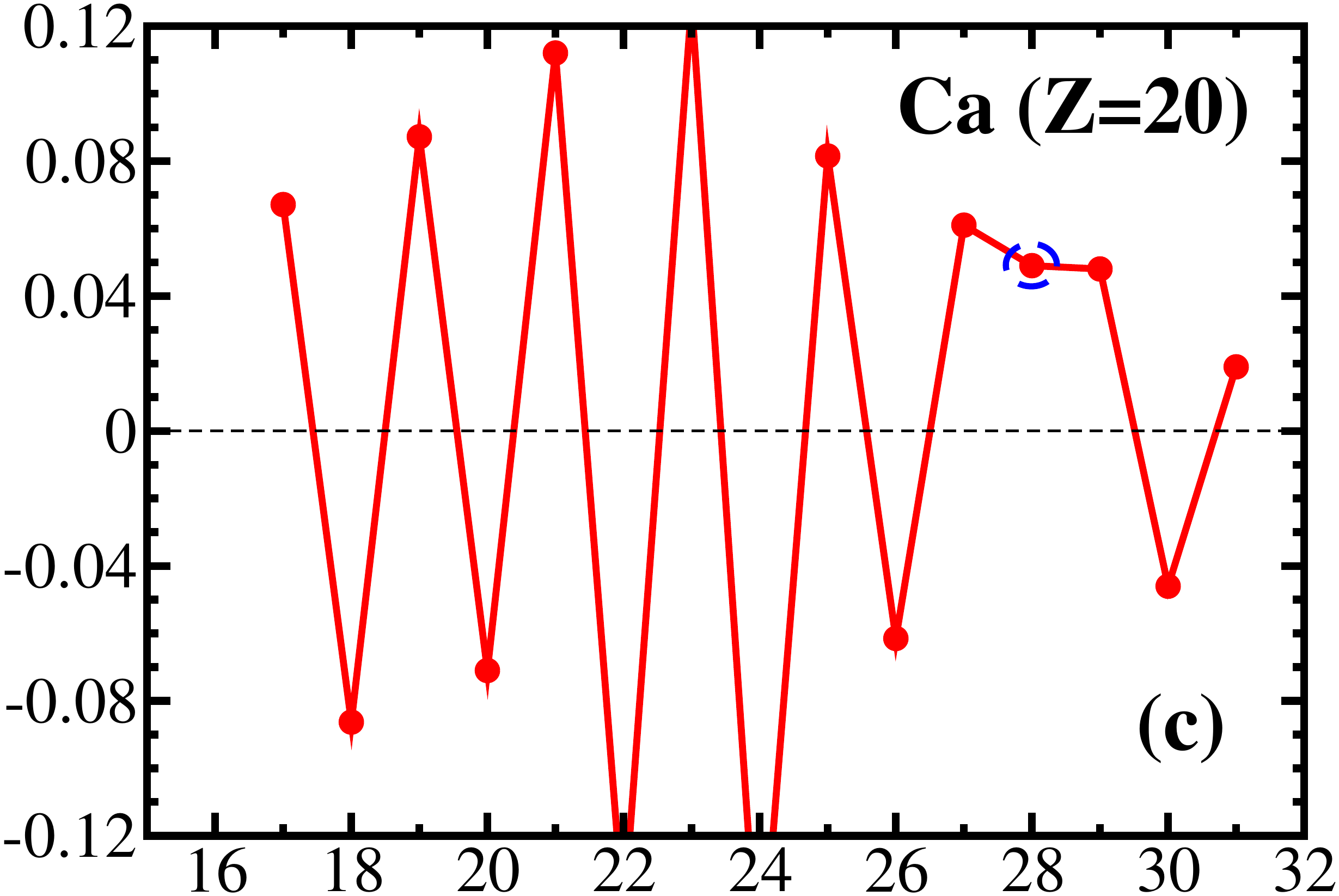}
\includegraphics[width=4.12cm]{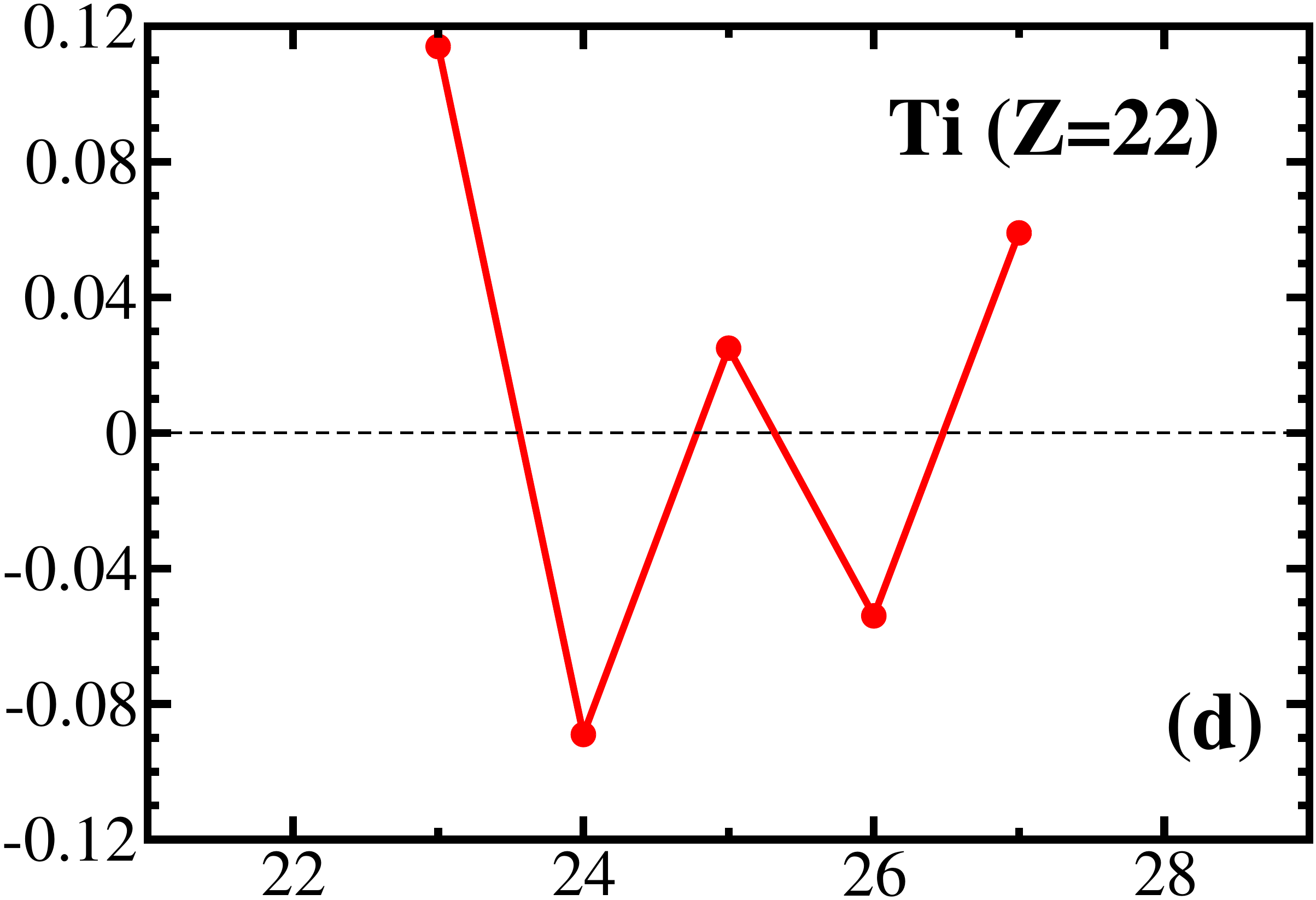}
\includegraphics[width=4.62cm]{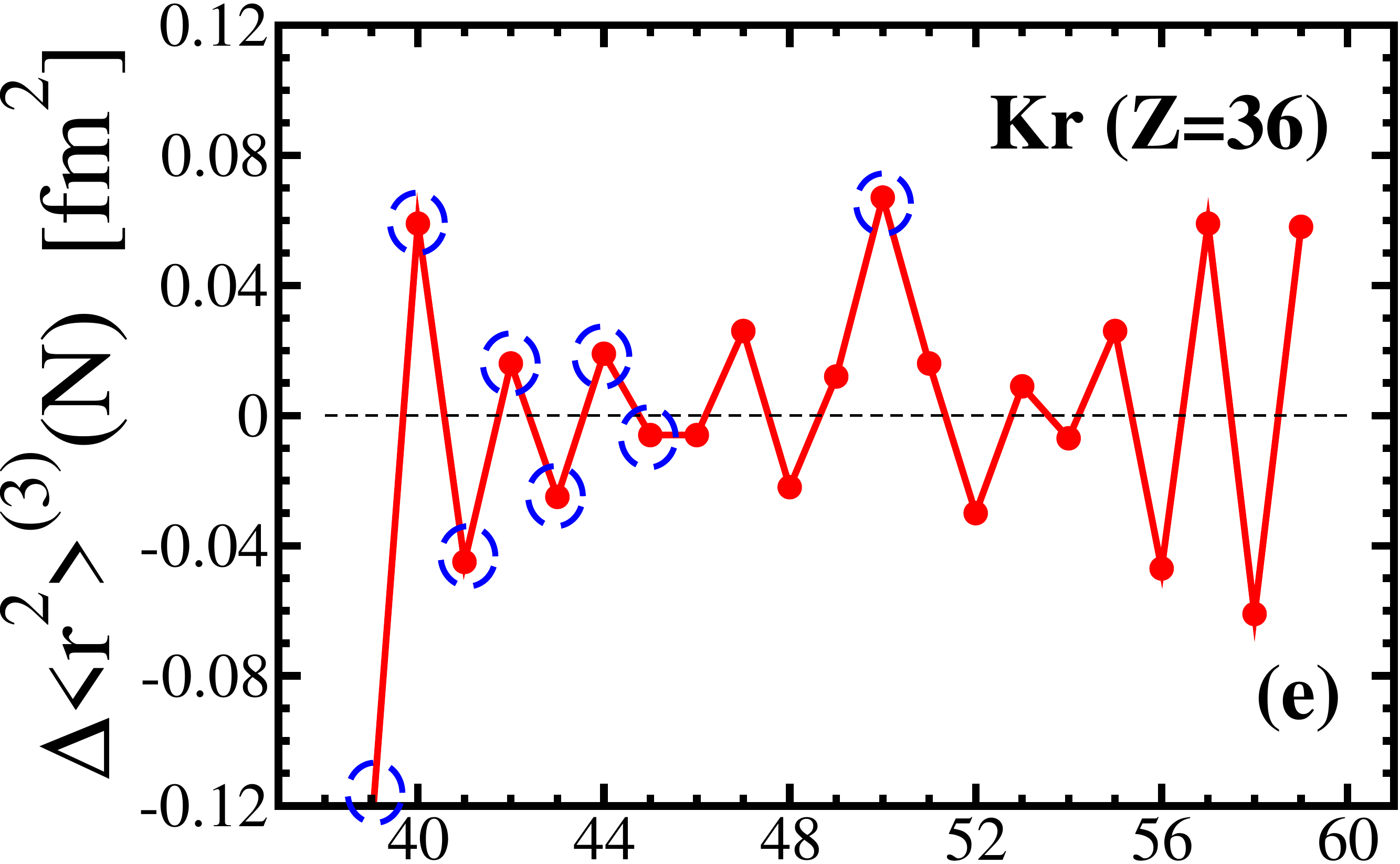}
\includegraphics[width=4.15cm]{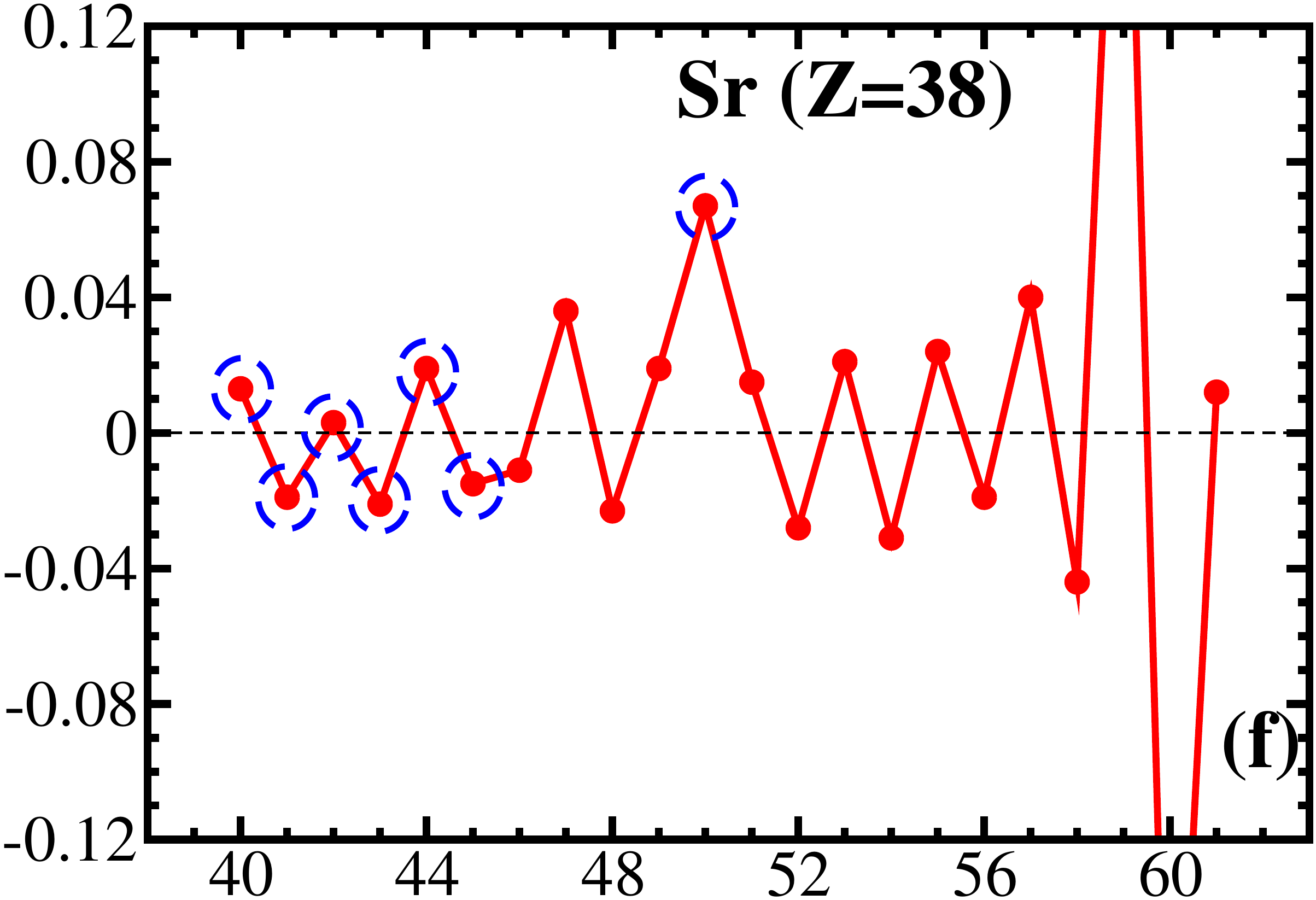}
\includegraphics[width=4.18cm]{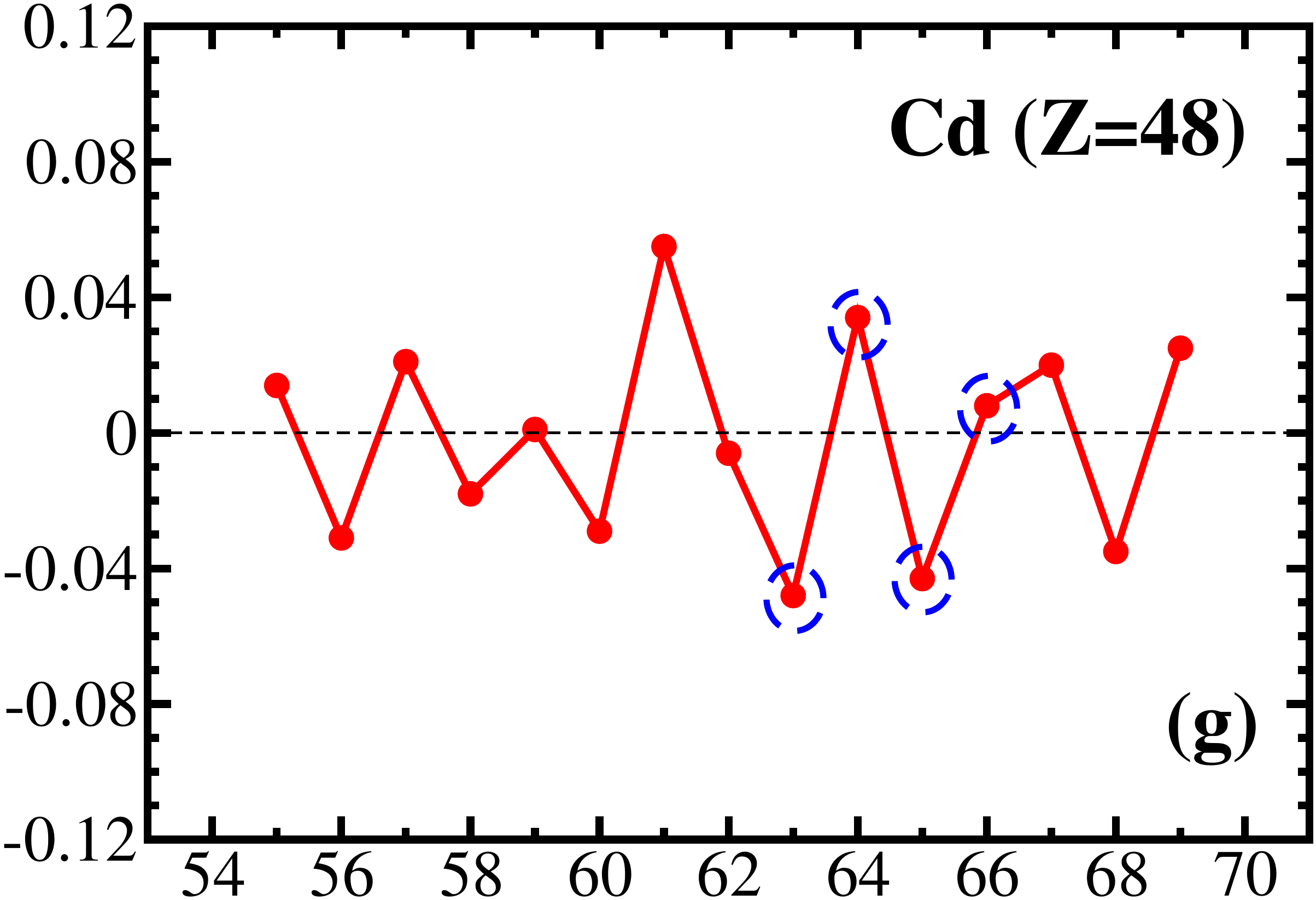}
\includegraphics[width=4.12cm]{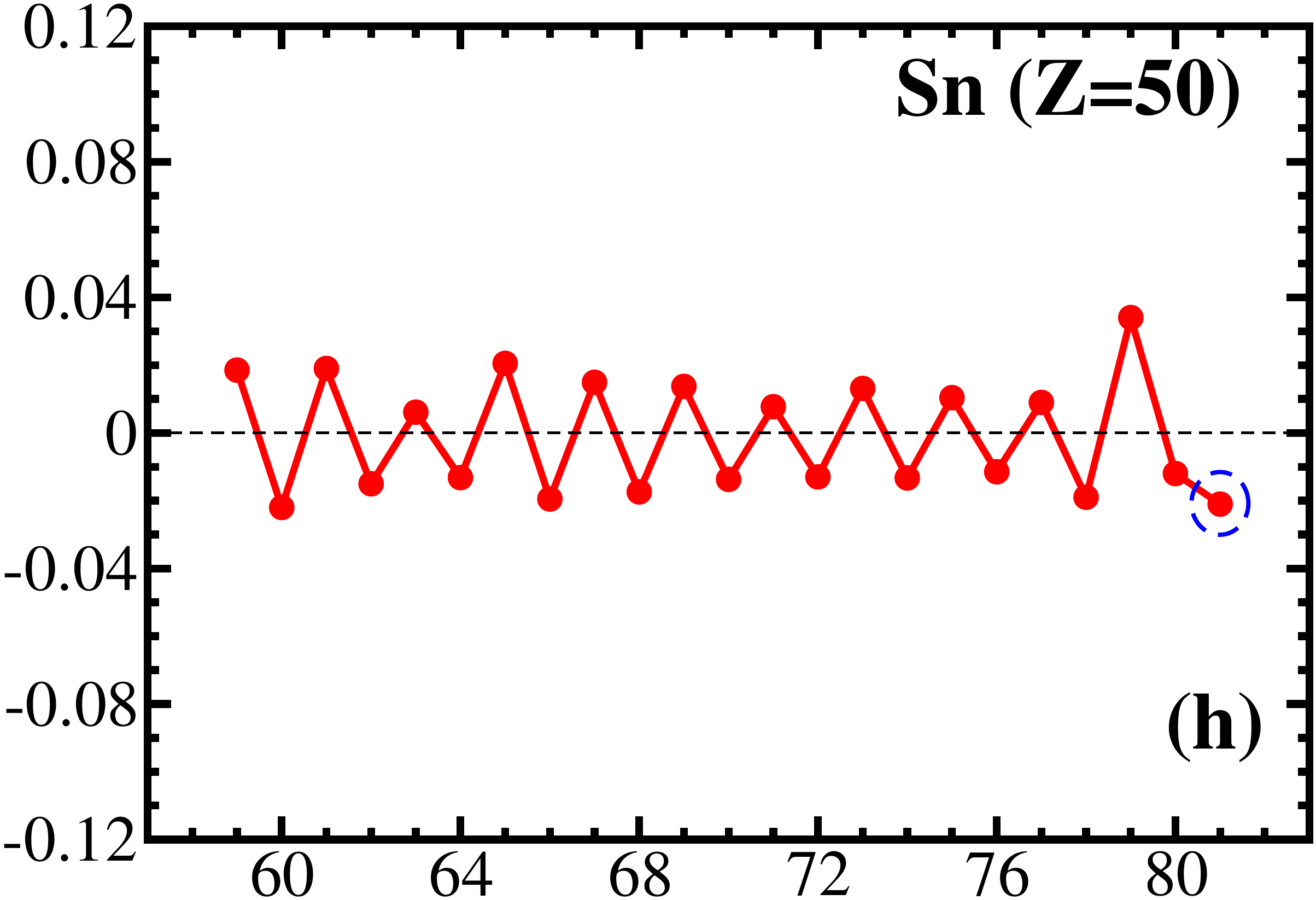}
\includegraphics[width=4.62cm]{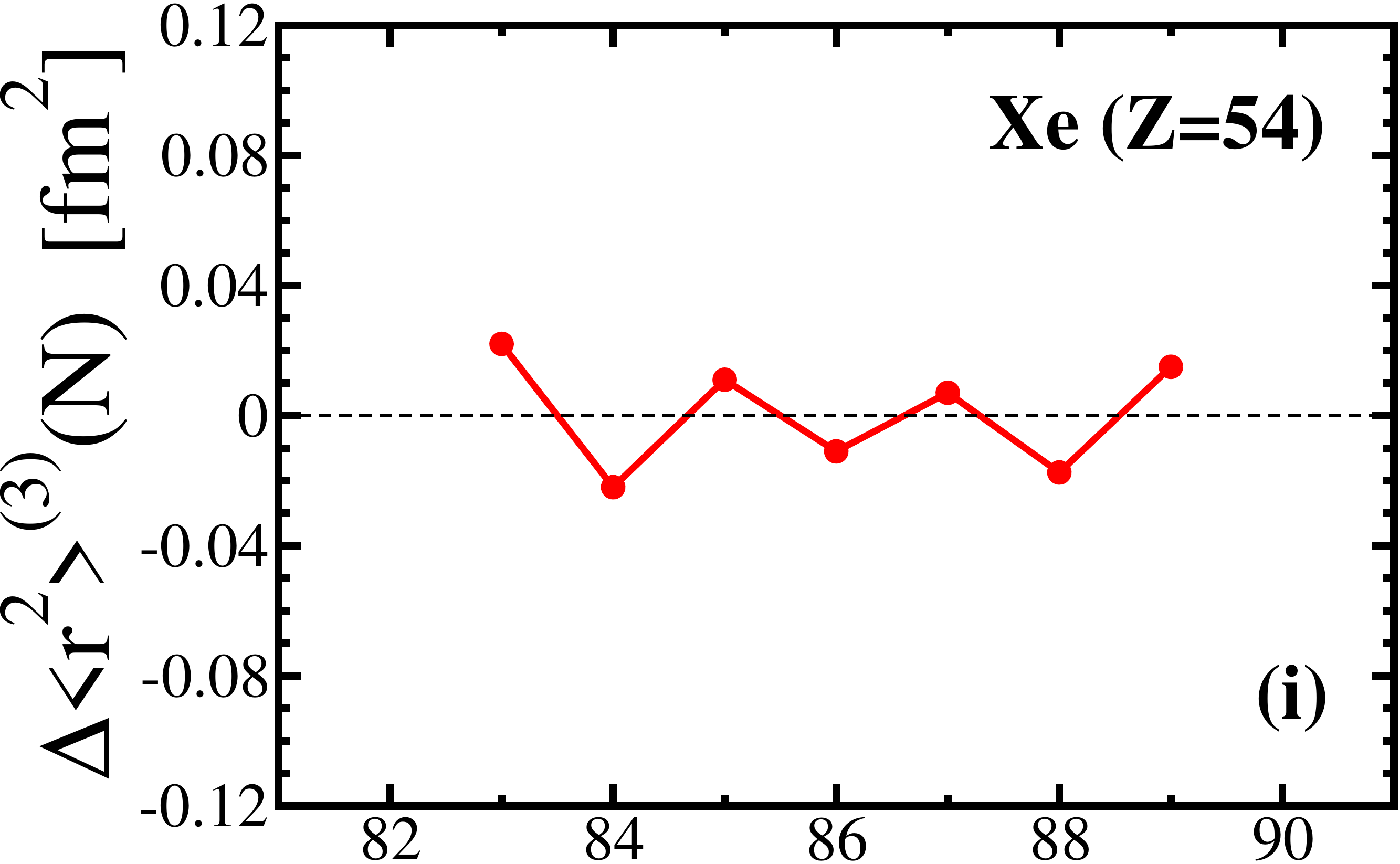}
\includegraphics[width=4.15cm]{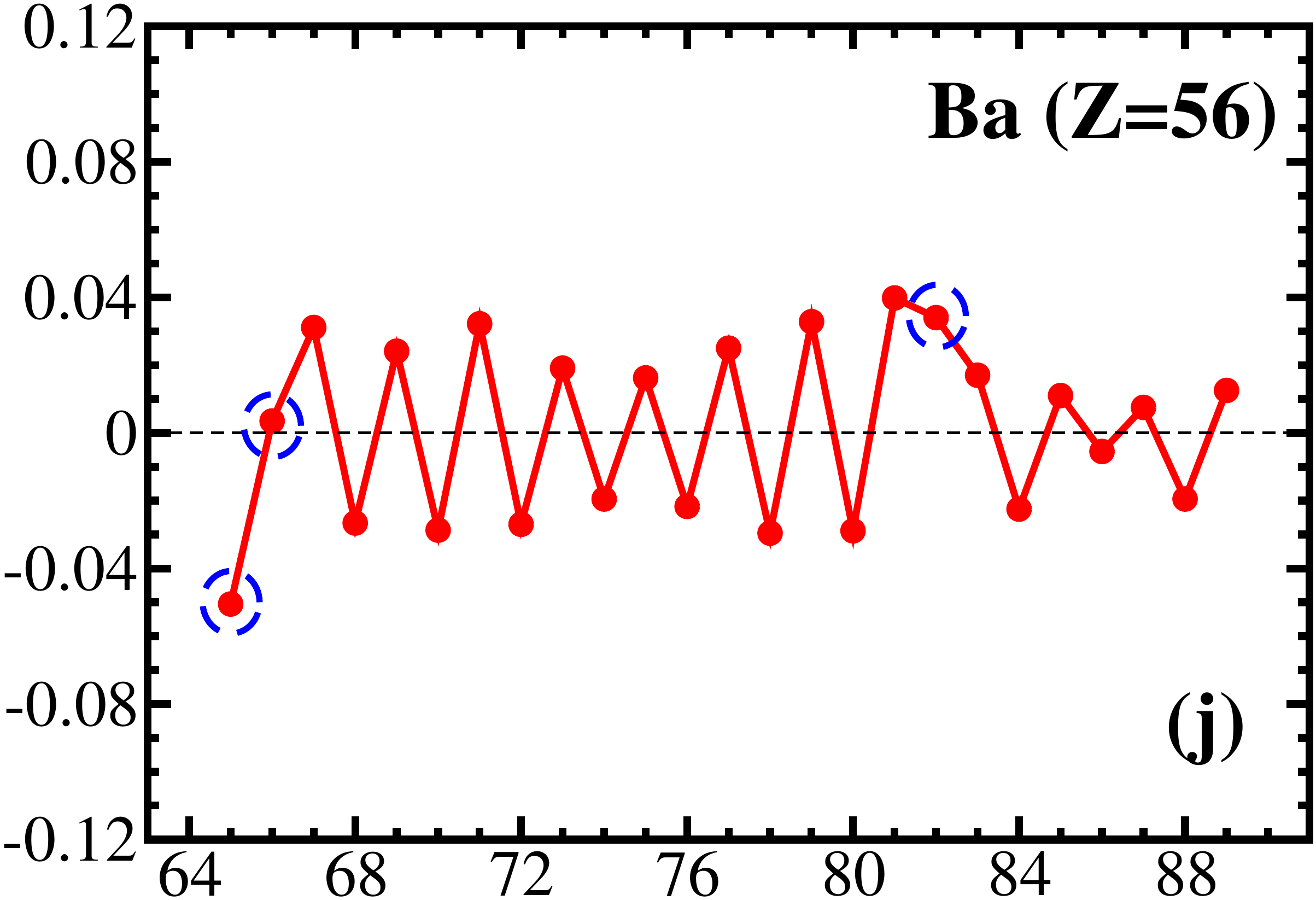}
\includegraphics[width=4.18cm]{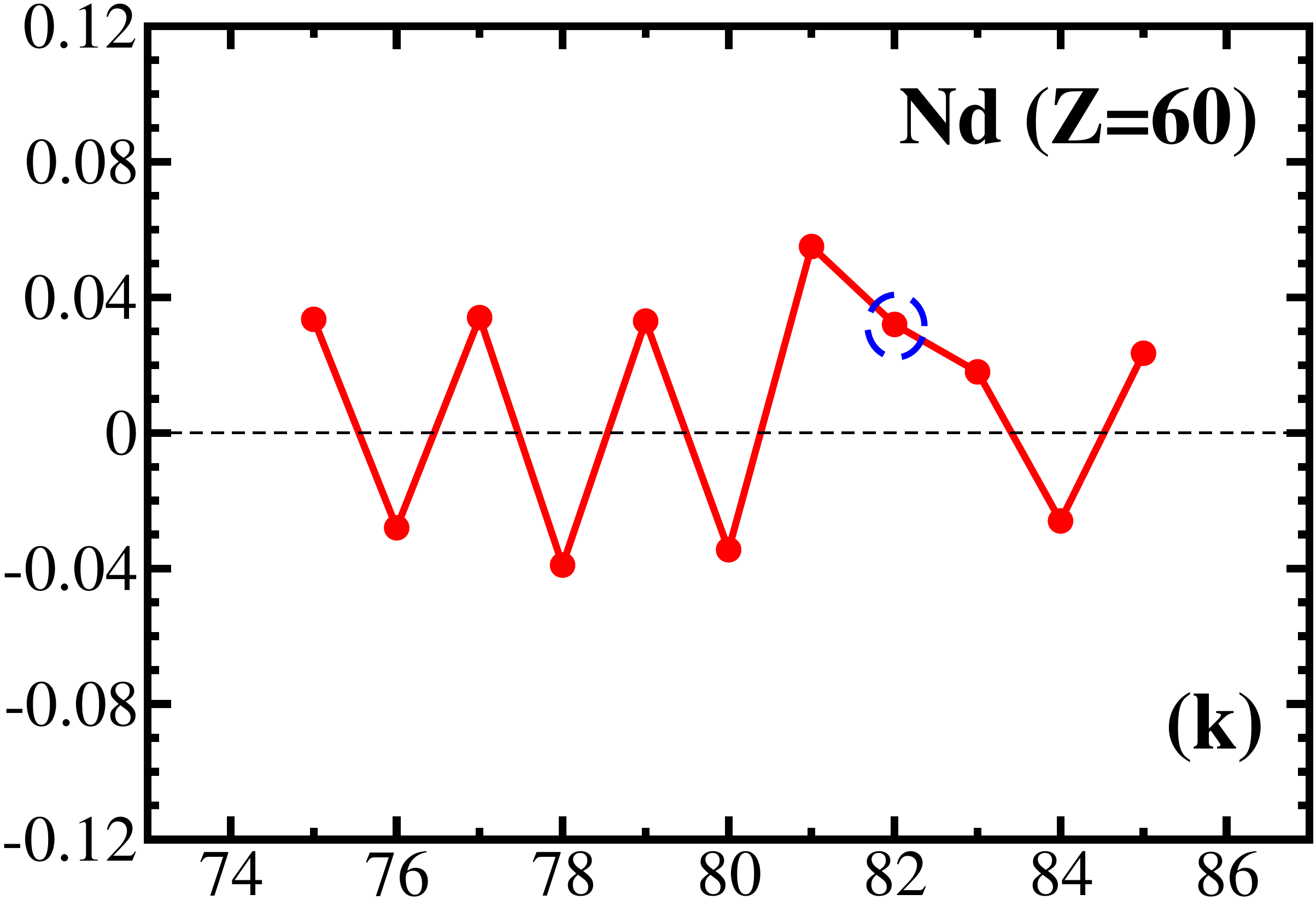}
\includegraphics[width=4.12cm]{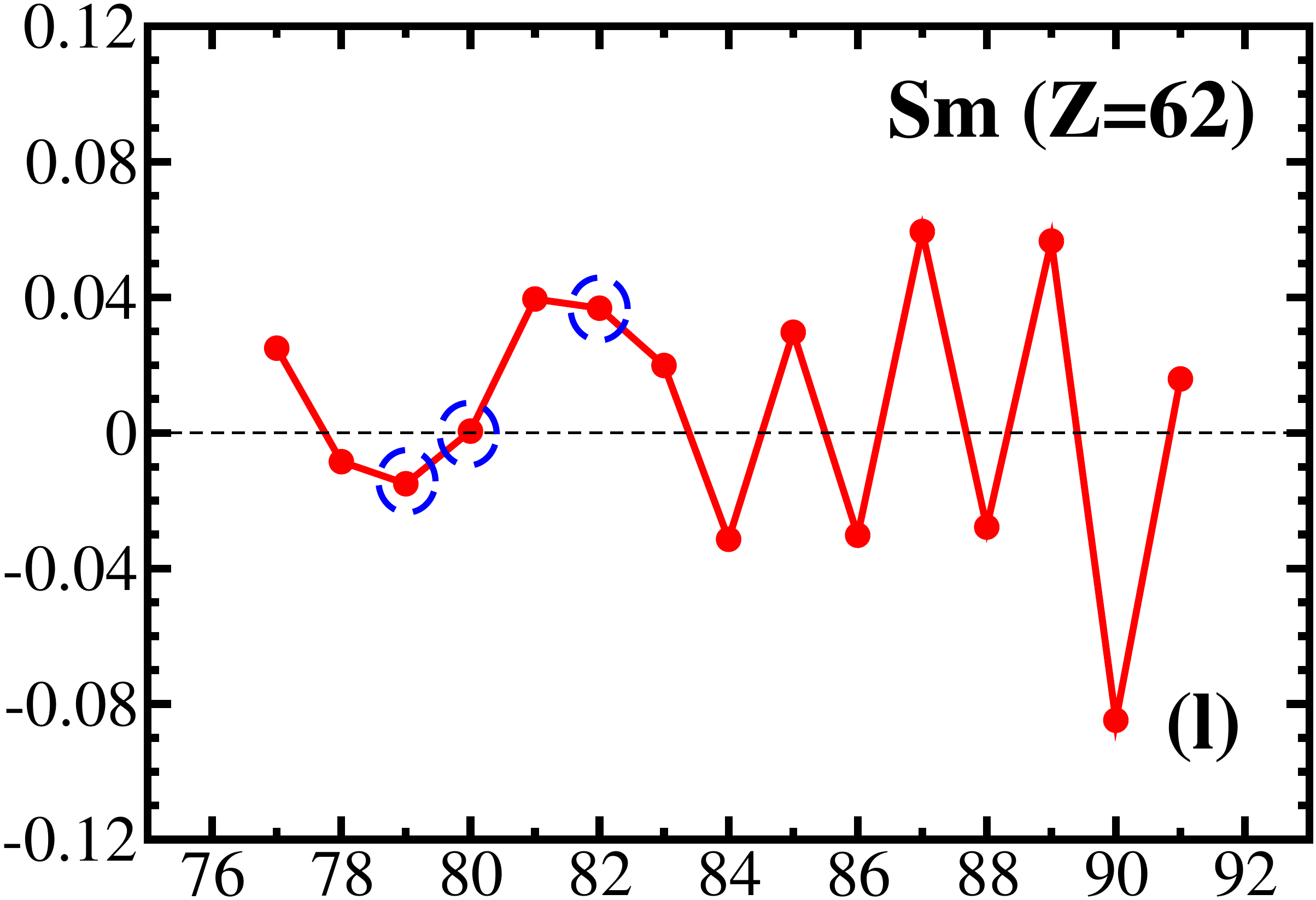}
\includegraphics[width=4.62cm]{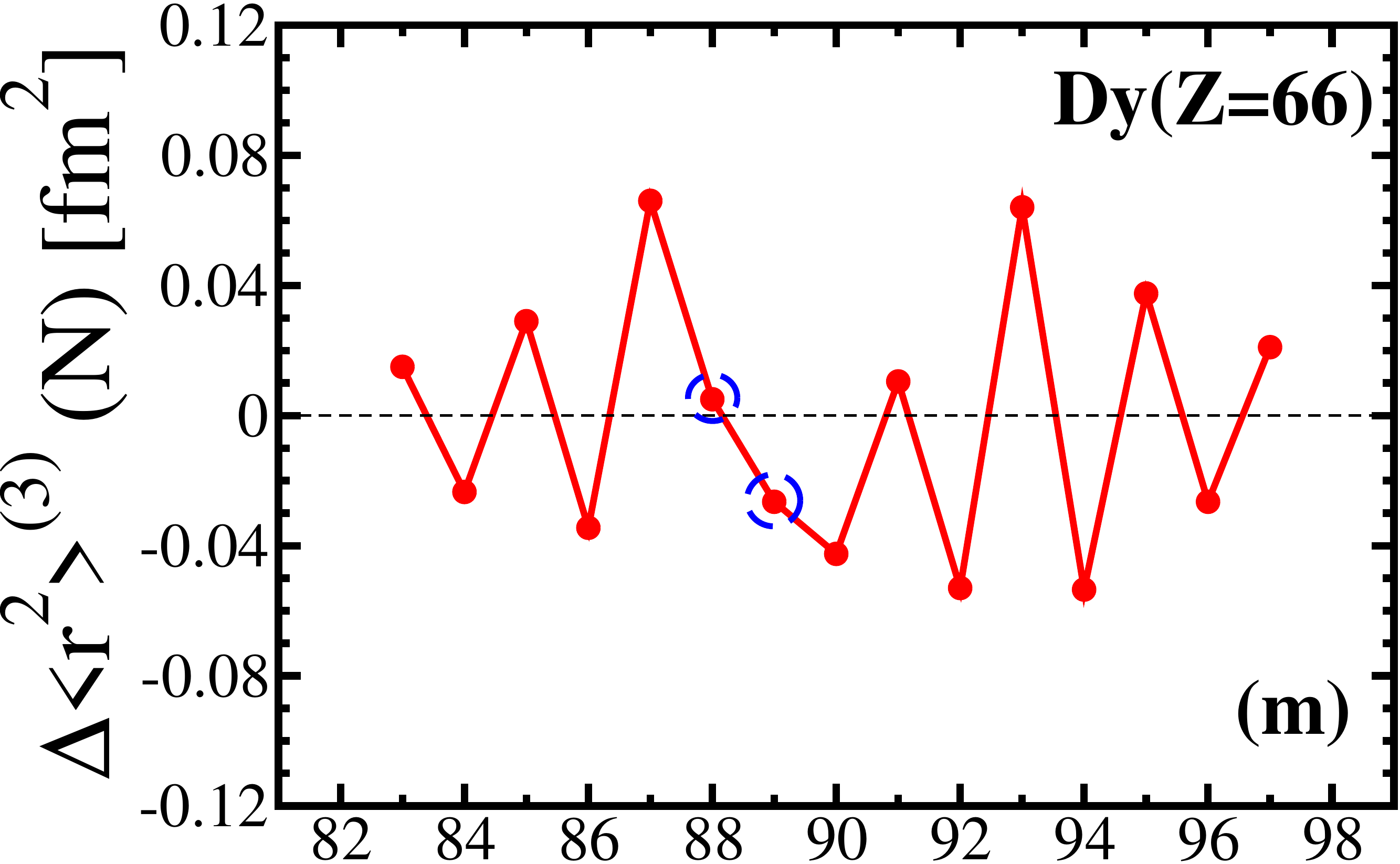}
\includegraphics[width=4.15cm]{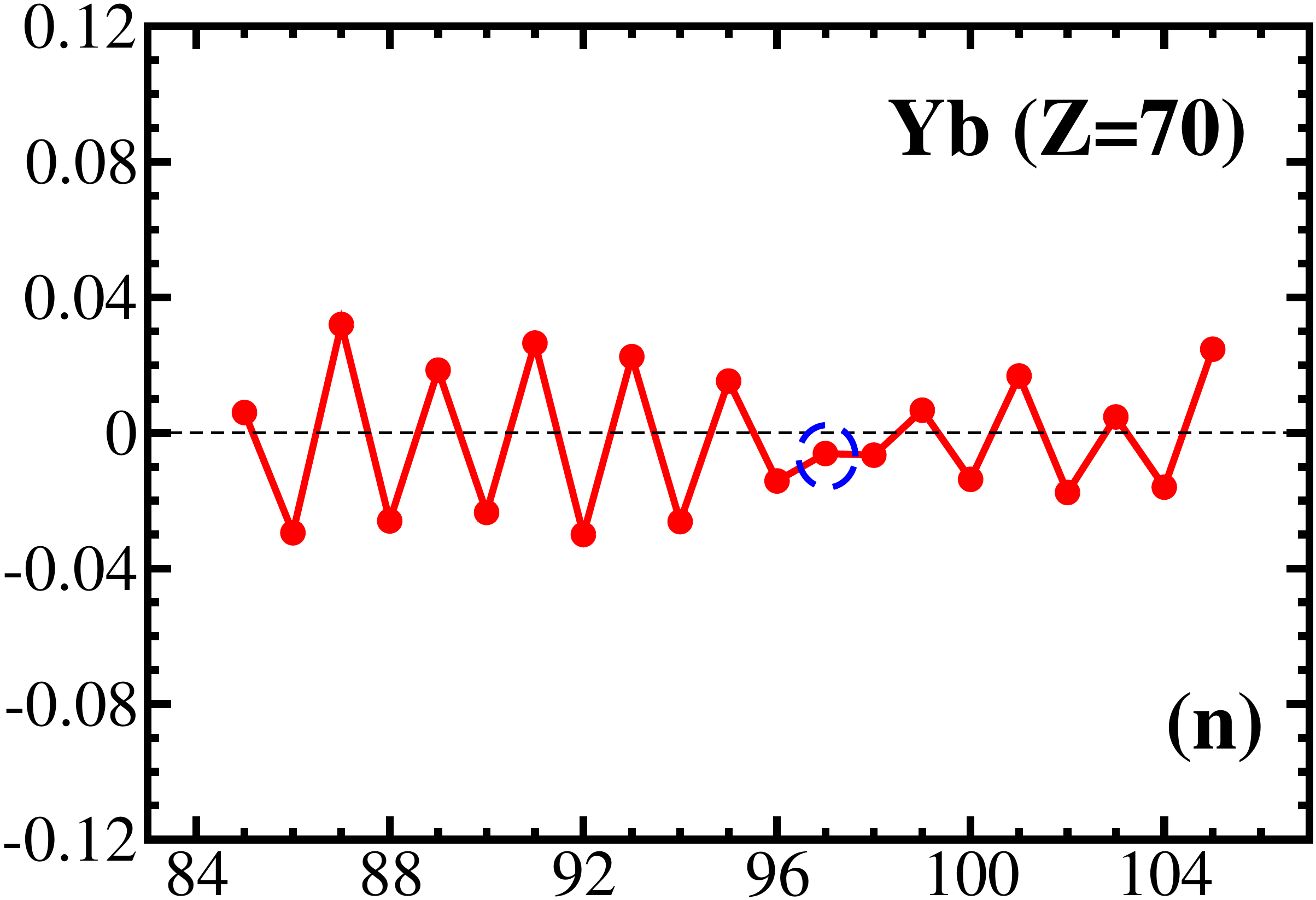}
\includegraphics[width=4.18cm]{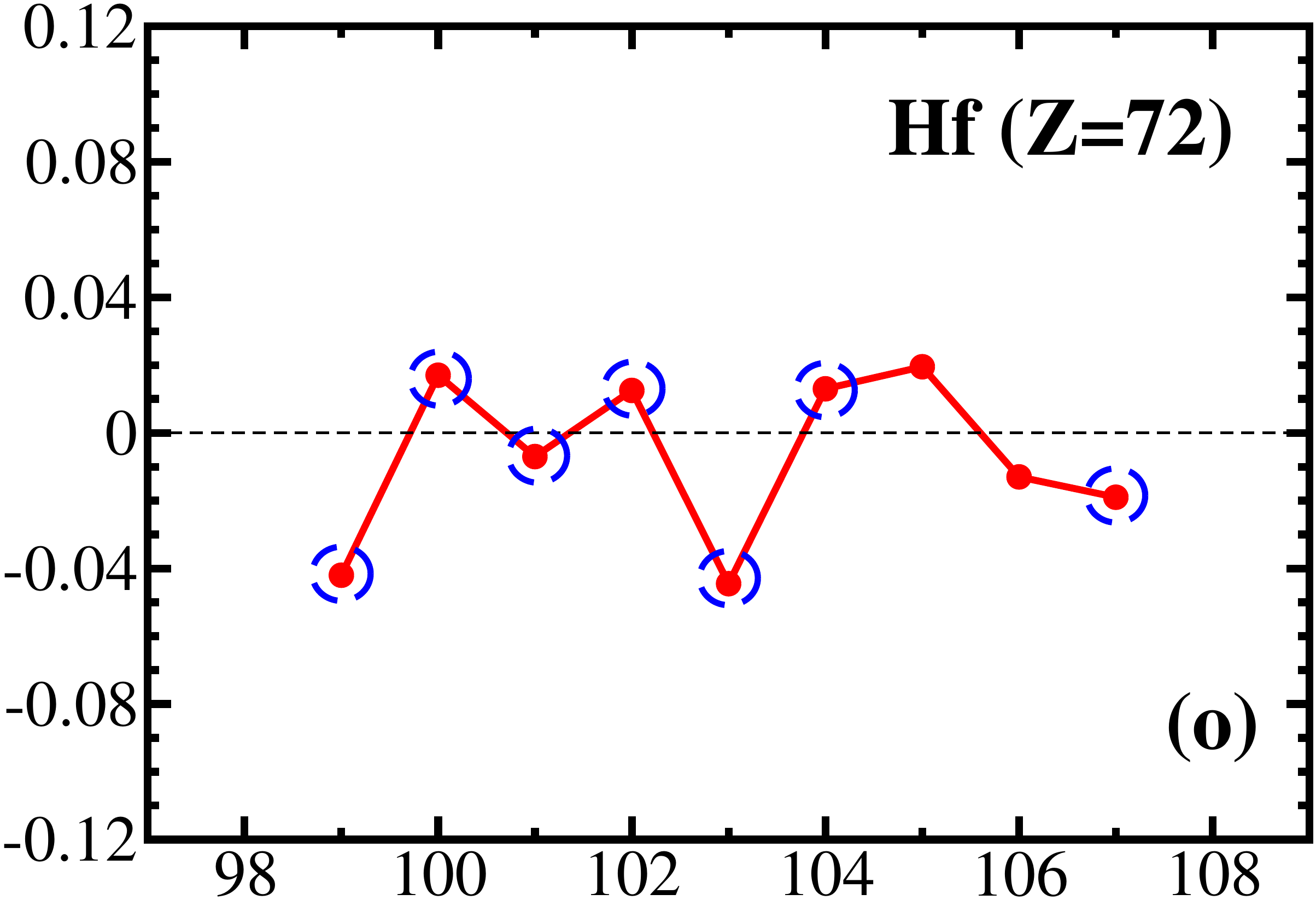}
\includegraphics[width=4.12cm]{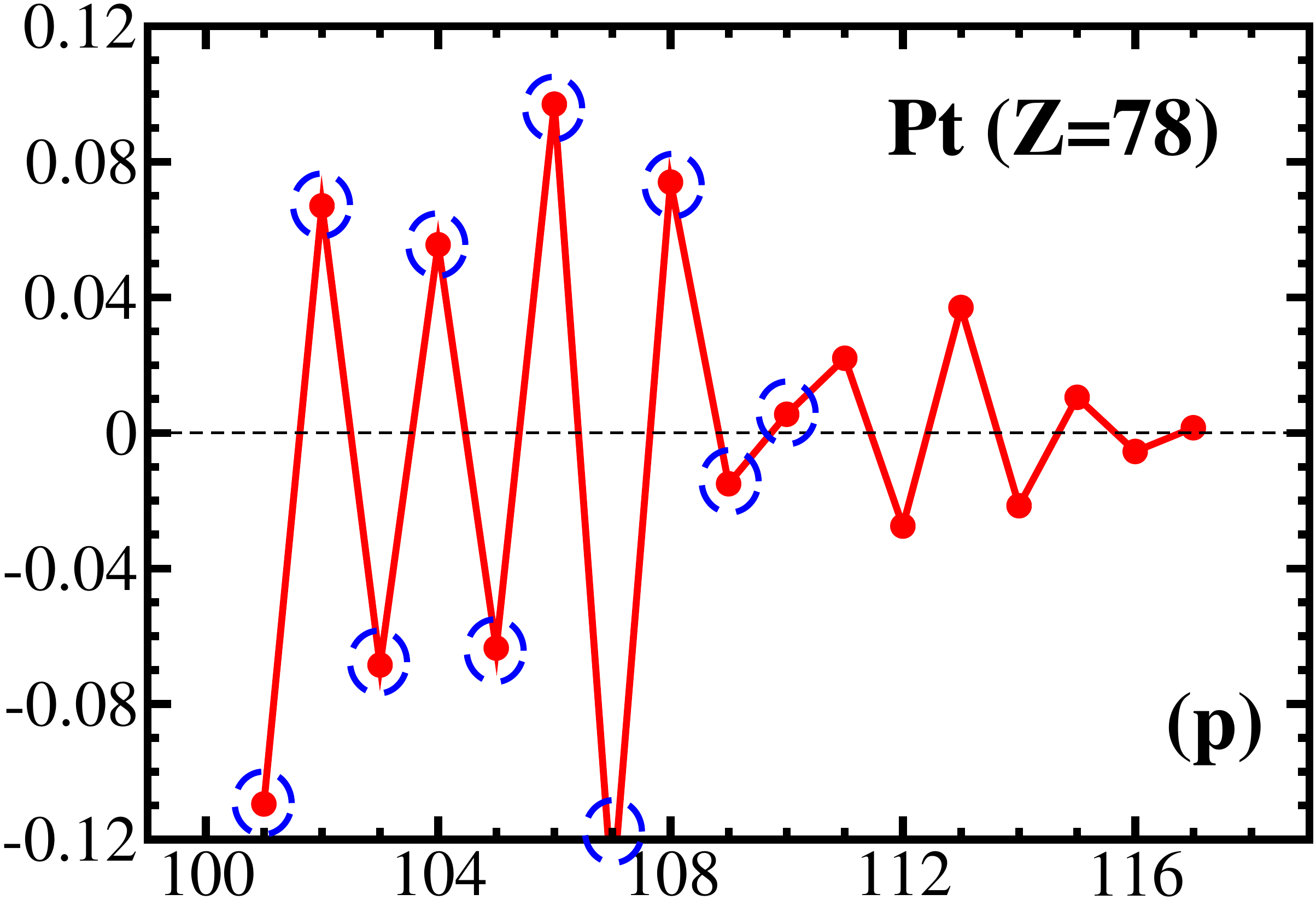}
\includegraphics[width=4.62cm]{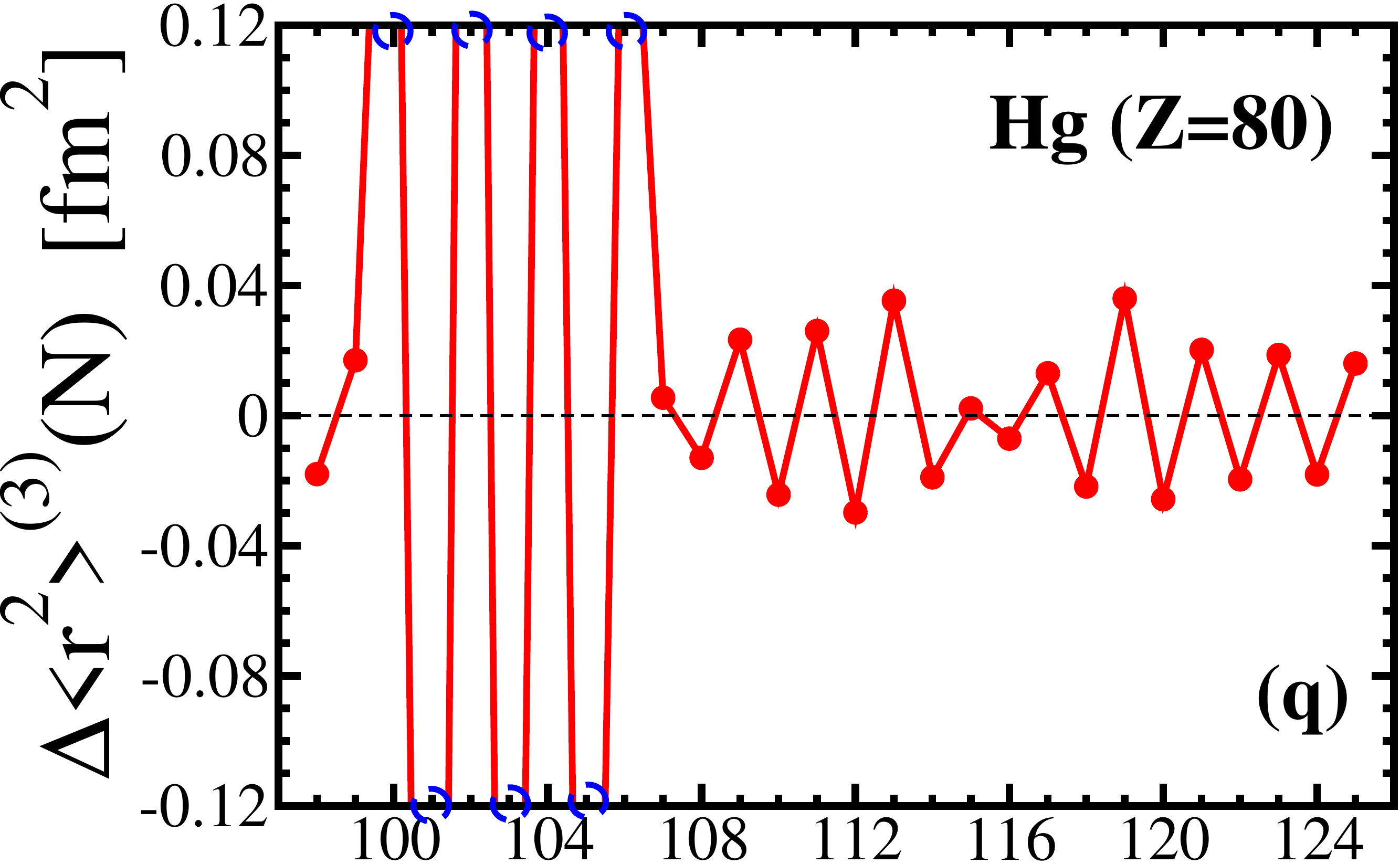}
\includegraphics[width=4.15cm]{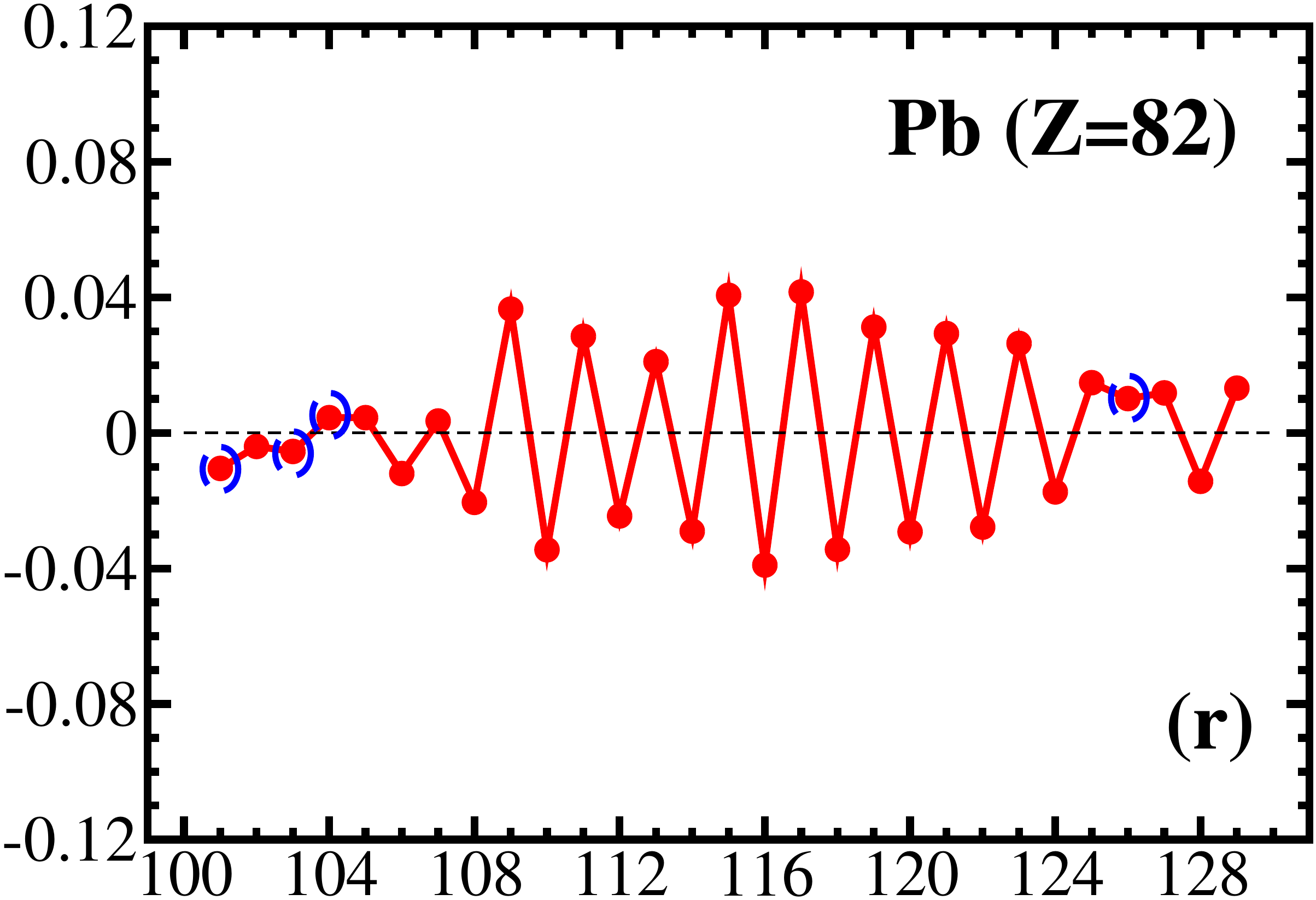}
\includegraphics[width=4.18cm]{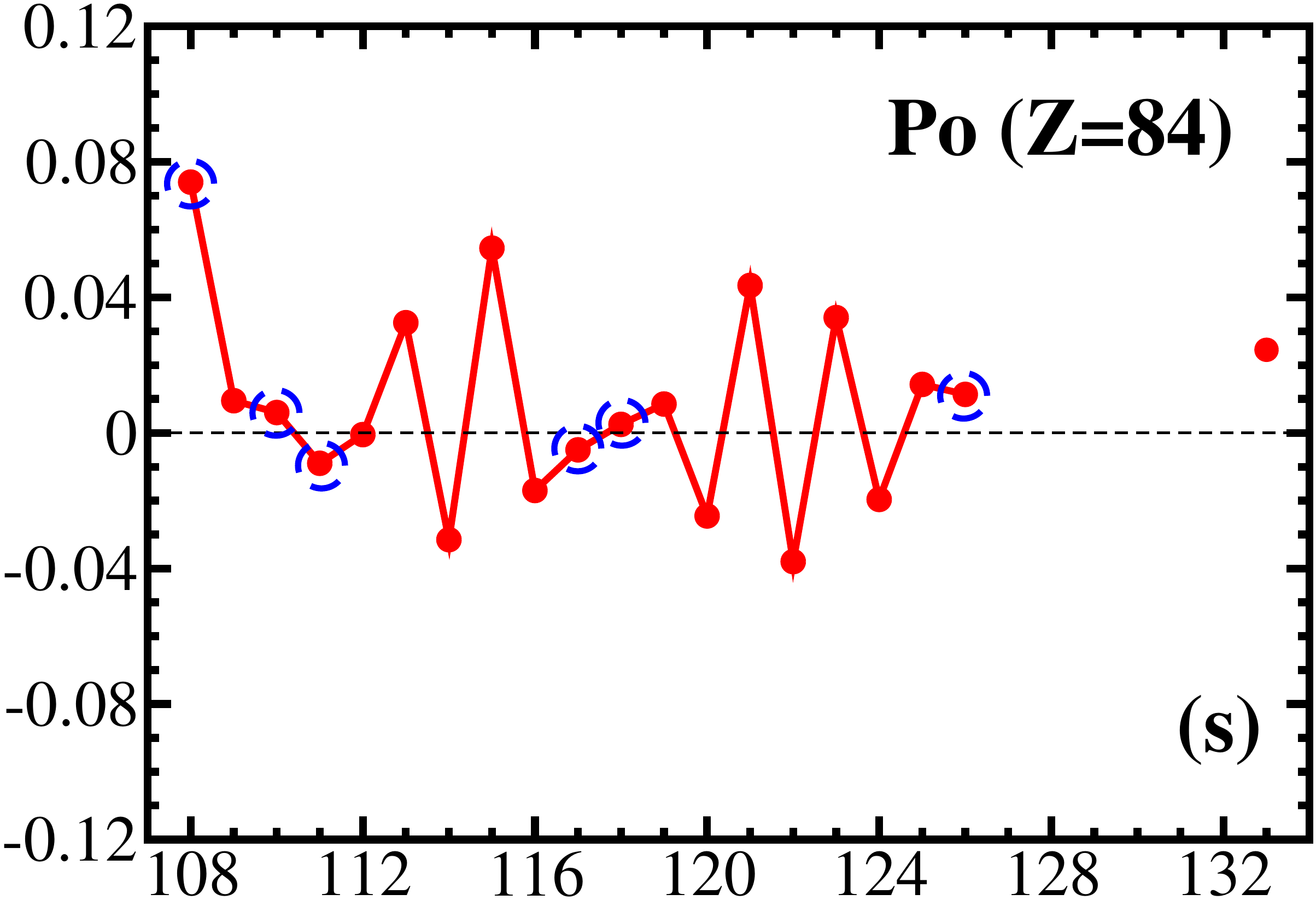}
\includegraphics[width=4.12cm]{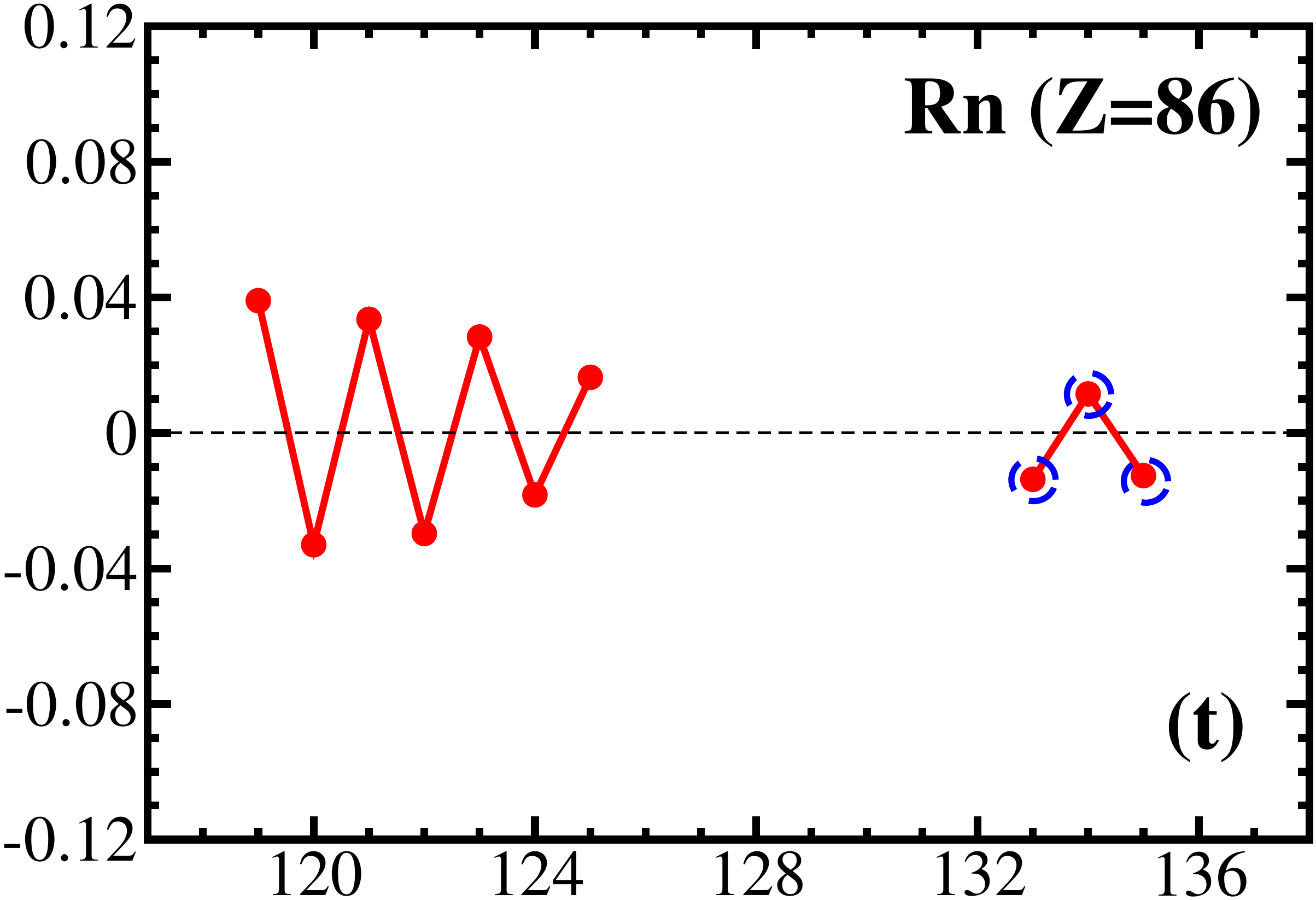}
\includegraphics[width=4.62cm]{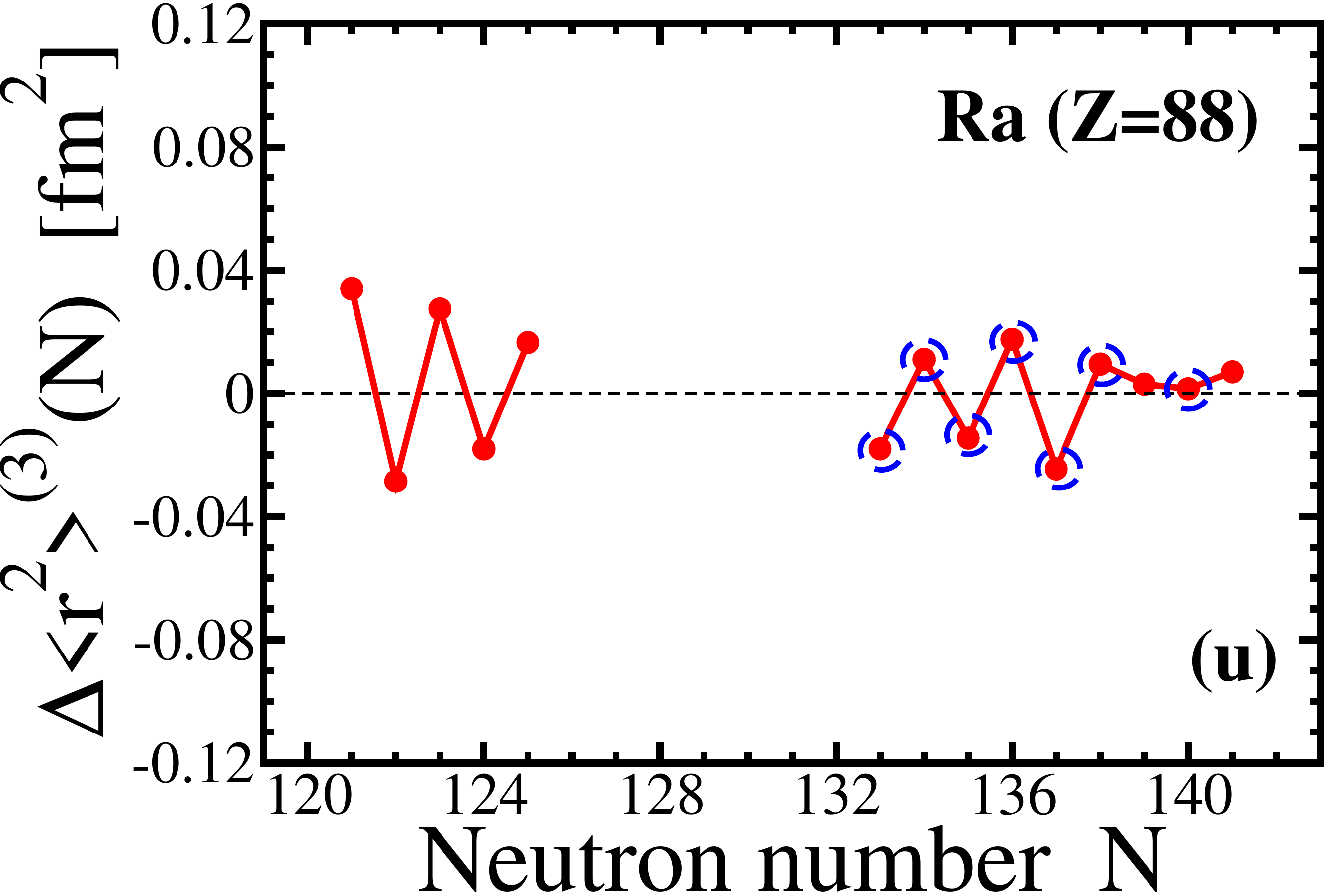}
\includegraphics[width=4.15cm]{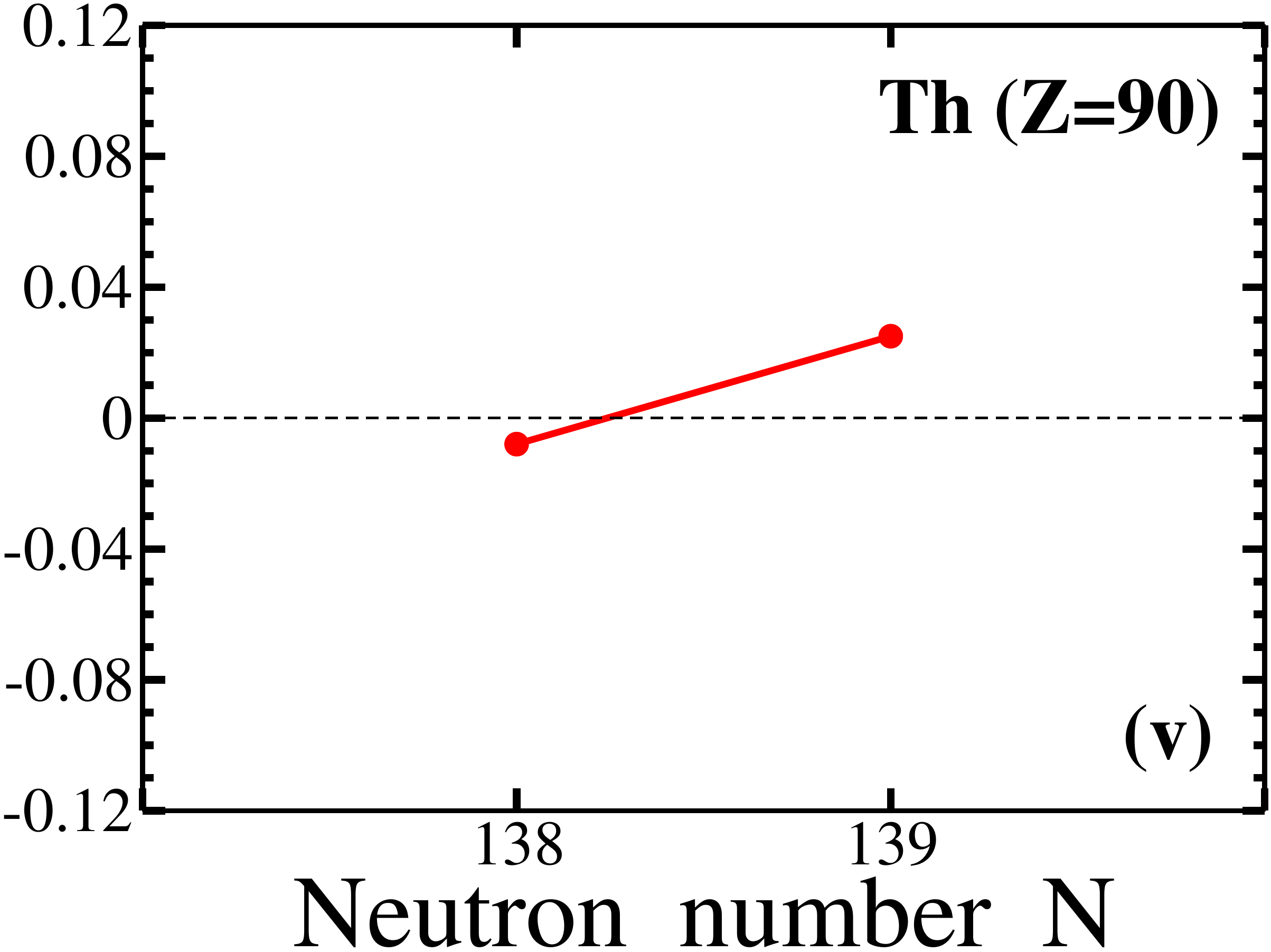}
\includegraphics[width=4.18cm]{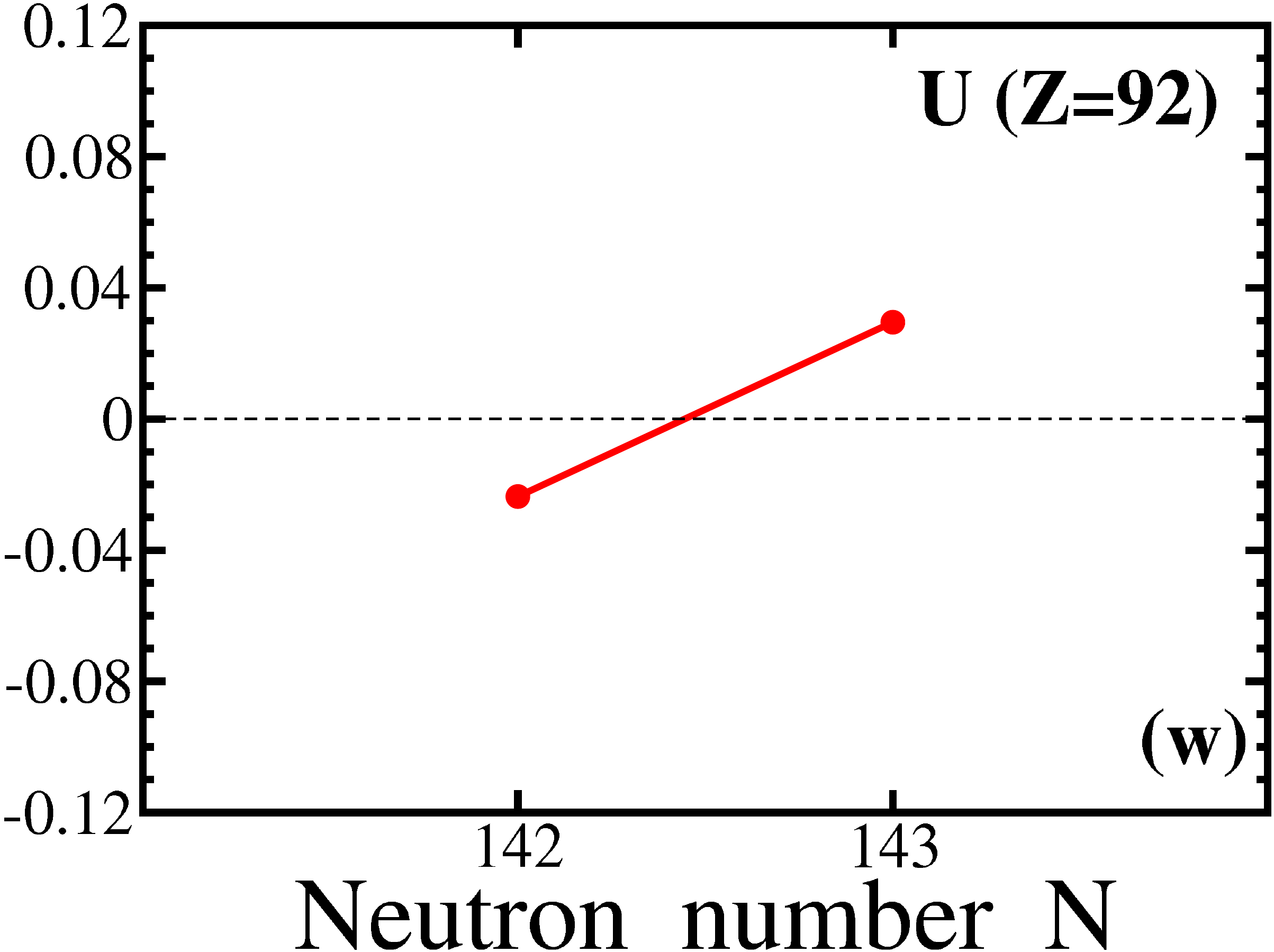}
\includegraphics[width=4.12cm]{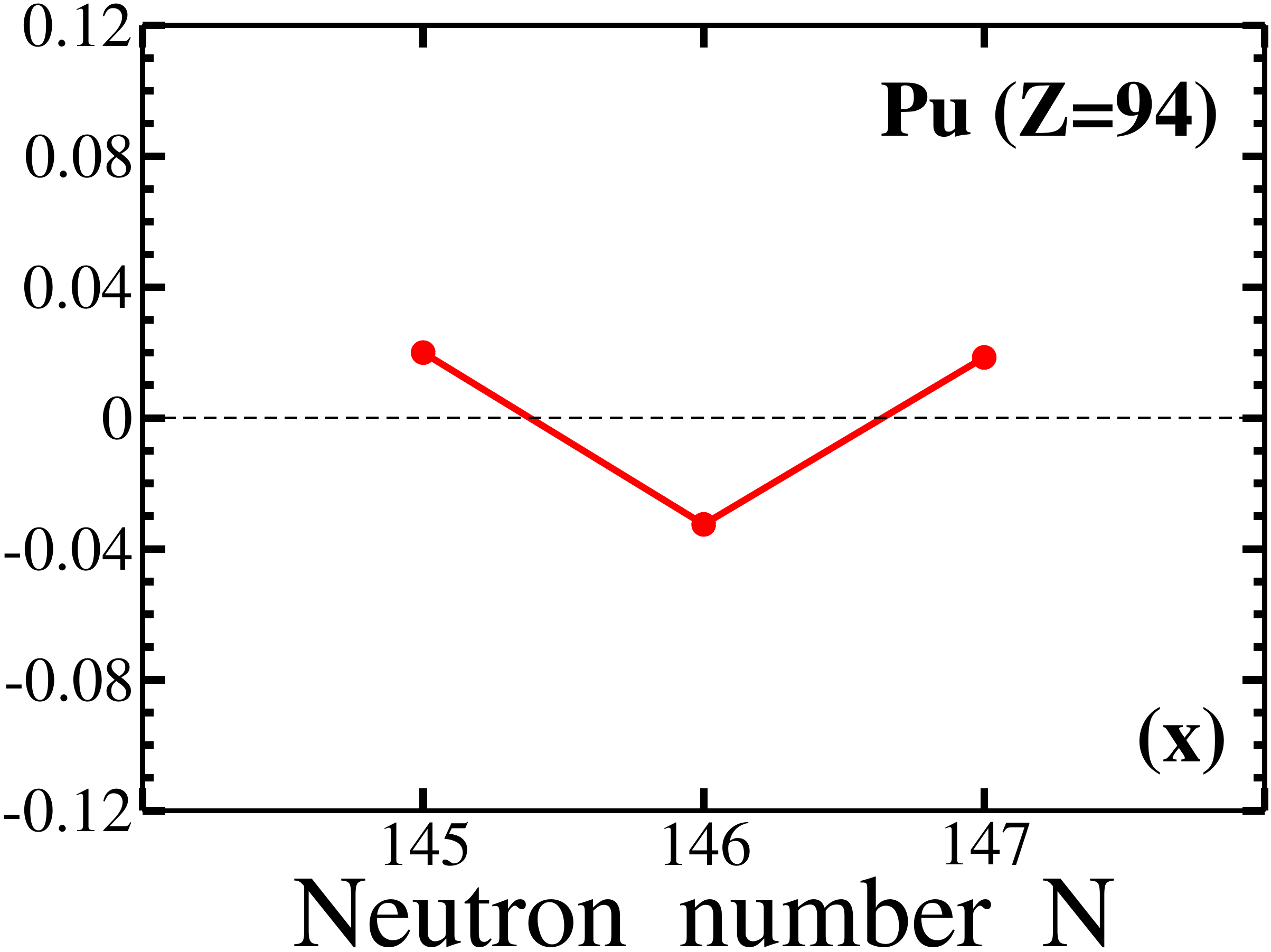}
\caption{Experimental OES of charge radii of even-$Z$ nuclei. The data
points  are encircled by blue dashed circles when they deviate from the regular
pattern [regular OES], namely, from $\Delta \left < r^2\right>^{(3)}(N)>0$ for odd values
of $N$ and from $\Delta \left < r^2\right>^{(3)}(N)<0$ for even values of $N$.
These encircled points correspond to inverted OESs. The experimental data are taken from Ref.\ \cite{AM.13}.
Note that, with the exception of the Ca isotopes, we use the same range of $\Delta \left < r^2\right>^{(3)}$
on the vertical  axis of all panels. Until specified otherwise, the experimental data are taken
from Ref.\ \cite{AM.13}. Only for the Ca isotopes they are mostly taken from
Ref.\ \cite{Ca-radii.2016}, from Ref.\ \cite{39-41Ca-radii.96} for $^{39,41}$Ca 
and from Ref.\ \cite{Ca-radii-prot-rich.19} for $^{36,37,38}$Ca. The experimental data for the radii
are taken from Ref.\ \cite{Hg-radii-low-N.19} for the Hg isotopes with $N<106$ and 
from \cite{Po-radii.13,Po-216-218-radii.15} for the Po isotopes.
\label{OES-even-Z}
}
\end{figure*}

\begin{figure*}[htb]
\centering
\includegraphics[width=4.62cm]{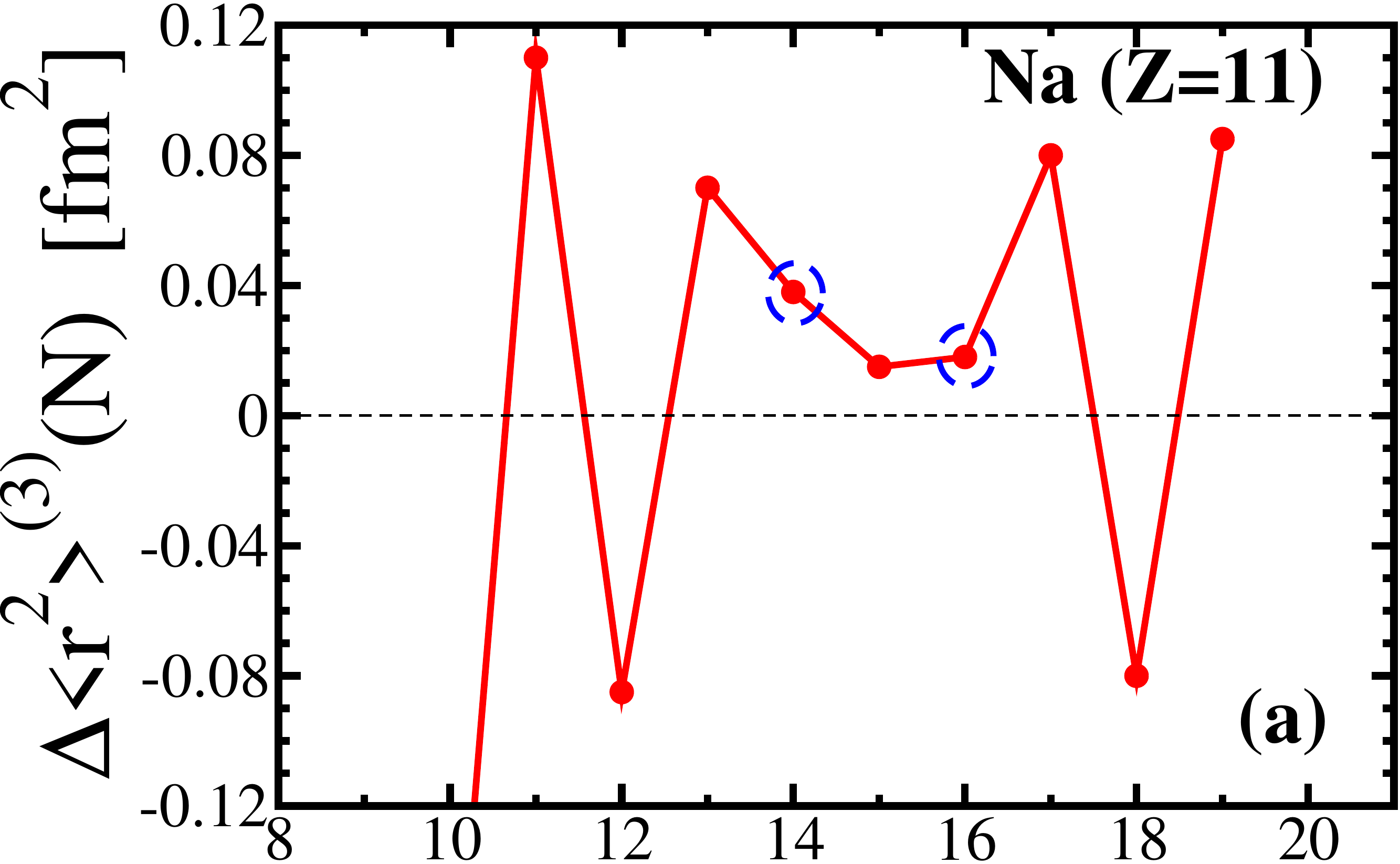}
\includegraphics[width=4.15cm]{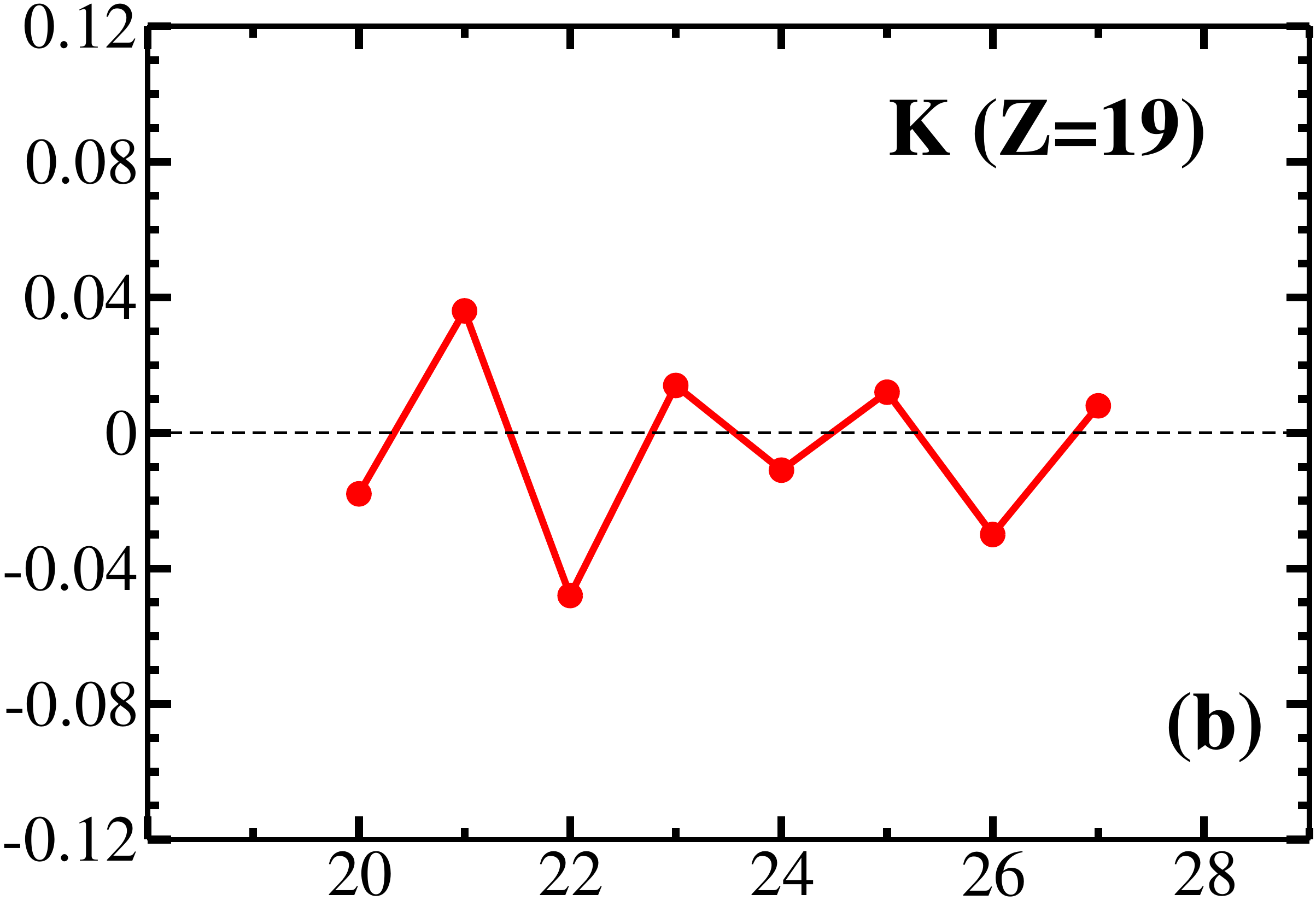}
\includegraphics[width=4.18cm]{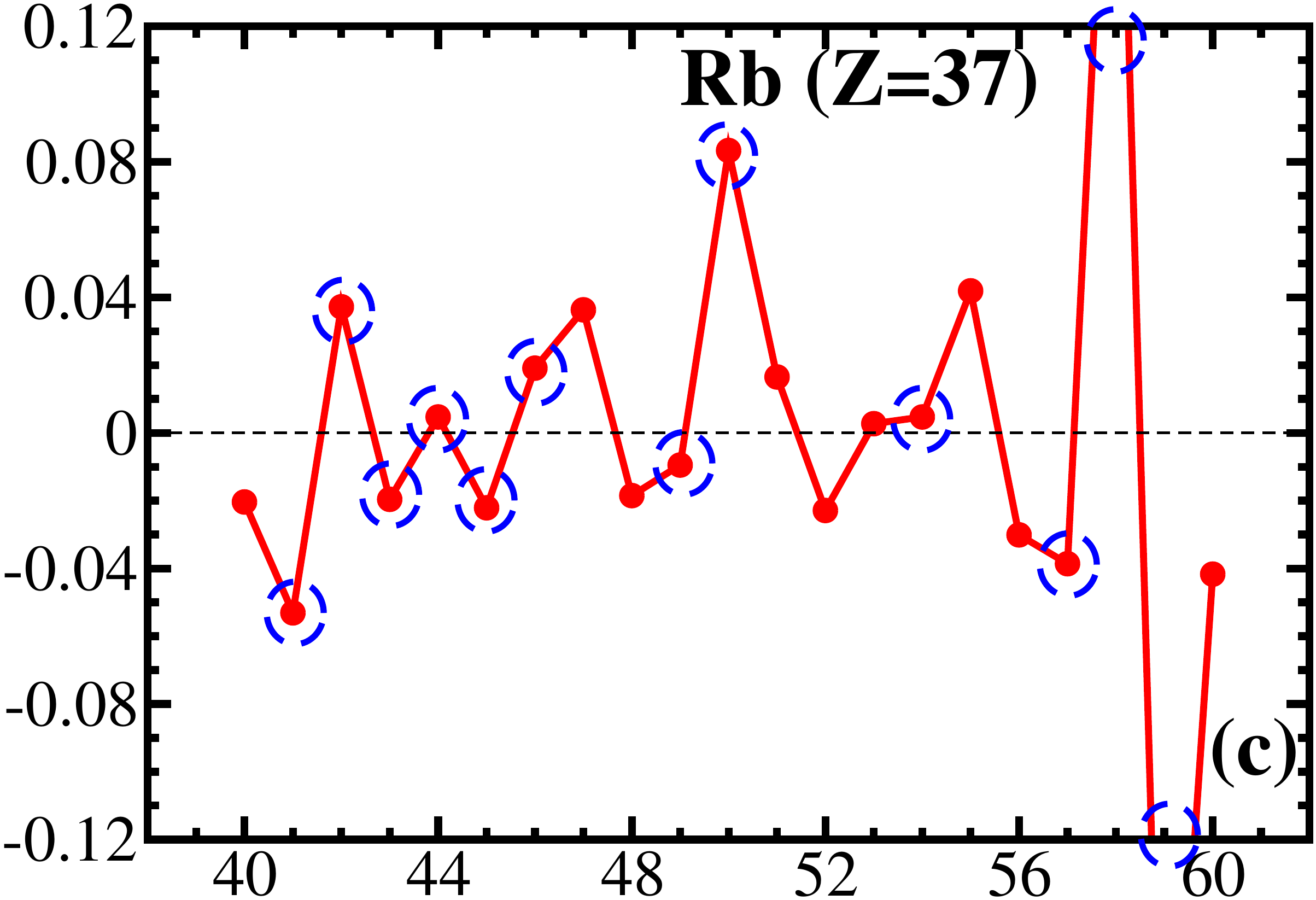}
\includegraphics[width=4.12cm]{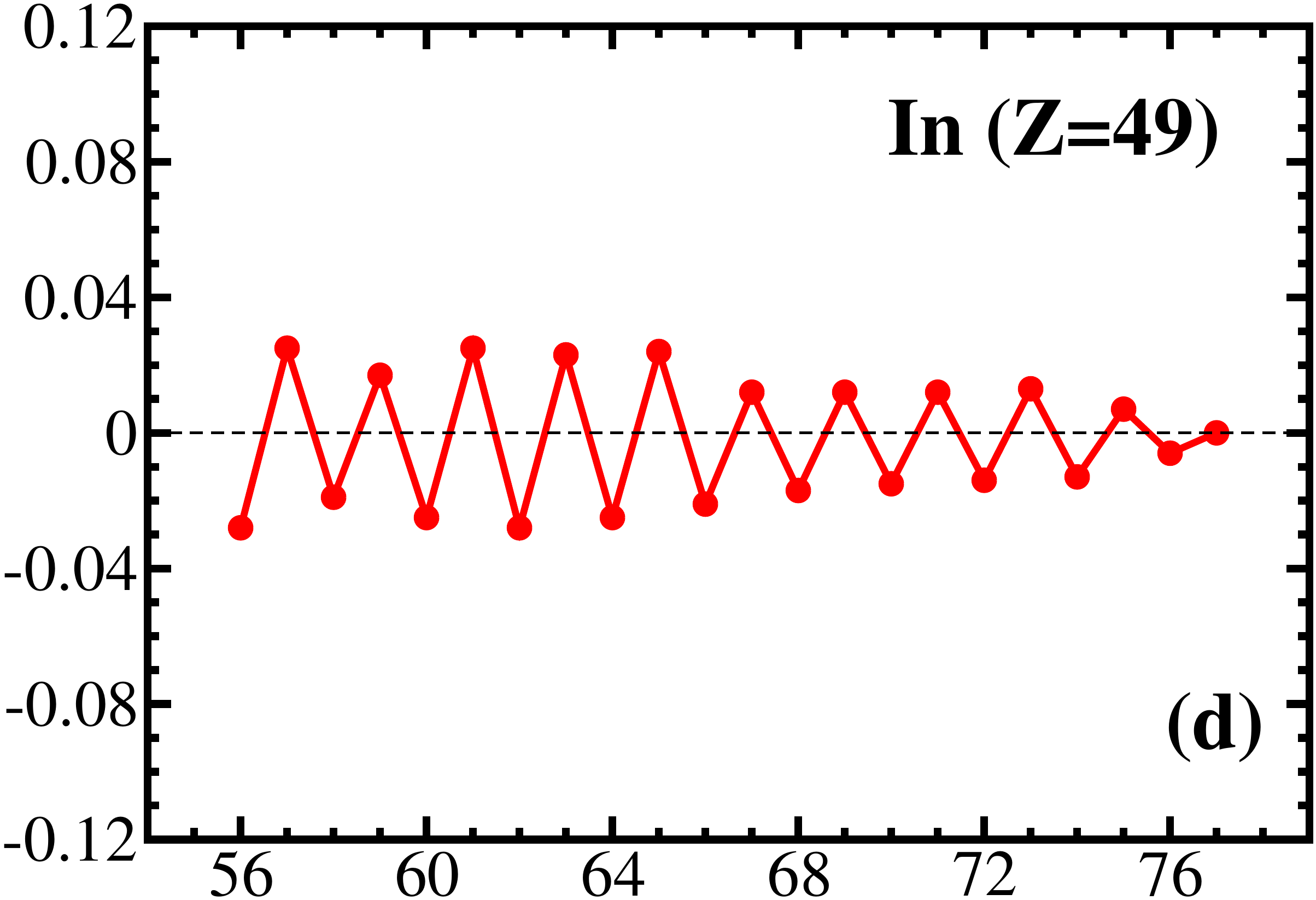}
\includegraphics[width=4.62cm]{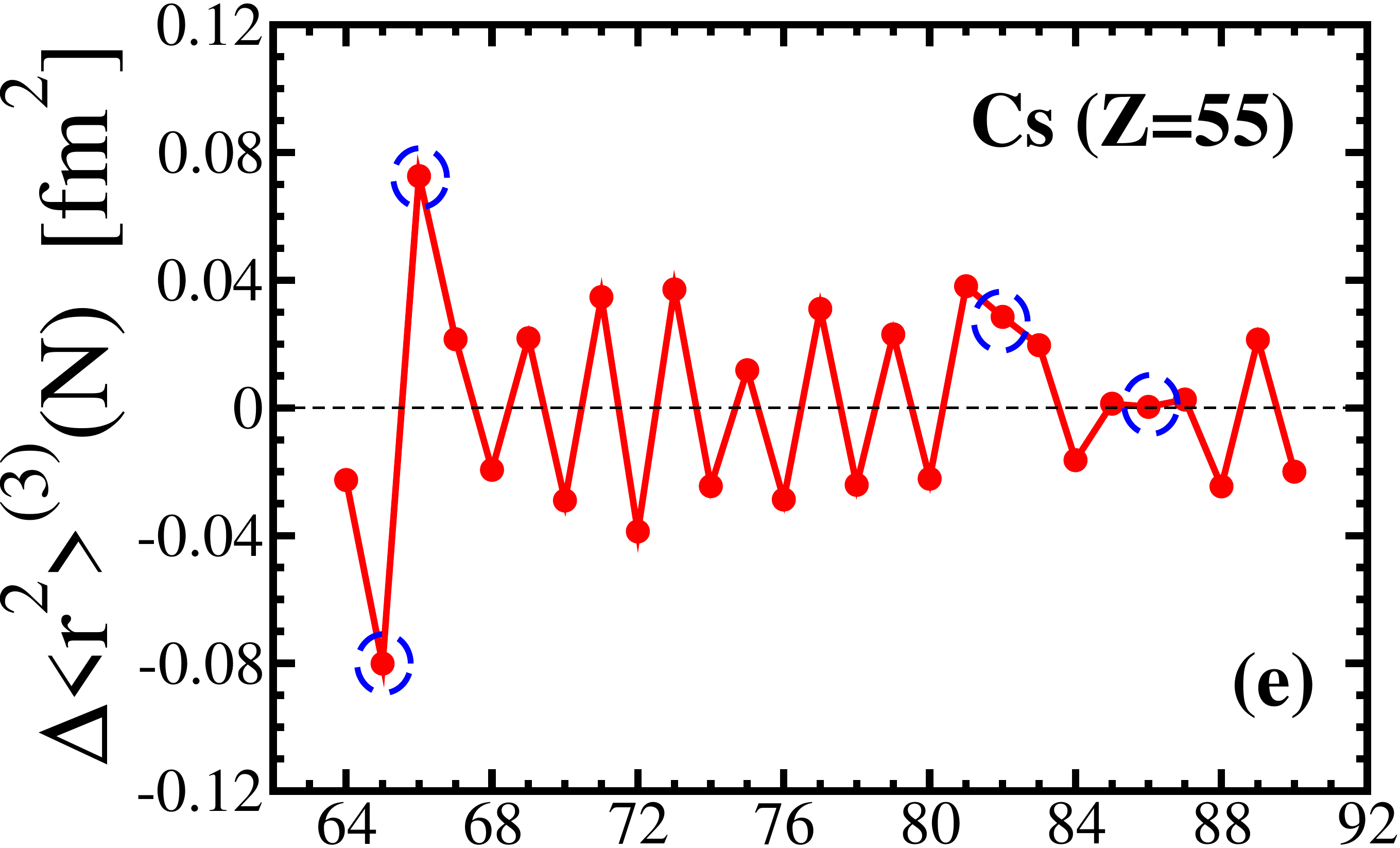}
\includegraphics[width=4.15cm]{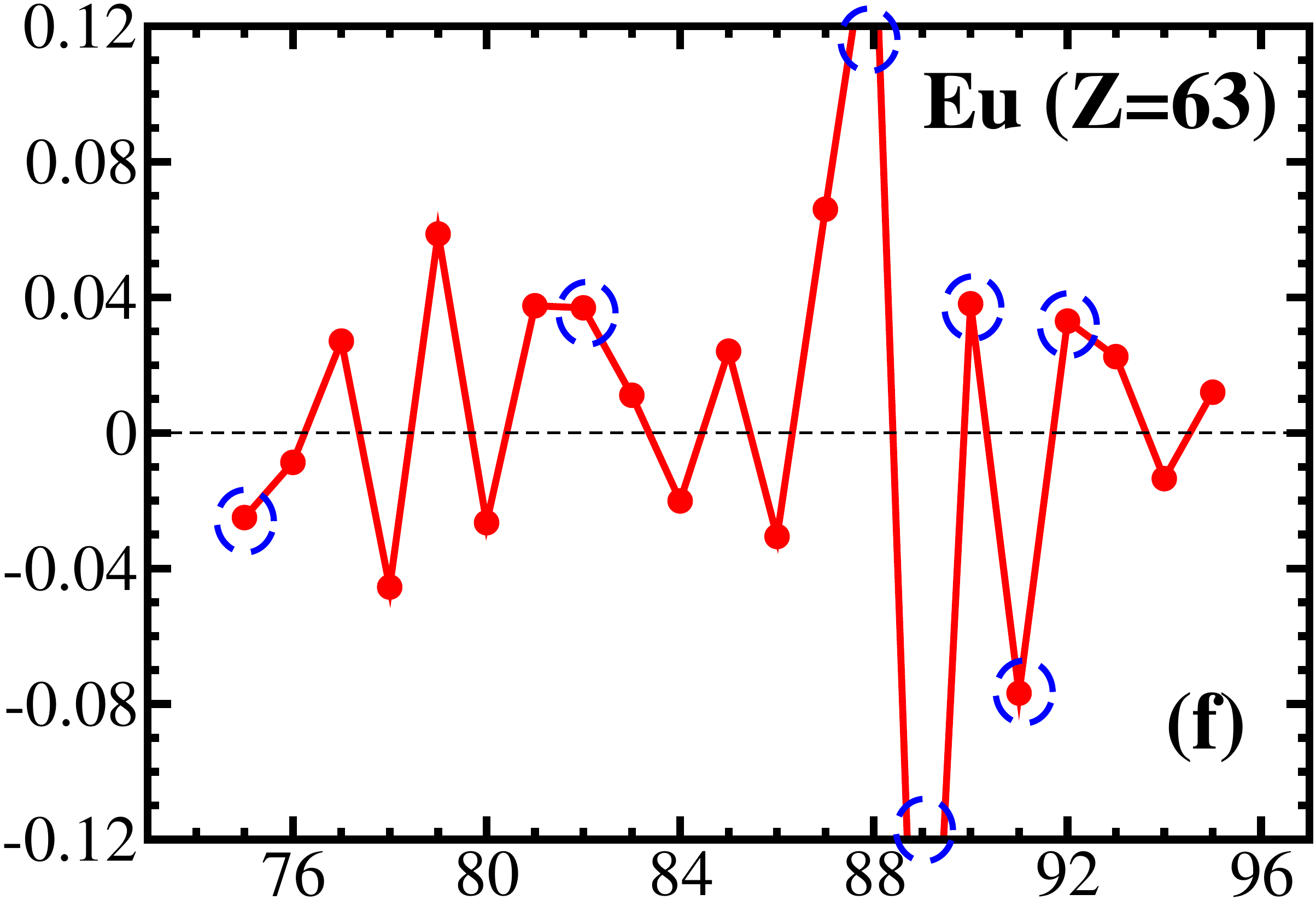}
\includegraphics[width=4.18cm]{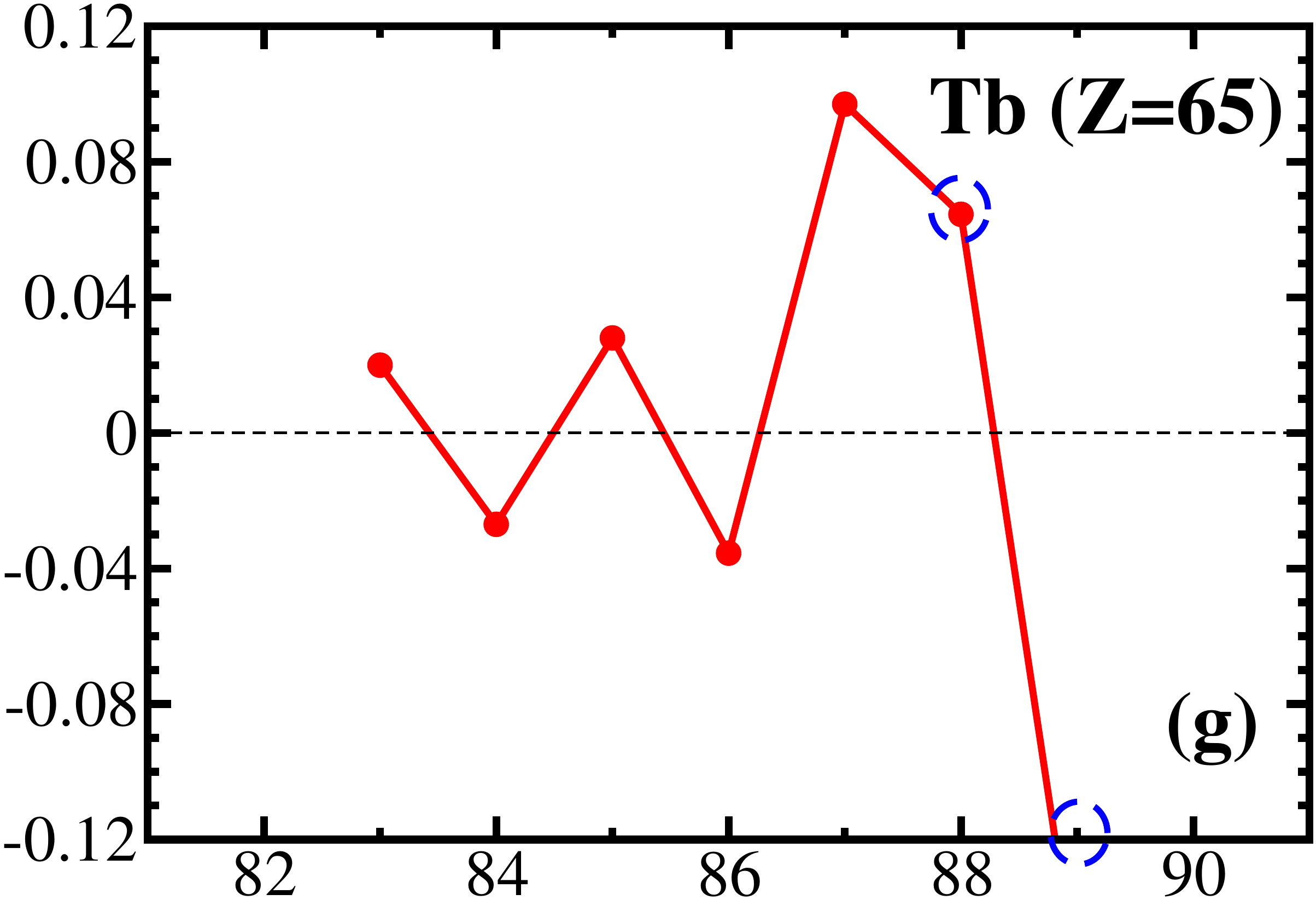}
\includegraphics[width=4.12cm]{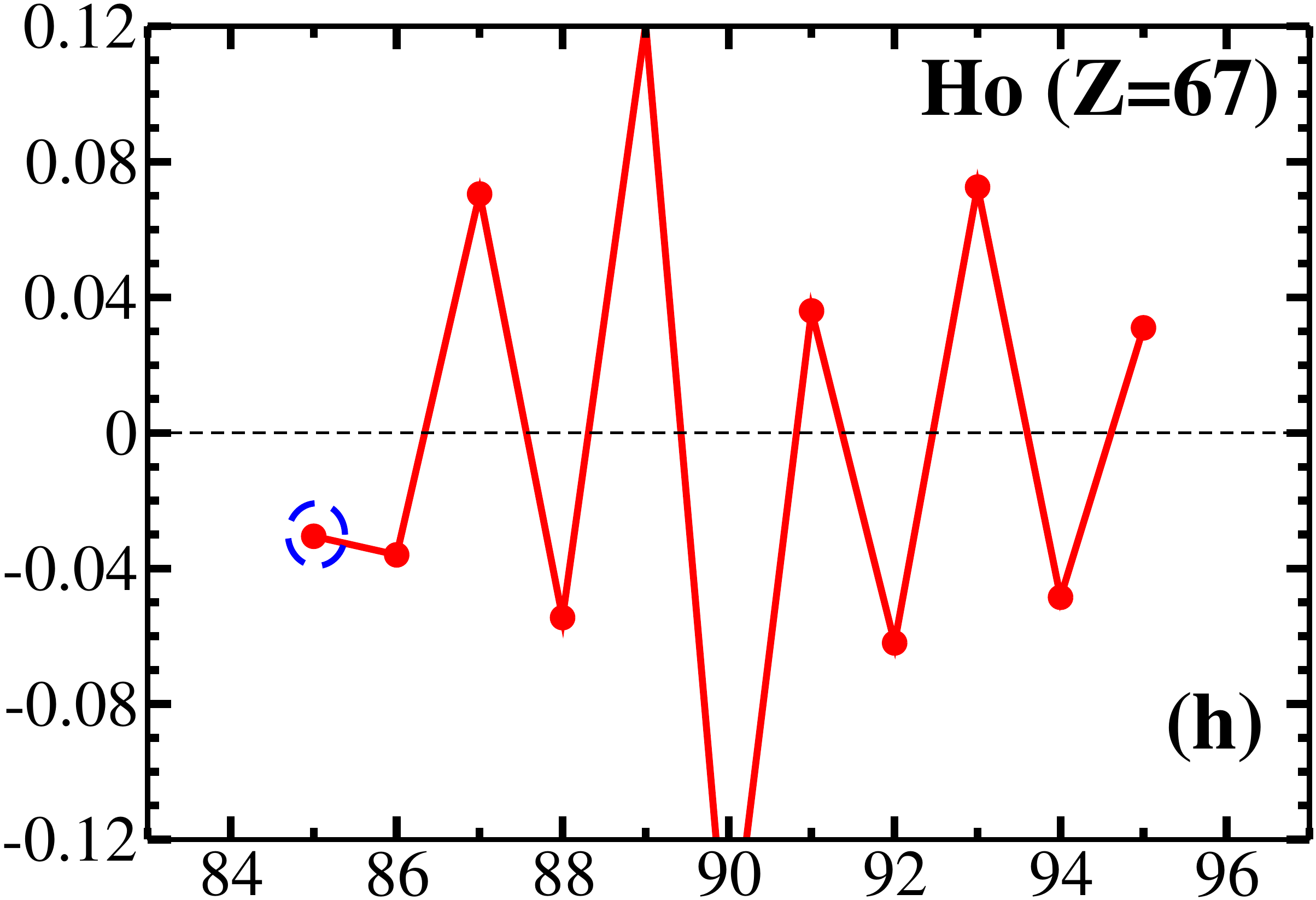}
\includegraphics[width=4.62cm]{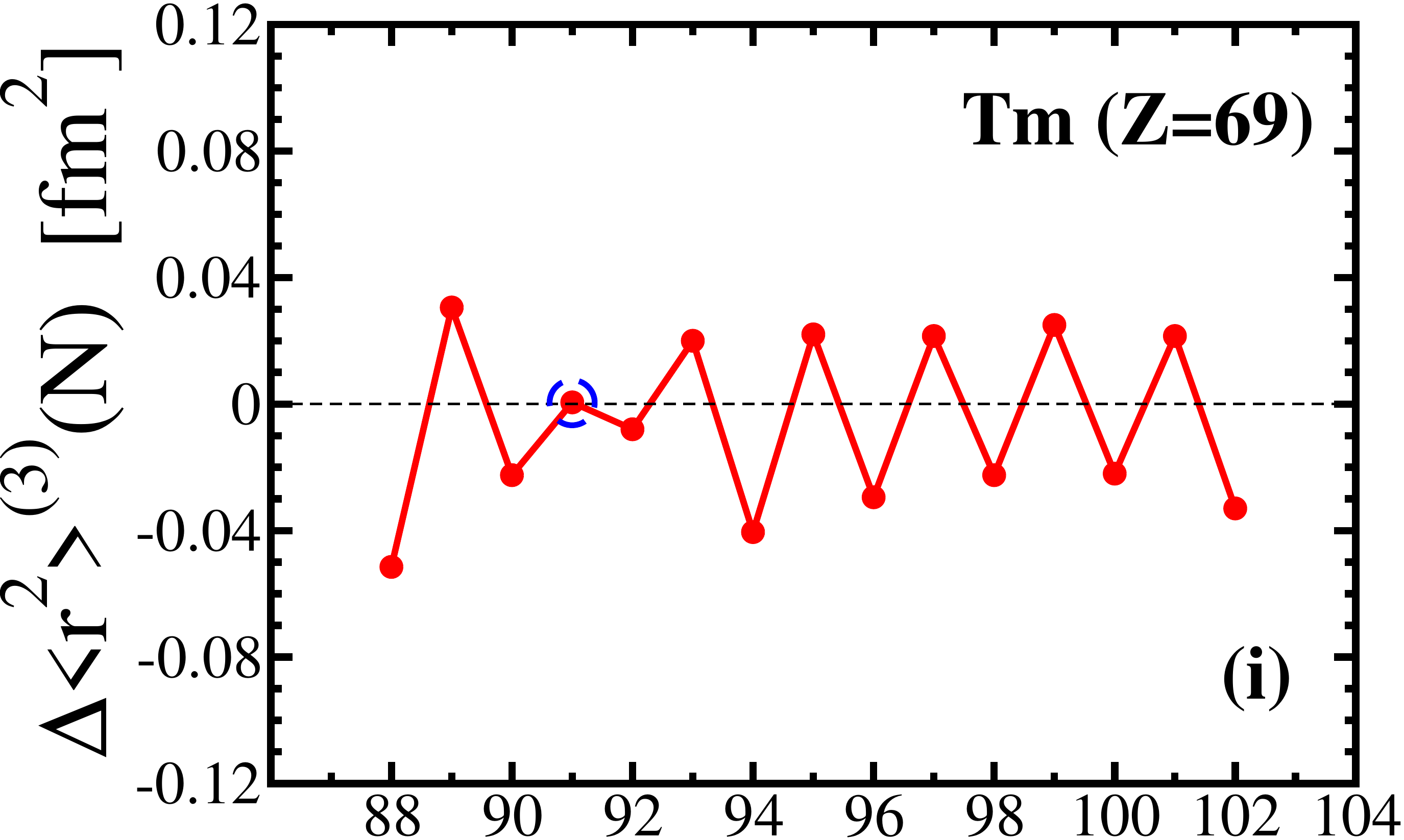}
\includegraphics[width=4.12cm]{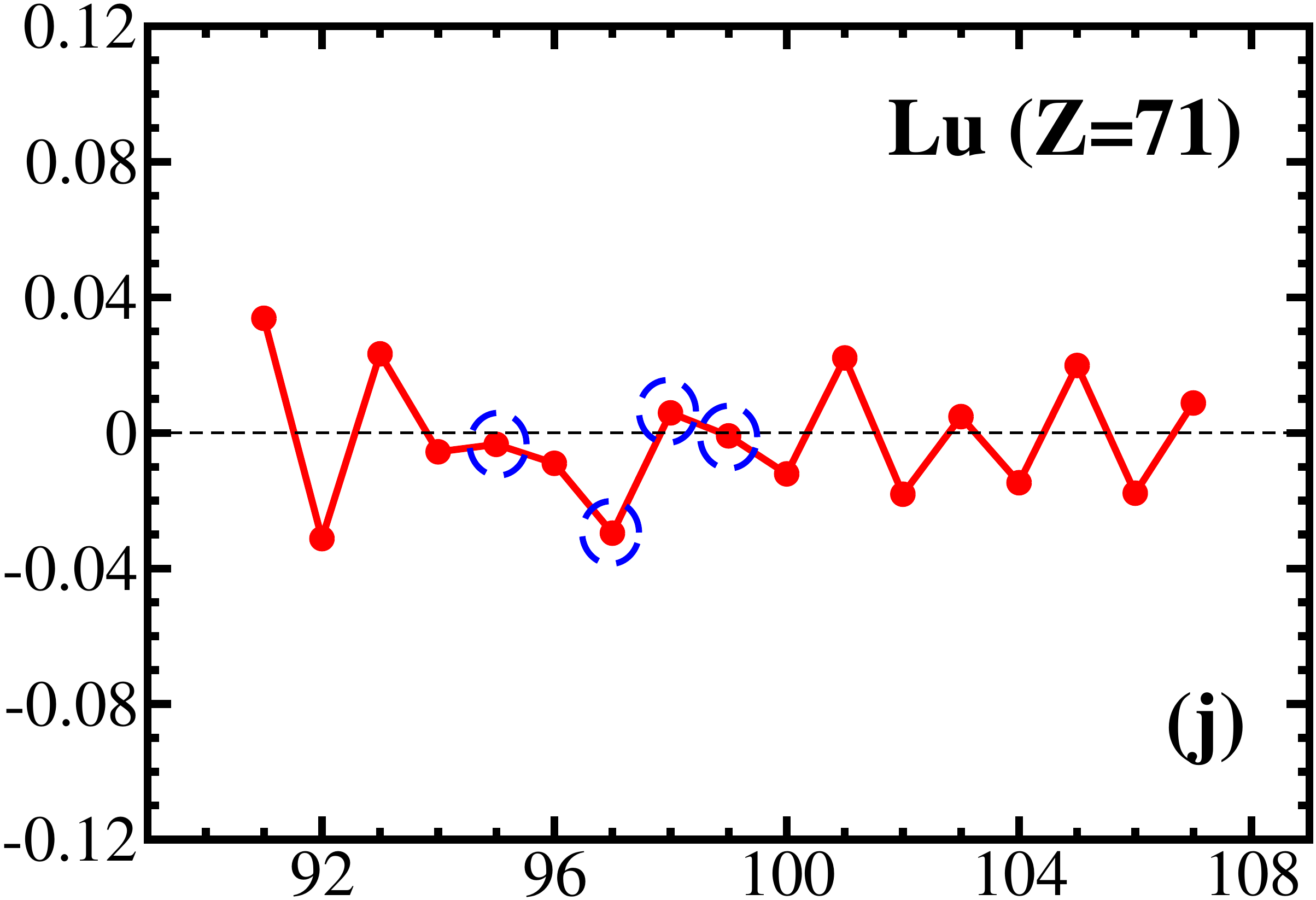}
\includegraphics[width=4.15cm]{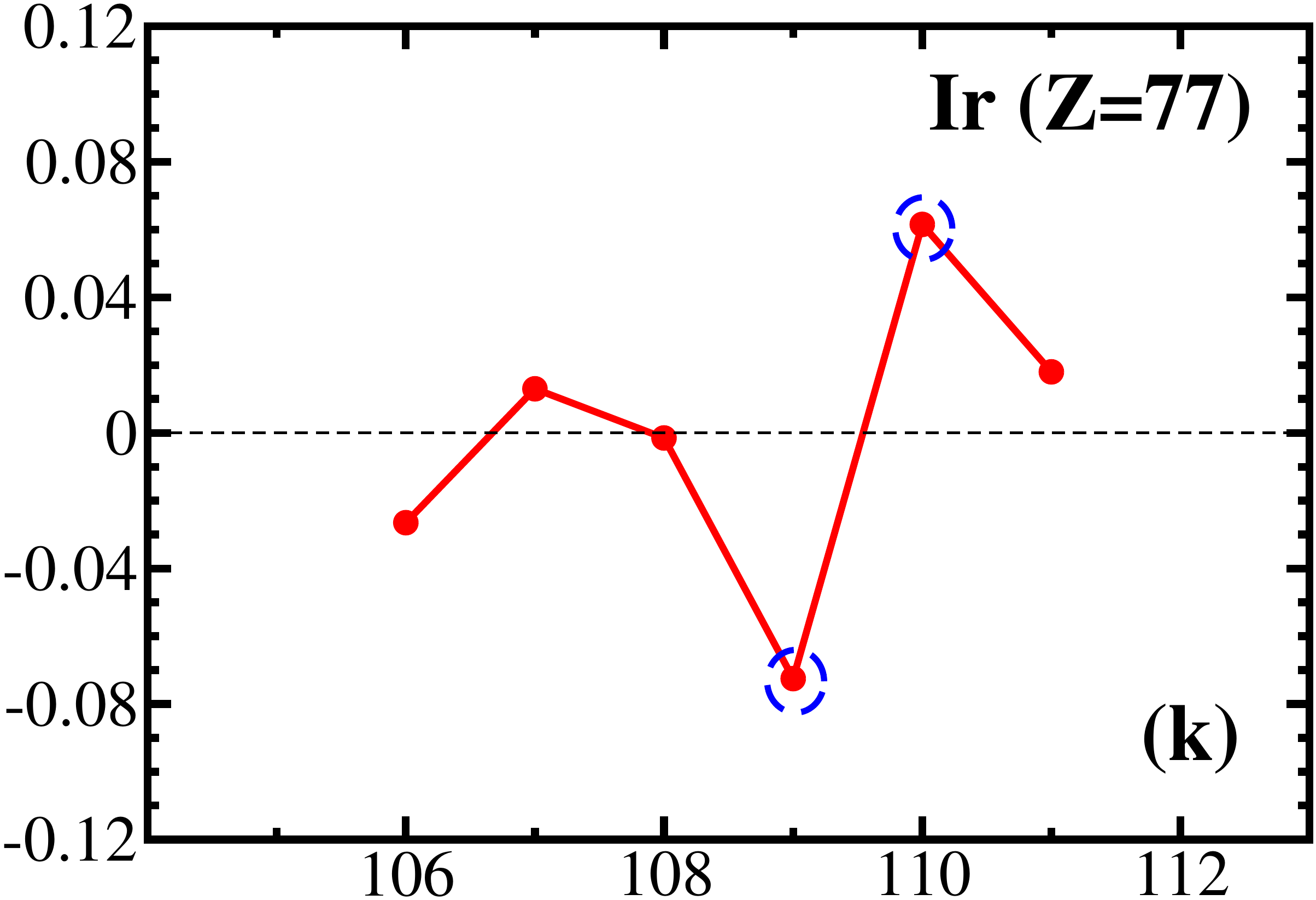}
\includegraphics[width=4.12cm]{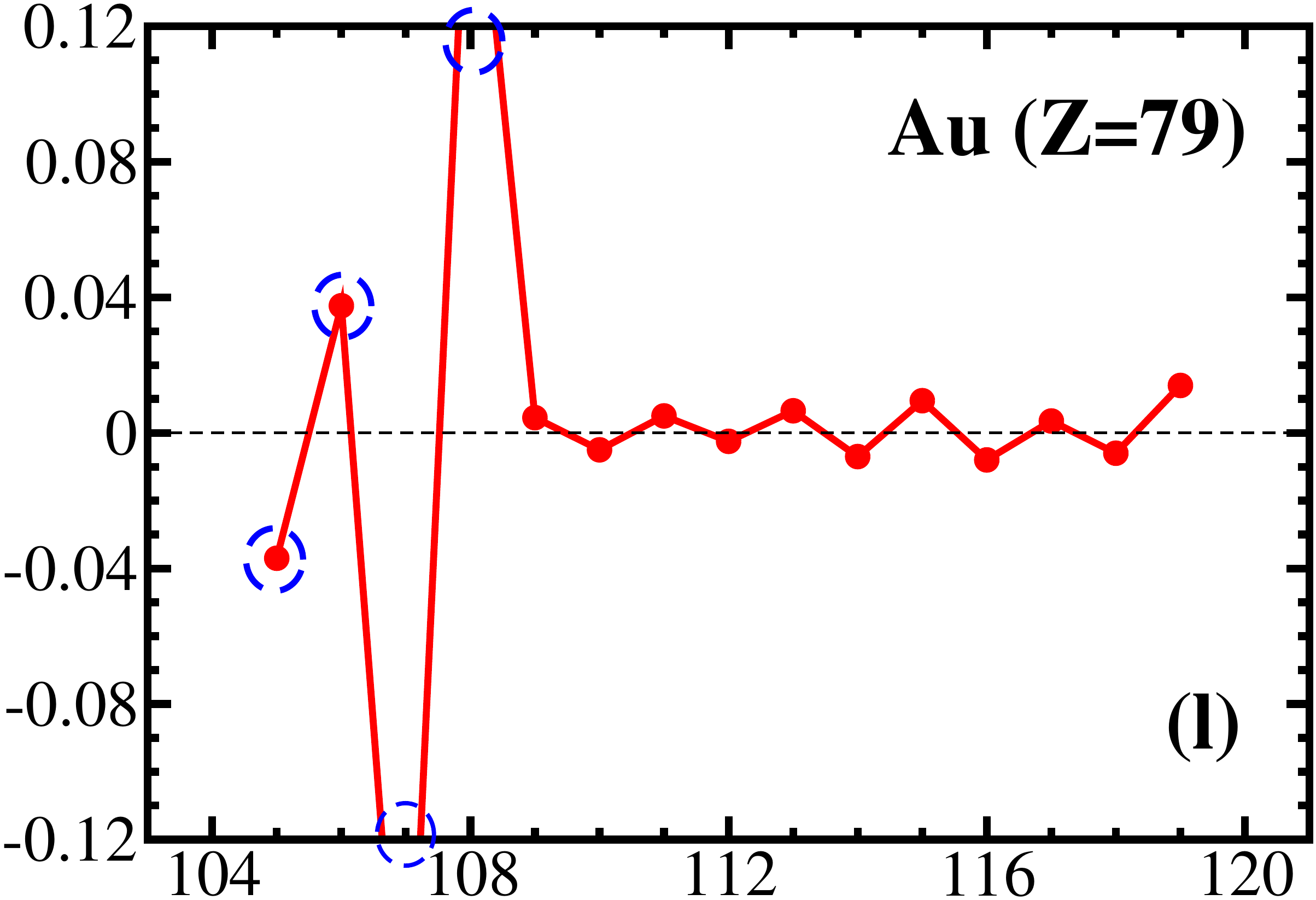}
\includegraphics[width=4.60cm]{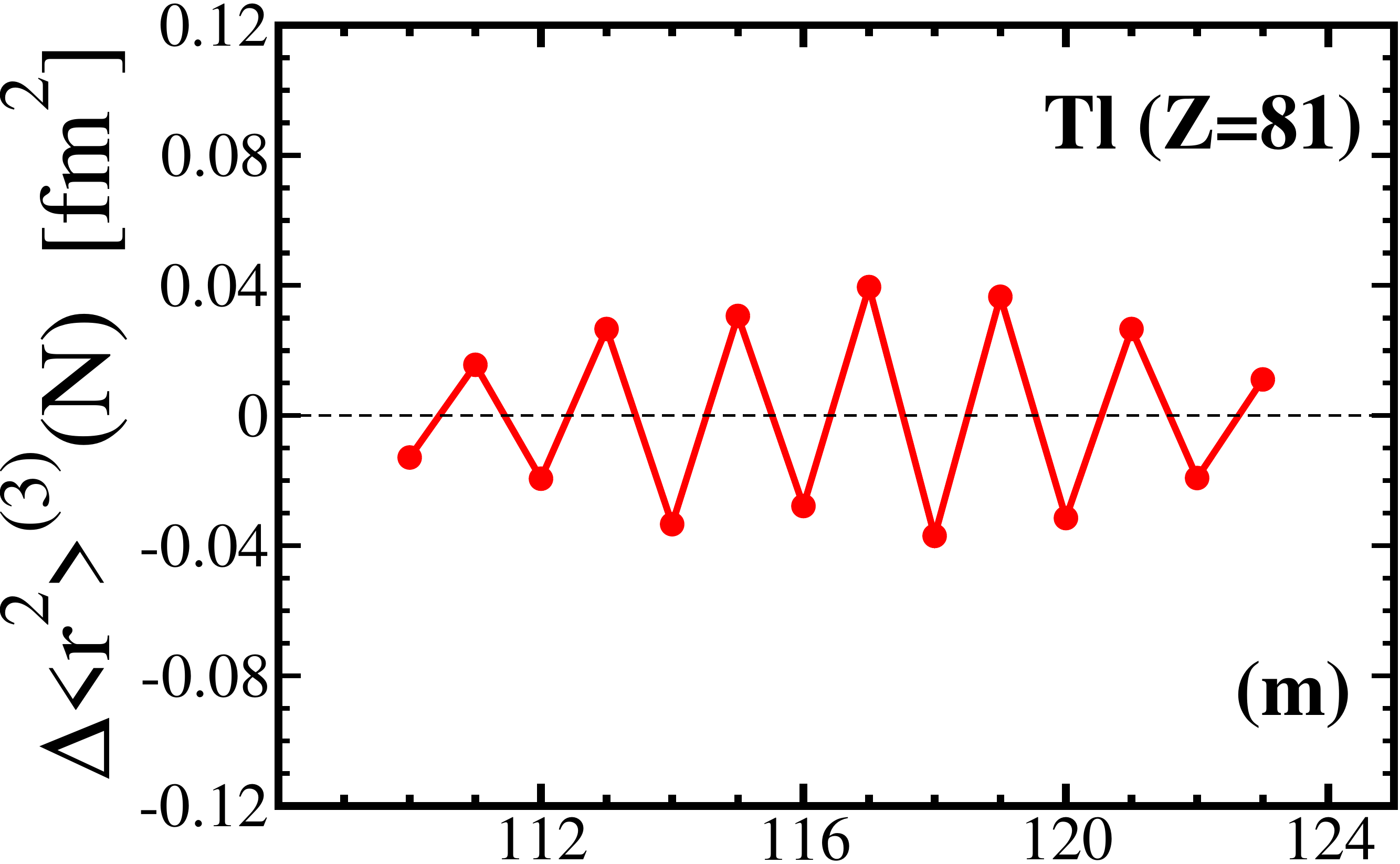}
\includegraphics[width=4.25cm]{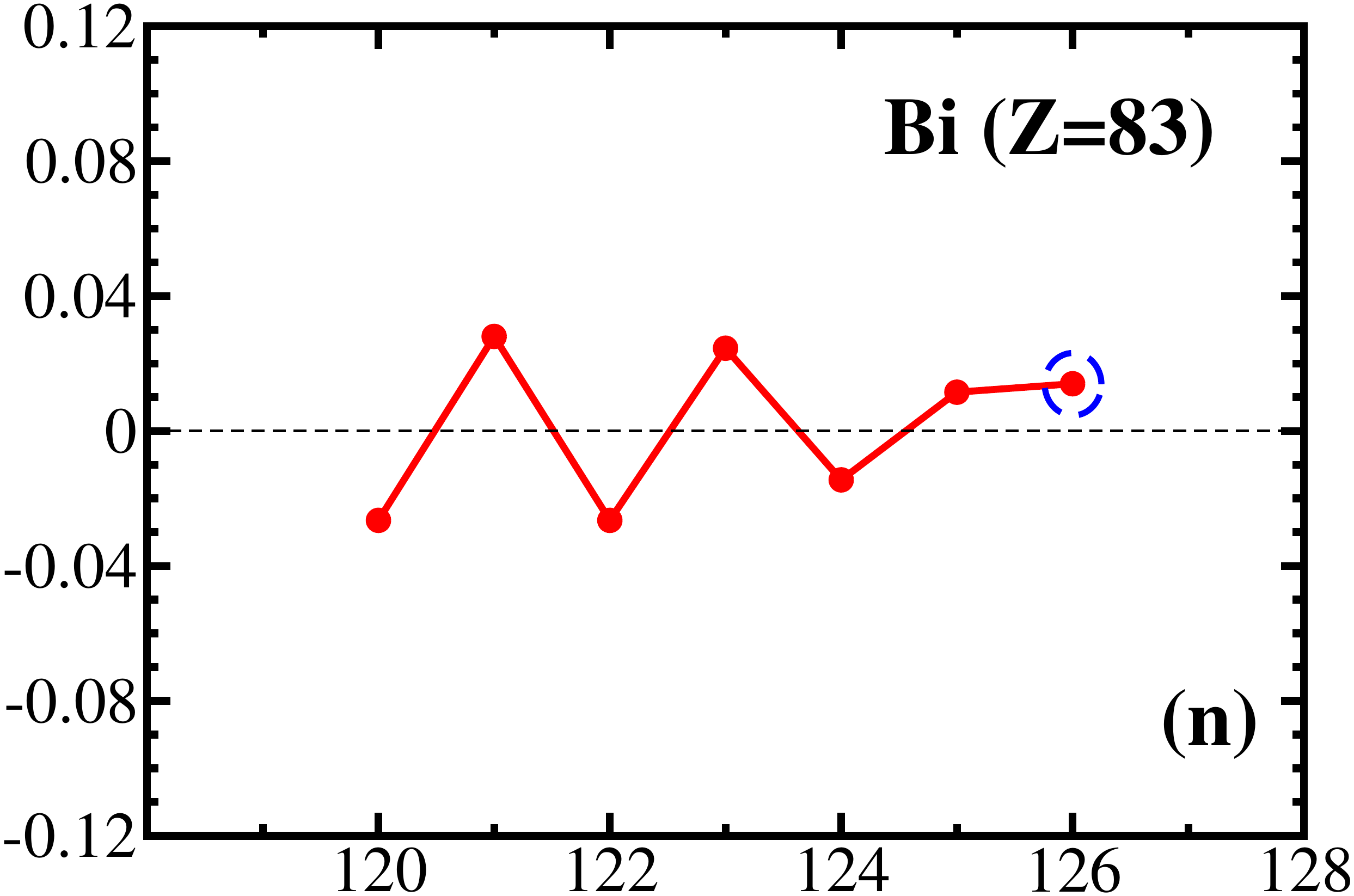}
\includegraphics[width=4.12cm]{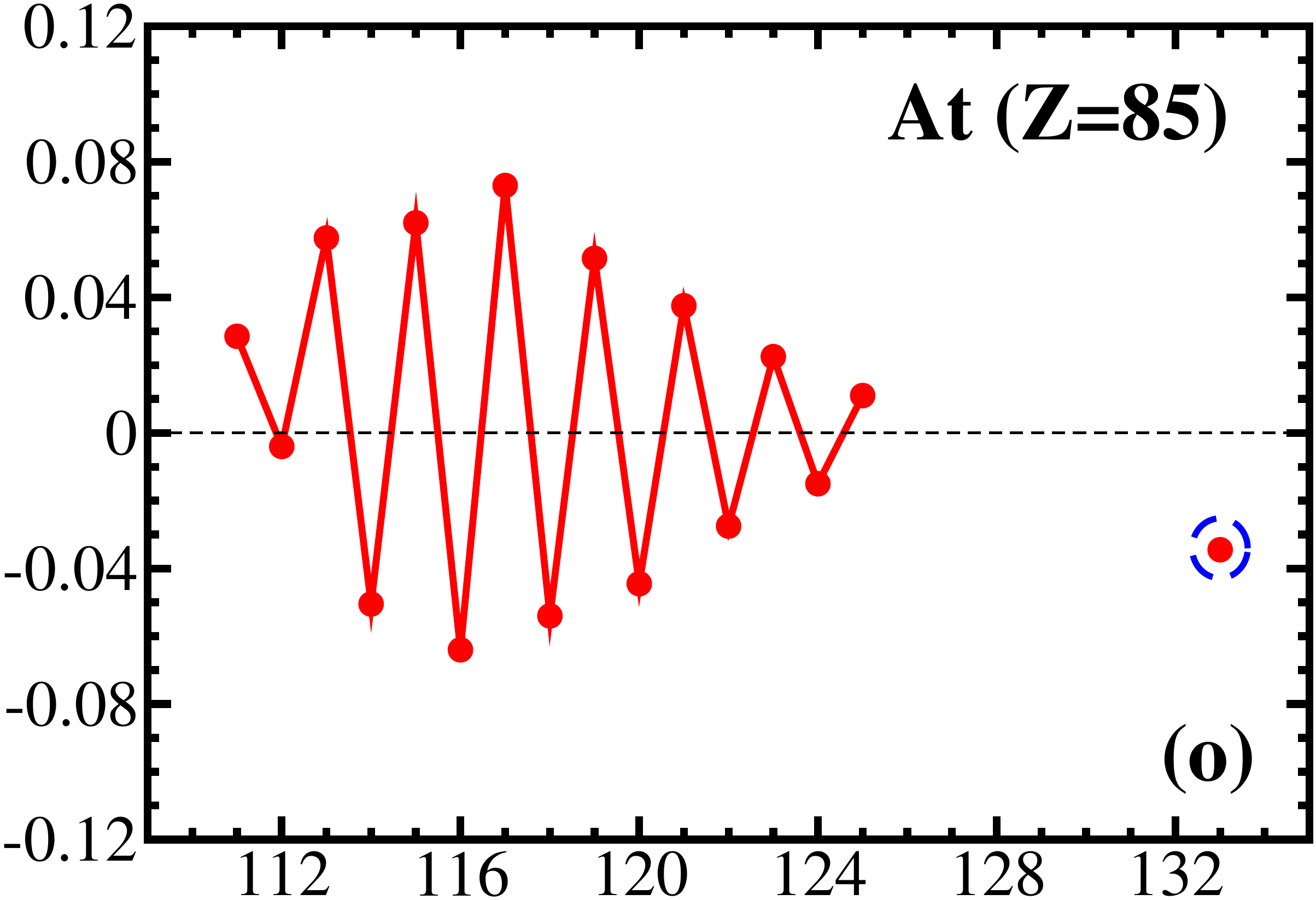}
\includegraphics[width=4.12cm]{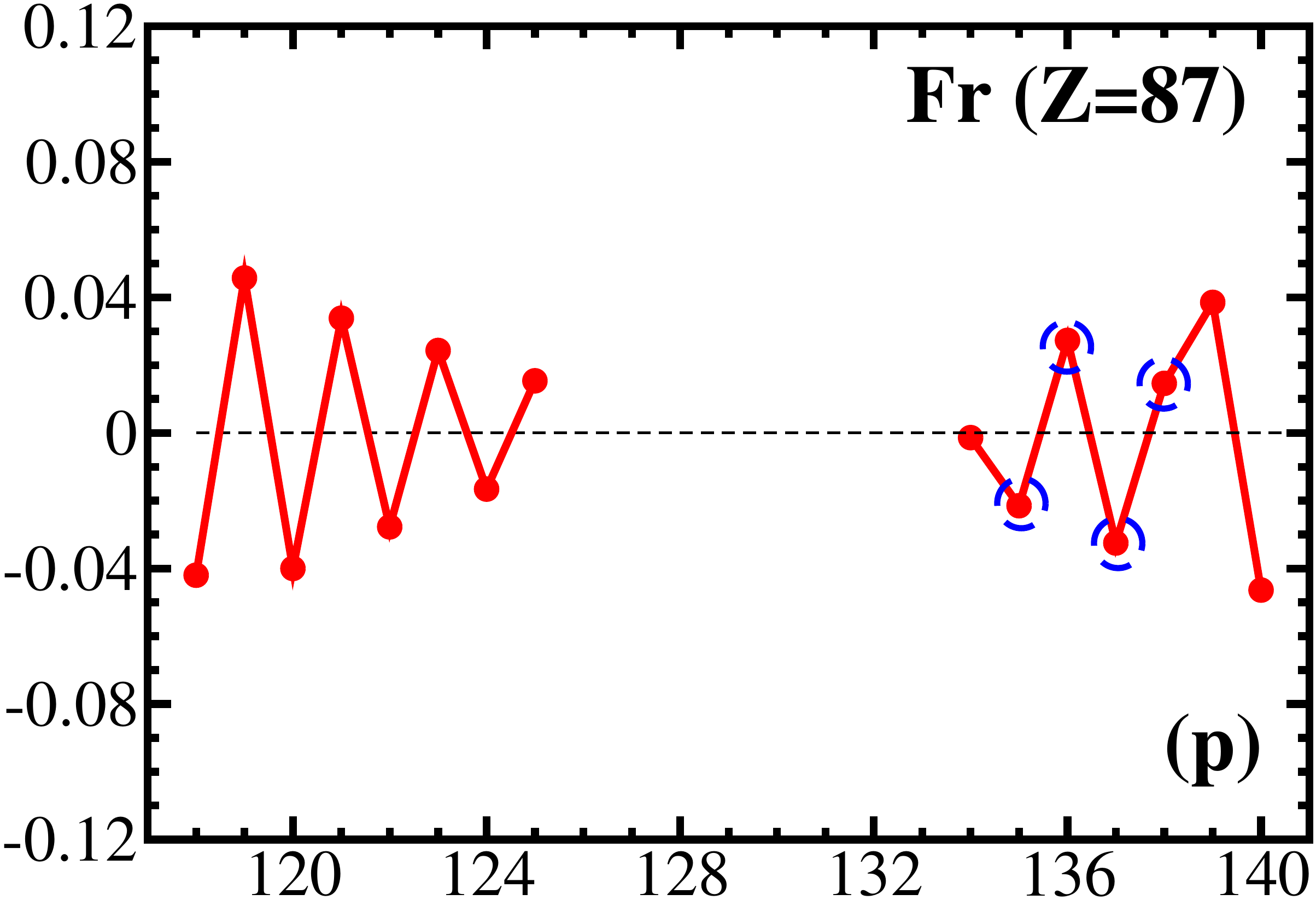}
\includegraphics[width=4.64cm]{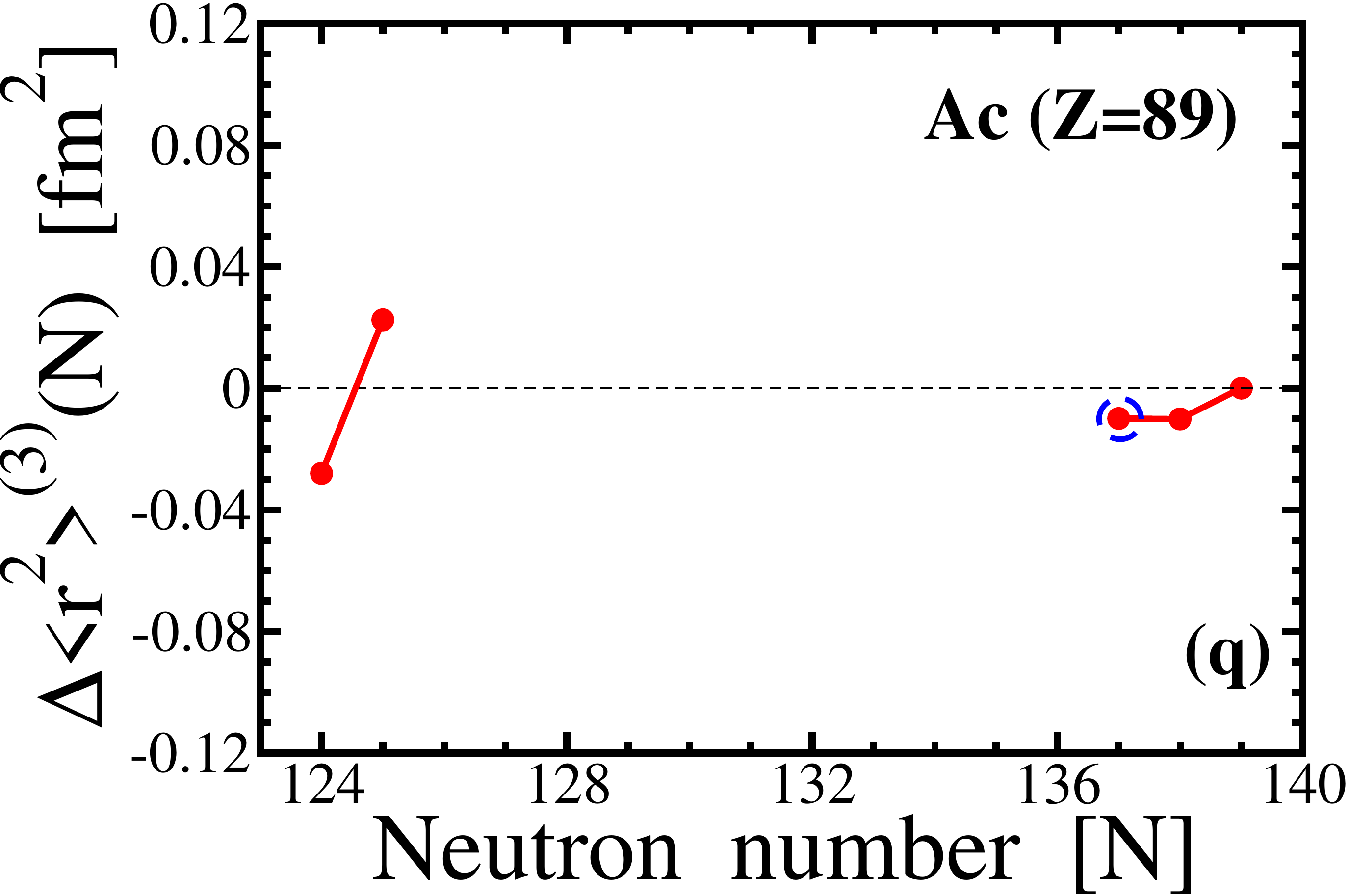}
\caption{The same as Fig.\ \ref{OES-even-Z} but for odd-$Z$ nuclei. The
experimental data are taken from Refs.\ \cite{AM.13} (compilation),
\cite{Fr-radii.15} (Fr isotopes), \cite{At-radii.19,At-radii.18} (At isotopes)
and \cite{Ac-radii.18,Ac-radii-NatureComm.17} (Ac isotopes).
\label{OES-odd-Z}
}
\end{figure*}

\begin{figure}[htb]
\centering
\includegraphics[width=8.5cm]{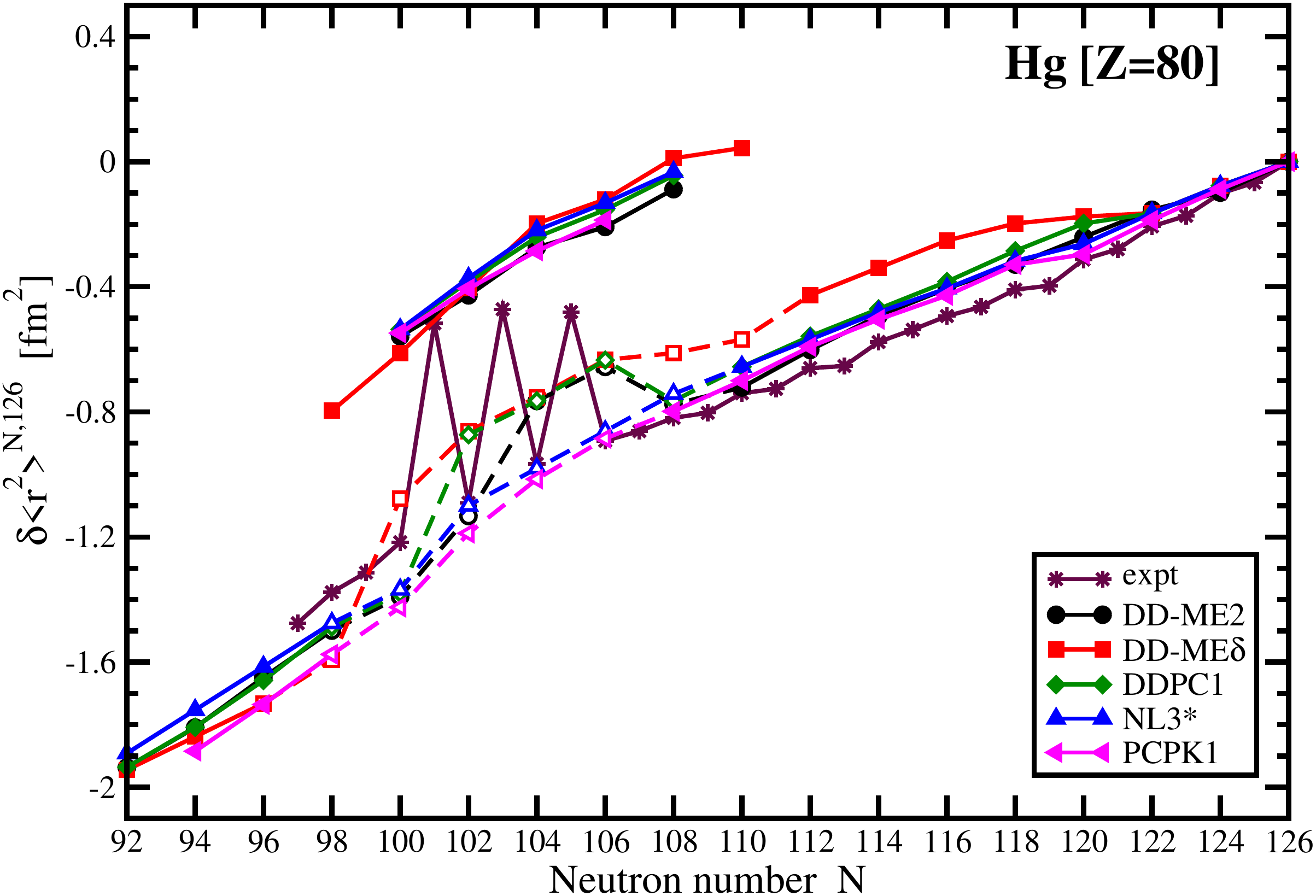}
\caption{Calculated and experimental $\delta \left<r^2 \right>^{N,126}$ values of the
Hg isotopes. The values corresponding to the lowest in energy solutions are shown
by solid lines, while those for excited ones by dashed lines.
}
\label{Hg-staggering}
\end{figure}

\section{Odd-even staggering in charge radii}
\label{sec-OES}

  The compilation of all available experimental data on OES in charge
radii is presented in Figs.\ \ref{OES-even-Z} and \ref{OES-odd-Z}.  In most of the
cases, the charge radius of an odd-$N$ nucleus is smaller than the average of
its even-$N$ neighbors.  This corresponds to positive and negative values
of the $\Delta \left< r^2 \right>^{(3)}(N)$ indicators at odd- and even-$N$ values,
respectively.  However, in approximately 25\% of the cases (indicated by
dashed circles in Figs.\ \ref{OES-even-Z} and \ref{OES-odd-Z}), this order
is inverted. Then we speak about inverted OES in charge radii.
The origin of this inversion depends on the neutron number.

 The full or near-complete collapse of neutron pairing at magic neutron shell closures
at $N=28$ in the Ca isotopes [Fig.\ \ref{OES-even-Z}(c)], at $N=50$ in the Kr, Sr
[Figs.\ \ref{OES-even-Z}(e) and (f)] and Rb [Fig.\ \ref{OES-odd-Z}(c)], at
$N=82$ in the Sn, Ba, Nd, and Sm [Figs.\ \ref{OES-even-Z}(h), (j), (k) and (l)],
Cs and Eu [Figs.\ \ref{OES-odd-Z}(e)
and (f)] isotopes, at $N=126$ in the Pb  [Fig.\ \ref{OES-even-Z}(r)] and Bi
[Fig.\ \ref{OES-odd-Z}(n)] is one of such sources of the inversion of OES in charge radii.
Note that this kind of inversion is mostly localized at neutron
numbers corresponding to the shell closures in these isotopic chains.

  The transition from spherical or quasi-spherical nuclei to deformed ones taking
place with increasing neutron number  at $N\approx 88$ triggers the inversion of
OES in charge radii of the  Dy, Tb, Eu, and Tm isotopic chains [see Fig.\
\ref{OES-even-Z}(m) and Fig.\ \ref{OES-odd-Z}(g), (f) and (i)]. A similar transition
at $N\approx 58$  is responsible for the inversion of OES in the Rb  isotopic chain
[see Fig.\ \ref{OES-odd-Z}(c)].
Note that  not in all cases this kind of transition triggers the inversion of OES: the magnitude of
OES in charge radii is simply increased in the vicinity of these neutron numbers as  compared
with the ones for lower/higher $N$ values in isotopic chains of the Kr, Sr, Sm, and Ho [Fig.\
\ref{OES-even-Z}(e), (f)  and (m)  and Fig.\ \ref{OES-odd-Z}(g)].

   In addition, several other mechanisms of OES in charge radii and its
inversion have been suggested earlier. They will be discussed below using the results
obtained in the CDFT framework. However, the mechanism presented in Sec.\
\ref{sec-OES-PVC} is completely new.

\subsection{Shape coexistence as a source of OES in charge radii
                   and its inversion}
\label{sec-OES-shape}

   Significant odd-even staggering in the Hg charge radii exists
at $N=100-106$ [see Fig.\ \ref{Pb-region-radii}(a) and Fig.\ \ref{Hg-staggering}].
Several scenarios have been suggested for an explanation of this OES
(see overviews  in Sec. 4.7 of Ref.\ \cite{Otten.89} and in Ref.\ \cite{Hg-radii-low-N.19}),
but the one which agrees most with experimental data on OES in charge radii was
first suggested in Ref.\ \cite{FP.75}.  This paper, together with the analysis in the Skyrme
DFT (Ref.\ \cite{Hg-radii-low-N.19}) and CDFT (present paper) suggest the
following scenario: the even-$N$ isotopes should have a weakly deformed oblate
minimum (quasi-spherical in the language of Ref.\ \cite{Otten.89}) while the odd-$N$
nuclei in the region should have large prolate deformations. Two such minima (oblate
with $\beta_2  \approx -0.15$ and prolate with $\beta_2\approx 0.3$) coexists in the
isotopes of interest (see  Fig.\ \ref{PEC-Hg}). Note that the latter values are close to
experimental estimates of the deformations in $^{181,183,185}$Hg (see Ref.\
\cite{Otten.89}). Under such a scenario the evolution of the $\delta \left< r^2 \right>^{N,126}$
values for even-$N$ numbers is reasonably well described, especially with the NL3*
and PC-PK1 functionals (see Fig.\ \ref{Hg-staggering}).  In addition, the magnitude of OES
(as the difference of the charge radii in prolate and oblate minima) is not far away from
experimental values.

   The only caveat in this CDFT interpretation is the fact that the prolate minimum is
the lowest in energy in the nuclei for which OES in charge radii is observed (see Fig.\ \ref{PEC-Hg}).
However,  the oblate  minimum is only by approximately 1 MeV higher in energy than the prolate
one for most functionals.  The only exception is the DD-ME$\delta$ functional,
for which this difference is more significant. One should note that this energy difference between
the minima is extremely sensitive to the fine details of the functional and that most
non-relativistic models also fail to reproduce this difference
(see review in Sec. IVD of Ref.\ \cite{Hg-radii-low-N.19} and Ref.\ \cite{GR.82}).
Note also that the PECs in Hg nuclei with $N\leq 100$  show that the oblate minimum
becomes the lowest in energy, and the prolate minimum starts to disappear. This
is consistent with the disappearance of OES in charge radii seen in the experiment
at low $N$ (see Fig.\ \ref{Hg-staggering}).

 Since the deformation and thus the charge radii are larger in odd-$N$ isotopes
as compared with even-$N$ ones, the OES of charge radii in the light Hg isotopes is
inverted\footnote{One can easily imagine a situation where the absolute values
of deformation (and thus charge radii) are smaller in odd-$N$ isotopes as compared
with even-$N$ ones. This will lead to a regular OES. Thus, a sensitive energy balance
between two local minima with different deformations and deformation driving properties
of the unpaired orbital in the odd-$N$ nucleus will define whether OES is regular or inverted.}
 [see Fig.\ \ref{OES-even-Z}(q)]. Note that this is the largest OES of charge
radii in the whole nuclear chart. A similar, but somewhat smaller, inverted OES is
observed in the neighboring Pt (for $N=101-110)$ [see Fig.\ \ref{OES-even-Z}(p)]
and Au (for $N=105-108$) [see Fig.\ \ref{OES-odd-Z}(l)] isotopic chains. Considering
the magnitude of OES as well as its localization in neutron number, it is quite
likely that it has the same origin as in the Hg isotopes. It may be that the small
inverted OES seen at $N=101-104$ in the Pb isotopes [see Fig.\ \ref{OES-even-Z}(r)]
has a similar origin.

\subsection{Pairing as a source of OES in charge radii}
\label{sec-OES-pairing}

  As illustrated in previous examples, the charge radii are increasing nearly linearly with
increasing neutron  number when the single-particle states of the same spherical  neutron
subshell (let us call it a $j$-shell) are occupied. This trend is schematically
illustrated as a dashed line in Fig.\ \ref{OES-origin}.  In the calculations with pairing, the blocking effect
in odd-$N$ nuclei  leads to an additional redistribution of the occupation of the single-particle orbitals
and typically to a decrease of the charge radius of the odd-$N$ nucleus below the average given by two
even-even neighbors. As a consequence, the increase in charge radii $\Delta (r_{ch})^{qp}_{MF}$
on going from even-$N$ to odd-$A$ is typically smaller as compared both with the increase
corresponding to a linear increase defined by $r_{ch1}$ and $r_{ch3}$ and that obtained in the
calculations without pairing (see Fig.\ \ref{OES-origin}). This leads to a regular
OES in charge radii and, so far, the pairing has been considered to be its dominant source in isotopic
chains which do not undergo significant shape changes like those discussed in Sec.\
\ref{sec-OES-shape} (see Refs.\ \cite{Otten.89,FTTZ.00,RN.17}) .

  Let us consider an example of realistic calculations in the Sn isotopes. In the RHB calculations,
two different procedures labeled as LES (lowest in energy solution) and EGS
(experimental ground-state) are used for the blocking in odd-A
nuclei\footnote{They were
first employed in the studies of OES of differential radii in the Pb and Hg isotopic chains in Ref.\
\cite{Pb-Hg-charge-radii-PRL.21}.}, and these abbreviations label the results of the respective calculations. 
In the LES procedure, the lowest in energy configuration
is used. It has been applied in all earlier calculations of OES with non-relativistic DFTs \cite{FTTZ.00,RN.17}.
In the EGS procedure, the configuration with the spin and parity of the blocked state corresponding to
those of the experimental ground state is employed,  although it is not necessarily the lowest in energy.

\begin{figure*}[htb]
\centering
\includegraphics[width=16.0cm]{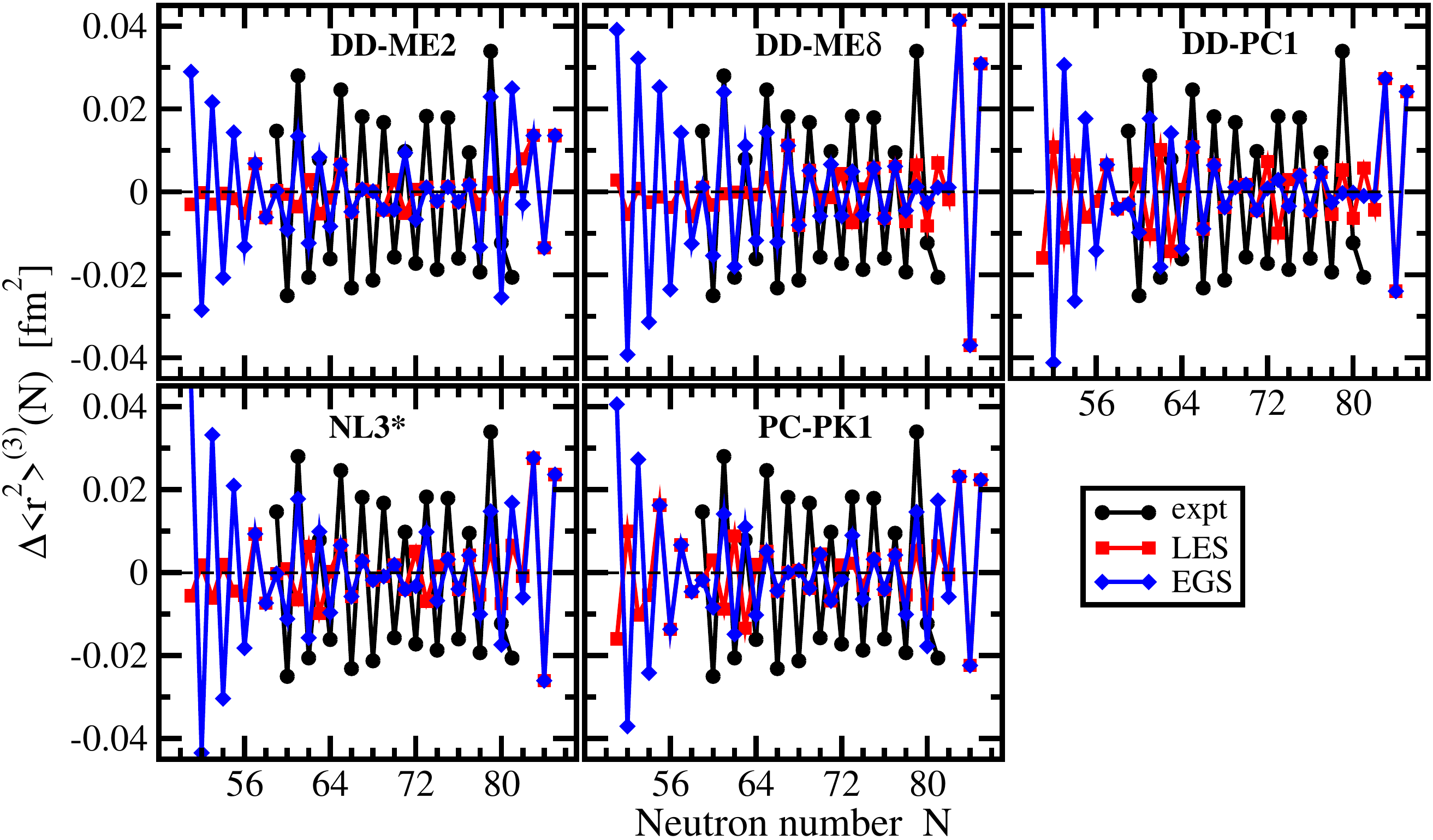}
\caption{The OES in charge radii of the Sn isotopes. The experimental data are taken from
Ref.\ \cite{AM.13}.
}
\label{Sn-OES}
\end{figure*}

 In Fig.\ \ref{Sn-OES} the results of calculations with different functionals are compared with experimental data.
One can see that the results of the RHB calculations with the LES procedure
significantly underestimate the magnitude of experimental OES and occasionally provide a
wrong phase of the OES. The use of the EGS procedure significantly improves the description of both
the phases and the magnitude of OES, especially for DD-ME$\delta$.
However, even in that case, the magnitude of OES is underestimated by a factor of approximately
two. This suggests that an important part of physics is still missing; it is addressed in
Sec.\ \ref{sec-OES-PVC}.  Note that the analysis of the Pb and Hg isotopic chains performed both
in relativistic and non-relativistic frameworks indicates that the EGS procedure is needed for a
proper description of OES in charge radii (see Ref.\ \cite{Pb-Hg-charge-radii-PRL.21}).

   A significant underestimate of OES in charge radii is also observed in Skyrme DFT calculations
with conventional functionals. The suggested resolution of this problem lies in the use of Fayans
functionals which include gradient terms in both surface term and pairing \cite{FZ.96,FTTZ.00}. The
latest functional of this type is Fy($\Delta r)$ \cite{RN.17}.  However, it overestimates the magnitude of
OES in the Sn \cite{Cd-isiotopes-radii.18} and other isotopic chains \cite{RN.17}.

\begin{figure}[htb]
\centering
\includegraphics[width=8.5cm]{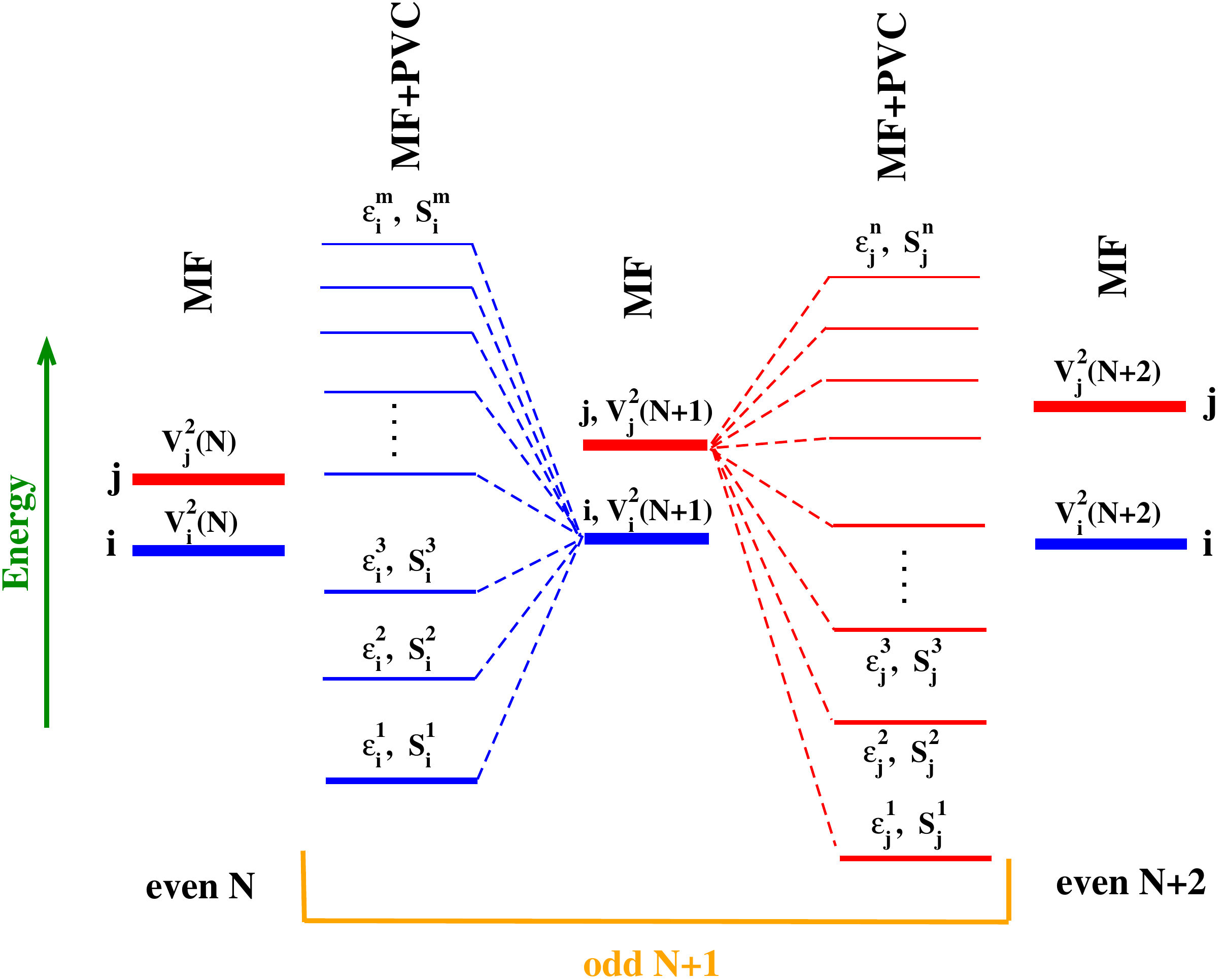}
\caption{Schematic illustration of the difference of the approximations (MF and MF+PVC)
used in the description of even-even and odd mass nuclei (see text for details). "MF" and
"MF+PVC" stands for "mean-field" and "mean-field + particle vibration coupling"
approximations, respectively.  The thickness of the horizontal lines is proportional
to spectroscopic factors $S_k^m$ in the MF+PVC columns.}
\label{EE-O-fragmen}
\end{figure}

\begin{figure}[htb]
\centering
\includegraphics[width=8.5cm]{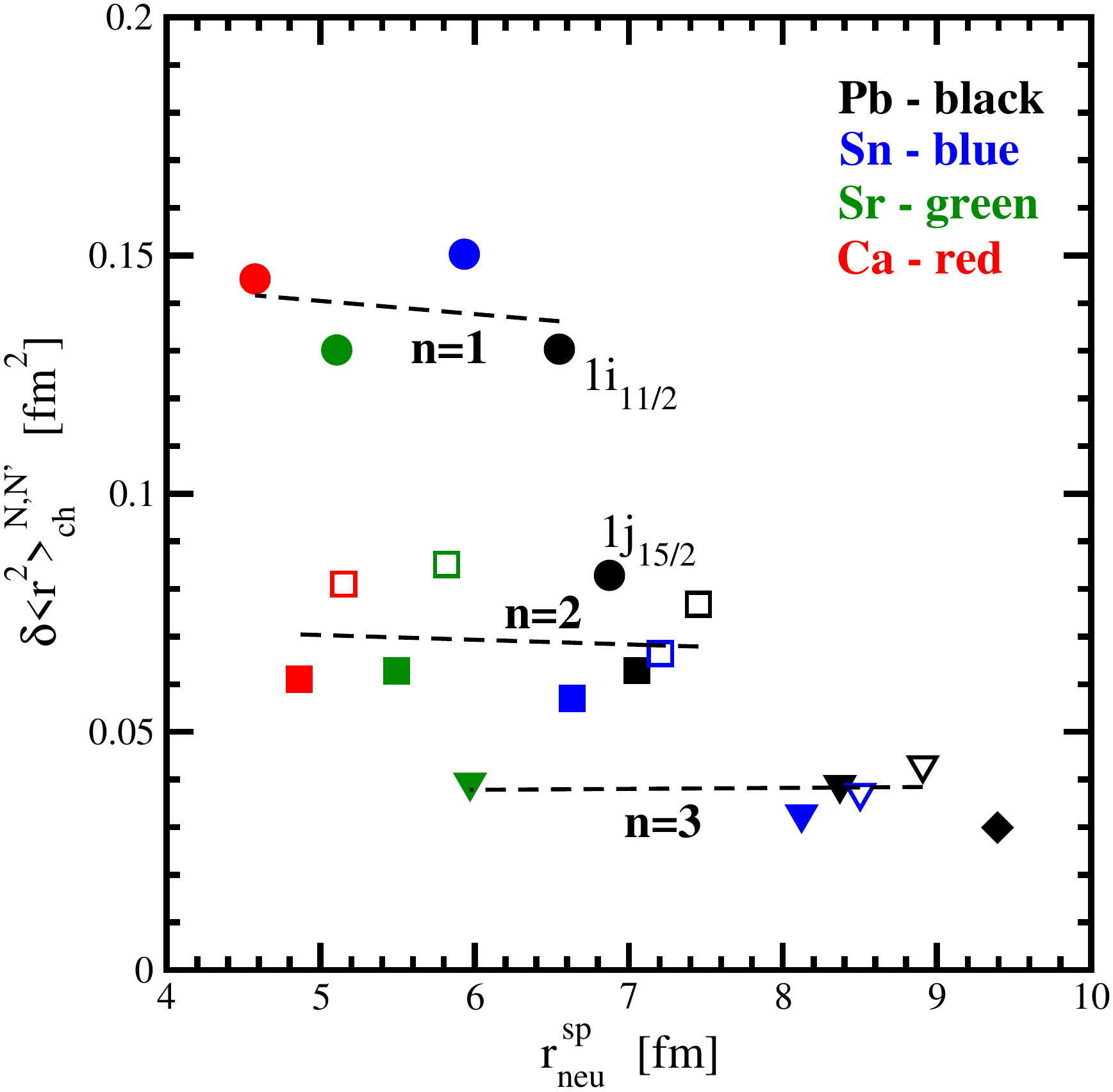}
\caption{The correlations between the differential charge radii $\delta \left< r^2 \right>^{N,N'}$
($N=N'+1$) generated by the occupation of a given single neutron orbital above the neutron shell 
closure located at $N'$ and the rms neutron radius $r_{neu}^{sp} = \sqrt{\left< r^2 \right>^{sp}}$ of this orbital. The 
results presented for the $^{48,49}$Ca $(N'=28)$, $^{88,89}$Sr $(N'=50)$,  $^{132,133}$Sn 
$(N'=82)$ and $^{208,209}$Pb $(N'=126)$ are based on the calculations without pairing and 
the NL3* CEDF. The calculations are restricted to spherical shapes. Circles, squares, triangles 
and diamonds are used for the orbitals with principal quantum numbers $n=1$, 2, 3 and 4, 
respectively. If  spin-orbit partner orbitals appear
above the shell closure, then solid (open) symbols of the same type are used for the
$j+1/2$ ($j-1/2$) ones. The following neutron orbitals are considered: $1i_{11/2}$,
$1j_{15/2}$, $2g_{9/2}$, $2g_{7/2}$, $3d_{5/2}$,  $3d_{3/2}$ and $4s_{1/2}$ in 
$^{208,209}$Pb (see Fig.\ \ref{208Pb-single-particle}(b)), $1h_{9/2}$, $2f_{7/2}$, $2f_{5/2}$,
$3p_{3/2}$ and $3p_{1/2}$ in $^{132,133}$Sn (see Fig.\ \ref{Sn132-sp-spectra}(b)), $1g_{7/2}$, $2d_{5/2}$,
$2d_{3/2}$ and $3s_{1/2}$ in $^{88,89}$Sr and $1f_{5/2}$, $2p_{3/2}$ and $2p_{1/2}$
in $^{48,49}$Ca (see Fig.\ \ref{Ca40-48-spe}(b)). Dashed lines show average trends for different values
of $n$.}
\label{sp-modific}
\end{figure}

\begin{figure}[htb]
\centering
\includegraphics[width=8.5cm]{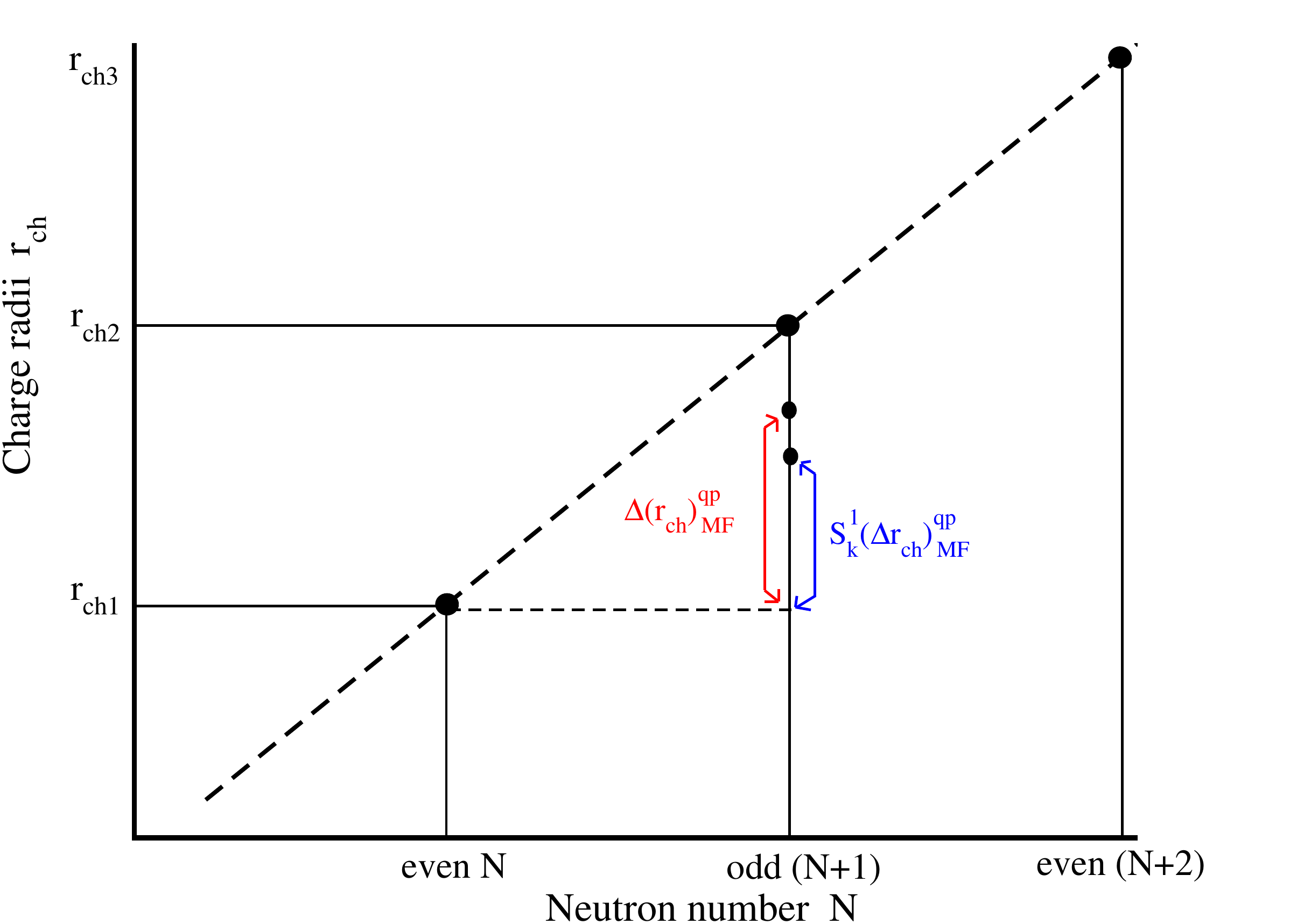}
\caption{Schematic illustration of the impact of pairing and particle-vibration coupling on
charge radii in odd-$N$ nuclei. The dashed straight line corresponds to a linear increase of charge
radii due to a sequential occupation of the single-particle subshell in the calculations without
pairing. $\Delta (r_{ch})^{qp}_{MF}$ corresponds to an increase of the charge radii in odd-$N$
as  compared with the one given in the system with even-$N$ neutrons when one neutron is added
in the calculations with pairing. $S_k^1 \Delta (r_{ch})^{qp}_{MF}$ provides a similar quantity for the
case when the depletion of the single-particle content due to fragmentation is additionally taken into
account (see text for details).}
\label{OES-origin}
\end{figure}

\subsection{Particle-vibration coupling as a source of OES in charge radii}
\label{sec-OES-PVC}

   There is a principal difference between the ground states in even-even and odd nuclei
which is schematically illustrated in Fig.~\ref{EE-O-fragmen} and which has been ignored
in the studies of differential charge radii before. It is related to a substantial fragmentation of
the wavefunction of the ground states
in odd-$A$ nuclei (especially in spherical nuclei) due to the coupling of single-particle motion with phonons
[particle-vibration coupling (PVC)].  In even-even nuclei, the correlations beyond mean-field can
affect the binding energies and equilibrium deformations [and thus the charge radii] of the ground states, but
they do not lead to a significant fragmentation of their wavefunctions in the nuclei with stiff parabola-like PECs 
(such as Pb and Sn isotopes [see, for example, Fig.\ 12 in the supplemental material]). Indeed, such correlations
are rather small in the ground states of the even-even spherical Pb and Sn nuclei \cite{RF.95,FTTZ.00,L.12}
and do not modify their charge radii substantially \cite{RF.95,FTTZ.00}.  Thus, the treatment of the ground states
of such nuclei at the mean-field level represents a reasonable approximation. In this  approximation the physical
observables of interest are defined by the single-particle properties and the occupation probabilities v$^2_{state}(N)$ 
of these states (see left and right columns of Fig.\  \ref{EE-O-fragmen}).

   In contrast, PVC in odd-$A$ nuclei leads to a substantial fragmentation of many single-particle states
(including the ground state) which is experimentally observed (see Refs.\ 
\cite{Broglia1969_NPA127-429,Hamamoto1969_NPA126-545,MBBD.85,Litvinova2006_PRC73-044328,Ring2009_PAN72-1285,
CSB.10,LA.11,AL.15} and references quote therein). As schematically shown in the middle part of Fig.\ \ref{EE-O-fragmen}, 
each mean-field state $k$ ($k=i$ or $j$ in this figure) with energy $\varepsilon_k$ is split into many levels due to
PVC, so the single-particle strength is fragmented over many levels. In the diagonal approximation for the
nucleonic self-energy, these levels have the same quantum numbers as the original mean-field state $k$,
but different energies $\varepsilon^{\nu}_k$ and spectroscopic factors $S^{\nu}_k$.  In the PVC model, the
spectroscopic factors, which are the real numbers between zero and one, play a role of the occupation
probabilities of these fragmented states  satisfying the sum rule $\sum_\nu S_k^{\nu}$=1.  For the states
in the vicinity of the Fermi surface, one dominant level  with $0.5 \leq S_k^{\nu} \leq 1.0$ and many other levels 
with small $S_k^{\nu}$ are usually obtained. Both in experiment and in the calculations, the dominant
single-particle state is typically the lowest in energy among the set of fragmented states originating from
the mean-field state $k$. However, for the mean-field states located far away from the Fermi surface, one observes
a very strong splitting over many levels with much smaller and comparable spectroscopic
factors.

   The detailed global analysis of the impact of the occupation of neutron single-particle
orbitals in the vicinity of spherical neutron shell closures presented in Fig.\ \ref{sp-modific}  
reveals a strong correlation between the principal quantum number $n$ of the single
neutron orbital occupied above the neutron shell closure and the impact of the 
occupation of this orbital on differential charge radii.  One can see that in a given
isotopic chain the largest impact on $\delta \left< r^2 \right>^{N,N'}$ is provided by 
the occupation of the orbital with the lowest $n$.  This feature has already been revealed
for the Pb isotopes in Ref.\ \cite{GSR.13} but the present study generalizes it to a larger
set of the nuclei and exposes new features.  For example, it uncovers that for the $n=1$
orbitals this feature is strictly speaking  true only for the occupation of the orbitals located 
above the shell closure the spin-orbit partner orbitals of which are fully occupied below this 
shell closure. Indeed, this is the case  for the $"j-1/2"$ type $1i_{11/2}$, $1h_{9/2}$, $1g_{7/2}$ 
and $1f_{5/2}$ orbitals the $"j+1/2"$  spin-orbit partners  of which, namely, $1i_{13/2}$, 
$1h_{11/2}$, $1g_{9/2}$  and $1f_{7/2}$, are fully occupied below the  $N'=126$, 82, 50 and 28 
neutron shell closures in the Pb, Sn, Sr and Ca nuclei under study.  However, the
differential charge radius in the Pb isotopes provided by the  occupation of the $n=1$ $1j_{15/2}$ 
orbital located above the $N'=126$ shell closure (see Fig.\ \ref{208Pb-single-particle}(b)) is 
only approximately half of that provided by the $1i_{11/2}$  one and it  is not far away from 
differential charge radii generated by occupation of the $n=2$ orbitals (see Fig.\ \ref{sp-modific}). 
In  addition, Fig.\ \ref{sp-modific} shows that there are no clear correlations between 
$\delta \left< r^2 \right>^{N,N'}$ and  $r_{neu}^{sp}$.  It is also interesting that the 
differential charge radii for a given $n$ only weakly depend both on the mass of the nucleus 
and  $r_{neu}^{sp}$. In addition, within the spin-orbit doublet the occupation of the lower 
lying partner orbital with $j+1/2$ provides smaller differential charge radii than the occupation of the 
higher lying partner orbital with $j-1/2$ (see Fig.\ \ref{sp-modific}). This is because the latter 
ones have large neutron radii as compared with former ones. 

   As a consequence, the single-particle content of unpaired neutron states in odd-$A$ nuclei
plays an important role in our understanding of OES  in charge radii since it defines the pull
on charge densities (see also the discussion in the introduction). One of the ways to modify 
this content is via the pairing interaction (see Sec.\ \ref{sec-OES-pairing}).  Another is via the 
fragmentation of the single-particle states by means of PVC. Indeed it reduces substantially 
(down to 60-90\% \cite{Litvinova2006_PRC73-044328,LA.11,AL.15}) their full single-particle 
content and this fact is experimentally confirmed.   A strict way to calculate the OES effect in 
charge radii in the presence of beyond mean field effects  would be to perform quasiparticle 
random phase approximation (QRPA) calculations in even-even $(Z,N)$ nucleus and then
PVC calculations in odd-$A$ $(Z,N+1)$ nucleus and then to define the differential charge
radius. Note that the PVC calculations in the latter nucleus use the $(Z,N)$ core with QRPA
correlations included and then adds particle-vibration coupling  
\cite{Litvinova2006_PRC73-044328,LA.11,AL.15}.  Existing calculations show that 
away from the vicinity of doubly magic shell closures even-even cores supplemented by QRPA 
correlations behave smoothly as a function of neutron number (see Ref.\ \cite{FTTZ.00})
and their contributions to charge radii are rather modest. Because of these reasons, the addition 
of the neutron is not expected to modify the proton part of the core in the PVC calculations.  The 
detailed  investigation of OES in differential charge radii requires a fully fledged PVC calculations 
which will be undertaken in a future.

   At this point, we want to estimate whether the depletion of the single-particle content of the
wave function in odd-$N$ nuclei due to fragmentation could lead to a right phase (defined as a
the sign of $\Delta \left< r^2 \right>^{(3)}(N)$ at given $N$) and magnitude of OES 
in charge radii. The basic assumptions behind the discussion below are the following. First, we use the fact 
that beyond mean field ground state correlations in even-even spherical nuclei and their impact on charge 
radii are rather small and that they behave smoothly as a function of neutron number (see Refs.\ 
\cite{RF.95,FTTZ.00,L.12}). Second, the wavefunction of the odd-neutron ground state in the odd-$N$ nucleus 
represents a superposition of single-particle and vibrational contributions. The pull on proton densities is 
provided predominantly by the former while the latter is not expected to provide a significant contribution 
to the differential charge radii. This is because these vibrational contributions (i) are the  superposition  of 
two-quasiparticle states the  wavefunctions of which have only a small overlap with that  of the ground 
states of even-even cores and (ii) even-even cores are very similar in neighbouring even-even and
odd-$A$ nuclei.

   In this kind of situation it is reasonable to expect that  (i) the average behavior of charge 
radii as a function of neutron number can be reasonably well approximated by the mean field [since anyway 
it does not provide a contribution to OES of charge radii] and  (ii) only the part of the single neutron in the 
odd-$N$ nucleus defined by the spectroscopic factor $S_k^1$  of the dominant single-particle level 
provides a pull on proton densities and thus the leading contribution to the oscillating part of 
$\Delta \left< r^2 \right>^{(3)}(N)$. The latter leads to a reduction of the increase of charge radii 
in going from the even-$N$ to the odd-$N$ nucleus from $\Delta (r_{ch})^{qp}_{MF}$ at the mean-field level to 
approximately $S_k^1 \Delta (r_{ch})^{qp}_{MF}$  when the fragmentation of the dominant single-particle 
level is taken into account (see Fig.\ \ref{EE-O-fragmen}).

\begin{figure}[htb]
\centering
\includegraphics[width=8.5cm]{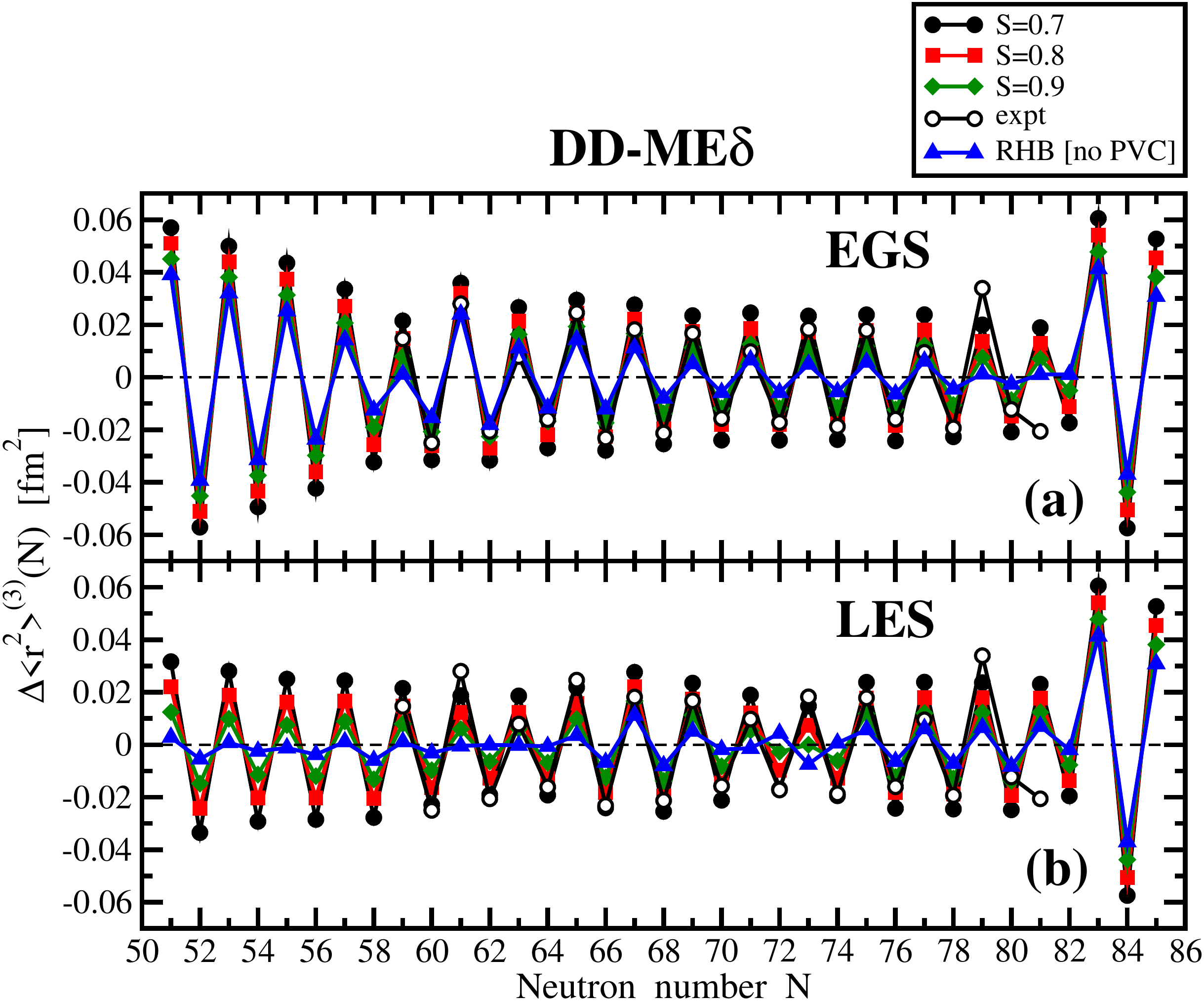}
\includegraphics[width=8.5cm]{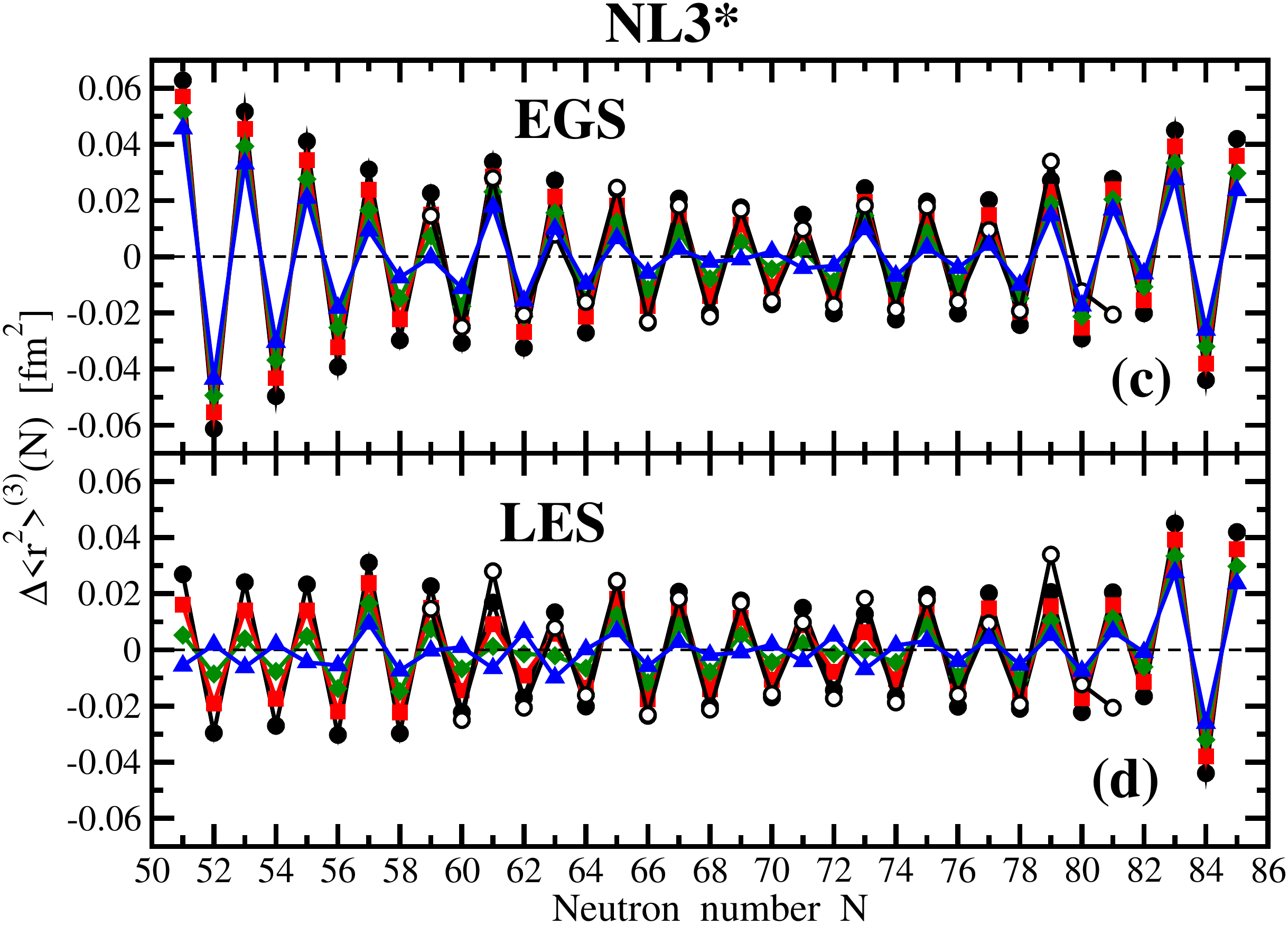}
\caption{OES in charge radii of the Sn isotopes corrected for the fragmentation
of the single-particle content of the dominant single-particle state in odd-$N$ nuclei
by PVC coupling within the framework of a schematic model discussed in the text
for different values of the spectroscopic factor $S=S_k^1$.}
\label{Sn-PVC-corrected}
\end{figure}

   The impact of this modification for representative values of the spectroscopic factors
$S_k^1 =0.9, 0.8$ and 0.7 is illustrated in Fig.\ \ref{Sn-PVC-corrected}. One can see that additional
fragmentation of the structure of the unpaired neutron in odd-$N$ nuclei 
leads to an increase of the magnitudes of OES in charge radii and correct phases of the OES both 
in the LES and EGS procedures. With the spectroscopic factors $S_k^1$  being in the vicinity of 
those calculated in Refs.\ \cite{Litvinova2006_PRC73-044328,LA.11,L.12,AL.15}, the calculated 
OES are close to experimental ones in the Sn isotopes (see Fig.\ \ref{Sn-PVC-corrected}).

     Note that  in some cases, PVC leads to a change of the relative order of the single-particle
states obtained at the mean-field level. Such a possibility is illustrated in Fig.\ \ref{EE-O-fragmen}.
In the odd-$N$ nucleus, the state $i$ is lower in energy than the $j$ state at the MF level, and both states
have single-particle nature. In contrast, in the MF+PVC case, the fragmented level with a dominant
single-particle $j$ state component and the energy $\varepsilon_j^1$ is lower in energy  than the
fragmented level with the dominant single-particle component $i$ and the energy $\varepsilon_i^1$.
This feature has been used in Ref.\ \cite{Pb-Hg-charge-radii-PRL.21} for a simultaneous explanation
of the kink in charge radii at  $N=126$ and the OES in charge radii of the Pb and Hg isotopes with  $N>126$.

\subsection{Other sources of the inversion of OES}
\label{sec-OES-other}

 The inverted OES is also observed in the At  (for $N=133$), Rn (for $N=133-135$),
Fr (for $N=135-138$), Ra (for $N=133-138$ and 140), and Ac (for $N=137$) isotopic
chains [see Fig.\ \ref{OES-even-Z}(t) and (u) and Figs.\ \ref{OES-odd-Z}(o), (p) and (s)].
It is frequently attributed to the effect of octupole deformation (see Refs.\
\cite{Otten.89,  At-radii.19,Ac-radii.19}.
For example, it was suggested in Refs.\ \cite{Ch.80,Lean.84} that the OES inversion originates
from  the fact that octupole deformation should be more pronounced in odd than in
even nuclei. This leads to a charge radius of the odd-$N$ nucleus being larger than
the average charge radius of the two even-even neighbors. However, non-relativistic Skyrme DFT
calculations for the Ac isotopes presented in Ref.\ \cite{Ac-radii.19} show that this is not necessary 
the case since such an inversion appears at some neutron numbers even in the calculations 
without octupole deformation.  Theoretical models also differ in the prediction of static octupole 
deformation in the Rn isotopes (see Table I in Ref.\ \cite{AAR.16} and Ref.\ \cite{CAANO.20}). 
However, experimental data presented in Refs.\ \cite{Rn-Ra-Th.99,220Rn-224Ra.13} strongly 
suggests that $^{218-222}$Rn nuclei behave like octupole vibrators and not like the nuclei with 
static octupole deformation. We also have to keep in mind that the picture of a static octupole 
deformation and the coupling to  dynamic octupole vibrations has much in common. Both models 
describe in many ways, but not completely, the  same physics of static and dynamic polarization 
effects (see, for instance, Ref.\  \cite{SGBGV.89}, where the same problem has 
been discussed in detail for static  pairing correlations and pairing vibrations).

In addition, there are other isotopic chains in which the OES is inverted either locally
(for a few neutron numbers only) or over a substantial range of neutron numbers.
These are isotopic chains of Kr, Sr, Rb, Ba, Sm, Cs, Yb, Hf, Eu, Lu, and Ir
 [see Figs.\ref{OES-even-Z}(e), (f), (j), (l), (n), and (o) and Figs.\ \ref{OES-odd-Z}(c), (e),
(f), (j) and (k)]. The octupole deformation is not present in the ground states of these
nuclei, so there should be other sources of the OES inversion different from octupole
deformation.  For example, it was speculated in Ref.\ \cite{Hr-Sr-OES.96} that OES
in charge radii of light Kr, Rb, and Sr nuclei is due to a polarization effect of the even-even core 
by the unpaired neutron, driving the odd-$N$ nuclei toward stronger quadrupole deformation [as compared
with the average given by even-$N$ neighbors]. However, this was not supported by any model calculation.

  Note that all above discussed cases involve deformed nuclei, and these isotopic
chains include both odd and odd-odd nuclei. Because of the complexity of the
description of such nuclei (see Refs.\ \cite{AA.10,AS.11}) a detailed investigation
of OES and its inversion in these isotopic chains goes beyond the
scope of the present paper, but it is planned for the future.

\section{Conclusions}
\label{Concl}

   A systematic global investigation of differential charge radii has been performed
within the CDFT framework for the first time. Theoretical results obtained with
conventional covariant energy density functionals and the separable pairing
interaction of Ref.\ \cite{TMR.09} are compared with experimental differential charge radii
in the regions of the nuclear chart where available experimental data crosses neutron
shell closures at $N = 28, 50, 82$, and 126. The main results can be summarized
as follows:

\begin{itemize}
\item
In spherical nuclei, the appearance of the kinks in the $\delta \left < r^2 \right>^{N,N'}$  
curves at neutron shell closures is defined predominantly in the particle-hole channel of the 
CDFT with details of the single-particle structure above shell closures playing an important role.
This conclusion is different from the one obtained in nonrelativistic Skyrme and 
Fayans DFTs in Ref.\ \cite{Gorges-Sn-radii.19} which indicates that pairing is the 
dominant contributor to the kink. In the RHB approach, the kinks are already present in 
the calculations without pairing.  Pairing acts only as an additional tool that mixes
different configurations and somewhat softens the evolution of charge radii
as a function of neutron number.

\item
  The relative energies of the single-particle states and the patterns of their
occupation with increasing neutron number have a significant impact on
the evolution of the $\delta \left < r^2 \right>^{N,N'}$ values even in  the
calculations with pairing included.  Considering existing inaccuracies in the
description of the energies of the single-particle states in the DFT
calculations, the predictive power of such models for $\delta \left < r^2 \right>^{N,N'}$
is expected to decrease in the regions of high densities of the single-particle states
of different origin.

\item
   The analysis of absolute differential radii of different isotopic chains and
their relative properties clearly indicate that such properties are reasonably well
described in model calculations in cases where the mean-field approximation is
justified. The analysis of potential energy curves provides the latter justification.
However, it turns out that it is more difficult to describe the clusterization
of the differential charge radii in the Sn and Ca regions for neutron numbers
above shell closures at $N=82$ and 28 since it depends on the details of the underlying
single-particle structure.

\item
There are regions of the nuclear chart where the description at the mean-field
level faces difficulties in reproducing experimental data.  In the CDFT 
calculations, these are the Ca
isotopes,  the $N<50$ and $N>58$ nuclei in the Sr region and the neutron-poor
nuclei in the Pb region. The latter two regions are characterized by shape coexistence,
and, in many cases, the assignment of the calculated excited prolate minimum to the experimental
ground-state allows understanding the trends of the evolution of differential charge
radii with neutron number. The inclusion of beyond mean-field effects
could possibly improve the description of charge radii is these systems. As follows from the
comparison of the calculated and experimental masses in Ref.\ \cite{AARR.14} and
from increased [as compared with heavy nuclei] OES in charge radii of very light nuclei
[see Figs.\ \ref{OES-even-Z}(a), (b) and (c) and Figs.\ \ref{OES-odd-Z}(a)], such effects
are expected to play a significant role in the properties of the ground states of light nuclei.
That was a  reason why light nuclei have been excluded from the analysis in the present
paper.

\item
  It is usually assumed that pairing is the dominant contributor to OES in charge radii.
Our analysis paints a more complicated  picture and suggests a new 
additional mechanism where 
the fragmentation of the single-particle content of the ground state in odd-mass nuclei 
due to particle-vibration coupling provides a significant contribution to OES in charge 
radii.  Note that similarly, the pairing indicators, which depend on 
OES of binding energies, are also expected to be affected by particle-vibration coupling
with its impact to be especially pronounced in spherical nuclei (see Ref.\ \cite{TA.21}).

\item
    The PECs curves obtained in the calculations with the CEDF 
DD-ME$\delta$\footnote{Note that the DD-ME$\delta$ CEDF provides a reasonable 
global description of the ground state properties (see Ref.\ \cite{AARR.14}), but it fails to predict 
octupole deformed actinides  \cite{AAR.16} and generates fission barriers in superheavy nuclei 
which are too small to make them relatively stable \cite{AARR.17}.  Thus, it is not recommended 
for applications to nuclei heavier than lead.} quite 
frequently deviate from those obtained with DD-ME2, DD-PC1, NL3* and PC-PK1.
This could also affect the results of beyond mean field calculations making 
them in some nuclei significantly dissimilar for DD-ME$\delta$ as compared with
above mentioned functionals.

This difference could be due to two factors, making DD-ME$\delta$ substantially different from the 
other conventional CDFTs:

First, DD-ME$\delta$ has less fit parameters, and therefore the adjustment of this CEDF could be 
less successful: DD-ME$\delta$ is the most microscopic CEDF. Only four parameters at the  saturation 
density are fitted to finite nuclei and the full density dependence of the parameters is derived from 
ab-initio calculations\ \cite{DD-MEdelta}. On the contrary, the other interactions contain an  additional 
2 (NL3*), 4 (DD-ME2), 5 (PC-PK1), and 6 (DD-PC1)  phenomenological parameters for the fine-tuning 
of different channels of CEDFs and their density dependence. Note that not all of these additional 
parameters  are independent (see Refs.\ \cite{AAT.19,TAAR.20}).

Second, in addition to the three spin-isospin channels represented by the $\sigma$-, $\omega$-, and 
$\rho$-meson, DD-ME$\delta$ also contains, as the only parameter set considered here, an isovector 
scalar channel represented by the $\delta$-meson. This influences the isospin dependence of the 
spin-orbit field and, therefore, that of the single-particle energies. However, it is practically impossible 
to adjust the parameters of the $\delta$-meson to experimental data because (i) there is very little data 
on the isospin dependence of single-particle energies and the largely unknown influence of tensor forces 
and of particle vibrational coupling \ \cite{AL.15} forbids the fitting to single-particle levels anyhow, 
and (ii) it has been shown in Ref. \cite{DD-MEdelta}, that the parameters of the $\delta$-meson cannot 
be determined by fitting to the usual bulk properties of finite nuclei, because here the changes in the 
parameters of the $\delta$-meson are completely compensated by corresponding changes in the 
remaining isovector channel, i.e. by the $\rho$-meson\ \cite{DD-MEdelta}. Therefore for the CEDF  
DD-ME$\delta$, in Ref.\ \cite{DD-MEdelta}, the strength and the density dependence of the $\delta$-nucleon 
vertex have been adjusted  to ab-initio results, i.e. to the isovector effective mass $m^*_p-m^*_n$, 
derived from relativistic Brueckner theory in Ref.\ \cite{VanDalen2007_EPJA31-29}. These relativistic 
Brueckner-Hartree-Fock (RBHF) calculations suffer from the fact that the Thompson equation 
has not been treated in full Dirac space, and the coupling to negative energy solutions is only 
treated approximately. Only recently the RBHF calculations for symmetric nuclear matter have been 
carried out in full Dirac space \cite{WANG-SB2021_PRC103-054319}, but corresponding solutions for 
asymmetric nuclear matter are still missing.

\end{itemize}

\section{ACKNOWLEDGMENTS}

 This material is based upon work supported by the U.S. Department of Energy,  
Office of Science, Office of Nuclear Physics under Award No. DE-SC0013037. 
PR acknowledges partial support from the Deutsche Forschungsgemeinschaft 
(DFG, German Research Foundation) under under Germany Excellence 
Strategy EXC-2094-390783311, ORIGINS.

\bibliography{references-30-PRC-global-charge-radii-my}
\end{document}